\documentclass[twocolumn,letterpaper,aps,prc,longbibliography,superscriptaddress,showpacs,nofootinbib,floatfix]{revtex4-1}
\usepackage{graphicx}	% Include figure filed
\usepackage{amsmath}
\usepackage{multirow}
\usepackage{xspace}	% Include xspace

\newcommand{\pt}{\mbox{$p_T$}\xspace}

\newcommand{\raa}{\mbox{$R_{AA}$}\xspace}

\newcommand{\Ncoll}{\mbox{$N_{\rm coll}$}\xspace}

\newcommand{\sqs}{\mbox{$\sqrt{s}$}\xspace}
\newcommand{\sqsn}{\mbox{$\sqrt{s_{_{NN}}}$}\xspace}
\newcommand{\sqsntwo}{\mbox{$\sqrt{s_{_{NN}}}=200$~GeV}\xspace}
\newcommand{\pp}{\mbox{$p$$+$$p$}\xspace}
\newcommand{\dau}{\mbox{$d$$+$Au}\xspace}

\newcommand{\auau}{\mbox{Au$+$Au}\xspace}
\newcommand{\pythia}{\mbox{\sc pythia}\xspace}
\newcommand{\geant}{\mbox{\sc geant3}\xspace}
\newcommand{\fonll}{\mbox{\sc fonll}\xspace}
\newcommand{\pbpb}{\mbox{Pb$+$Pb}\xspace}
\newcommand{\cucu}{\mbox{Cu$+$Cu}\xspace}

\newcommand{\D}{$D$\xspace}
\newcommand{\B}{$B$\xspace}
\newcommand{\DCAR}{\mbox{$\mathrm{DCA}_{T}$}\xspace}
\newcommand{\DCAZ}{\mbox{$\mathrm{DCA}_{L}$}\xspace}
\newcommand{\midy}{\mbox{$|y|<0.35$}\xspace}
\newcommand{\gev}{\mbox{GeV/$c$}\xspace}
\newcommand{\pte}{\mbox{$p_T^e$}\xspace}
\newcommand{\ptc}{\mbox{$p_T^c$}\xspace}
\newcommand{\ptb}{\mbox{$p_T^b$}\xspace}
\newcommand{\pth}{\mbox{$p_T^h$}\xspace}

\newcommand{\xvec}{\mathbf{x}}

\newcommand{\rvec}{\mathbf{R}}
\newcommand{\thetavec}{\boldsymbol\theta}
\newcommand{\prior}{\pi(\thetavec)}
\newcommand{\post}{p(\thetavec|\xvec)}
\newcommand{\like}{P(\xvec|\thetavec)}

\newcommand{\eptdata}{\mathbf{Y}^{\rm data}}
\newcommand{\dcadata}[1]{\mathbf{D}_{#1}^{\rm data}}
\newcommand{\epttheta}{\mathbf{Y}(\thetavec)}
\newcommand{\dcatheta}[1]{\mathbf{D}_{#1}(\thetavec)}
\newcommand{\My}{\mathbf{M}^{\mathbf{(Y)}}}
\newcommand{\Md}{\mathbf{M}_{j}^{\mathbf{(D)}}}

\begin{document}

\title{Single electron yields from semileptonic charm and bottom hadron 
decays in Au$+$Au collisions at $\sqrt{s_{NN}}=200$~GeV}

\newcommand{\abilene}{Abilene Christian University, Abilene, Texas 79699, USA}
\newcommand{\augie}{Department of Physics, Augustana University, Sioux Falls, South Dakota 57197, USA}
\newcommand{\banaras}{Department of Physics, Banaras Hindu University, Varanasi 221005, India}
\newcommand{\barc}{Bhabha Atomic Research Centre, Bombay 400 085, India}
\newcommand{\baruch}{Baruch College, City University of New York, New York, New York, 10010 USA}
\newcommand{\bnlcoll}{Collider-Accelerator Department, Brookhaven National Laboratory, Upton, New York 11973-5000, USA}
\newcommand{\bnlphys}{Physics Department, Brookhaven National Laboratory, Upton, New York 11973-5000, USA}
\newcommand{\caucr}{University of California-Riverside, Riverside, California 92521, USA}
\newcommand{\charlesczech}{Charles University, Ovocn\'{y} trh 5, Praha 1, 116 36, Prague, Czech Republic}
\newcommand{\chonbuk}{Chonbuk National University, Jeonju, 561-756, Korea}
\newcommand{\ciae}{Science and Technology on Nuclear Data Laboratory, China Institute of Atomic Energy, Beijing 102413, P.~R.~China}
\newcommand{\cns}{Center for Nuclear Study, Graduate School of Science, University of Tokyo, 7-3-1 Hongo, Bunkyo, Tokyo 113-0033, Japan}
\newcommand{\colorado}{University of Colorado, Boulder, Colorado 80309, USA}
\newcommand{\columbia}{Columbia University, New York, New York 10027 and Nevis Laboratories, Irvington, New York 10533, USA}
\newcommand{\czechtech}{Czech Technical University, Zikova 4, 166 36 Prague 6, Czech Republic}
\newcommand{\debrecen}{Debrecen University, H-4010 Debrecen, Egyetem t{\'e}r 1, Hungary}
\newcommand{\elte}{ELTE, E{\"o}tv{\"o}s Lor{\'a}nd University, H-1117 Budapest, P{\'a}zm{\'a}ny P.~s.~1/A, Hungary}
\newcommand{\ewha}{Ewha Womans University, Seoul 120-750, Korea}
\newcommand{\fsu}{Florida State University, Tallahassee, Florida 32306, USA}
\newcommand{\gsu}{Georgia State University, Atlanta, Georgia 30303, USA}
\newcommand{\hanyang}{Hanyang University, Seoul 133-792, Korea}
\newcommand{\hiroshima}{Hiroshima University, Kagamiyama, Higashi-Hiroshima 739-8526, Japan}
\newcommand{\howard}{Department of Physics and Astronomy, Howard University, Washington, DC 20059, USA}
\newcommand{\ihepprot}{IHEP Protvino, State Research Center of Russian Federation, Institute for High Energy Physics, Protvino, 142281, Russia}
\newcommand{\illuiuc}{University of Illinois at Urbana-Champaign, Urbana, Illinois 61801, USA}
\newcommand{\inrras}{Institute for Nuclear Research of the Russian Academy of Sciences, prospekt 60-letiya Oktyabrya 7a, Moscow 117312, Russia}
\newcommand{\instpasczech}{Institute of Physics, Academy of Sciences of the Czech Republic, Na Slovance 2, 182 21 Prague 8, Czech Republic}
\newcommand{\isu}{Iowa State University, Ames, Iowa 50011, USA}
\newcommand{\jaea}{Advanced Science Research Center, Japan Atomic Energy Agency, 2-4 Shirakata Shirane, Tokai-mura, Naka-gun, Ibaraki-ken 319-1195, Japan}
\newcommand{\jyvaskyla}{Helsinki Institute of Physics and University of Jyv{\"a}skyl{\"a}, P.O.Box 35, FI-40014 Jyv{\"a}skyl{\"a}, Finland}
\newcommand{\karoly}{K\'aroly R\'oberts University College, H-3200 Gy\"ngy\"os, M\'atrai\'ut 36, Hungary}
\newcommand{\kek}{KEK, High Energy Accelerator Research Organization, Tsukuba, Ibaraki 305-0801, Japan}
\newcommand{\korea}{Korea University, Seoul, 136-701, Korea}
\newcommand{\kurchatov}{National Research Center ``Kurchatov Institute", Moscow, 123098 Russia}
\newcommand{\kyoto}{Kyoto University, Kyoto 606-8502, Japan}
\newcommand{\labllr}{Laboratoire Leprince-Ringuet, Ecole Polytechnique, CNRS-IN2P3, Route de Saclay, F-91128, Palaiseau, France}
\newcommand{\lahorelums}{Physics Department, Lahore University of Management Sciences, Lahore 54792, Pakistan}
\newcommand{\lawllnl}{Lawrence Livermore National Laboratory, Livermore, California 94550, USA}
\newcommand{\losalamos}{Los Alamos National Laboratory, Los Alamos, New Mexico 87545, USA}
\newcommand{\lpc}{LPC, Universit{\'e} Blaise Pascal, CNRS-IN2P3, Clermont-Fd, 63177 Aubiere Cedex, France}
\newcommand{\lund}{Department of Physics, Lund University, Box 118, SE-221 00 Lund, Sweden}
\newcommand{\maryland}{University of Maryland, College Park, Maryland 20742, USA}
\newcommand{\mass}{Department of Physics, University of Massachusetts, Amherst, Massachusetts 01003-9337, USA}
\newcommand{\michigan}{Department of Physics, University of Michigan, Ann Arbor, Michigan 48109-1040, USA}
\newcommand{\muhlenberg}{Muhlenberg College, Allentown, Pennsylvania 18104-5586, USA}
\newcommand{\myongji}{Myongji University, Yongin, Kyonggido 449-728, Korea}
\newcommand{\nagasaki}{Nagasaki Institute of Applied Science, Nagasaki-shi, Nagasaki 851-0193, Japan}
\newcommand{\nara}{Nara Women's University, Kita-uoya Nishi-machi Nara 630-8506, Japan}
\newcommand{\natmephi}{National Research Nuclear University, MEPhI, Moscow Engineering Physics Institute, Moscow, 115409, Russia}
\newcommand{\newmex}{University of New Mexico, Albuquerque, New Mexico 87131, USA}
\newcommand{\nmsu}{New Mexico State University, Las Cruces, New Mexico 88003, USA}
\newcommand{\ohio}{Department of Physics and Astronomy, Ohio University, Athens, Ohio 45701, USA}
\newcommand{\ornl}{Oak Ridge National Laboratory, Oak Ridge, Tennessee 37831, USA}
\newcommand{\orsay}{IPN-Orsay, Univ. Paris-Sud, CNRS/IN2P3, Universit\'e Paris-Saclay, BP1, F-91406, Orsay, France}
\newcommand{\peking}{Peking University, Beijing 100871, P.~R.~China}
\newcommand{\pnpi}{PNPI, Petersburg Nuclear Physics Institute, Gatchina, Leningrad region, 188300, Russia}
\newcommand{\riken}{RIKEN Nishina Center for Accelerator-Based Science, Wako, Saitama 351-0198, Japan}
\newcommand{\rikjrbrc}{RIKEN BNL Research Center, Brookhaven National Laboratory, Upton, New York 11973-5000, USA}
\newcommand{\rikkyo}{Physics Department, Rikkyo University, 3-34-1 Nishi-Ikebukuro, Toshima, Tokyo 171-8501, Japan}
\newcommand{\saispbstu}{Saint Petersburg State Polytechnic University, St.~Petersburg, 195251 Russia}
\newcommand{\saopaulo}{Universidade de S{\~a}o Paulo, Instituto de F\'{\i}sica, Caixa Postal 66318, S{\~a}o Paulo CEP05315-970, Brazil}
\newcommand{\seoulnat}{Department of Physics and Astronomy, Seoul National University, Seoul 151-742, Korea}
\newcommand{\stonybrkc}{Chemistry Department, Stony Brook University, SUNY, Stony Brook, New York 11794-3400, USA}
\newcommand{\stonycrkp}{Department of Physics and Astronomy, Stony Brook University, SUNY, Stony Brook, New York 11794-3800, USA}
\newcommand{\tenn}{University of Tennessee, Knoxville, Tennessee 37996, USA}
\newcommand{\titech}{Department of Physics, Tokyo Institute of Technology, Oh-okayama, Meguro, Tokyo 152-8551, Japan}
\newcommand{\tsukuba}{Center for Integrated Research in Fundamental Science and Engineering, University of Tsukuba, Tsukuba, Ibaraki 305, Japan}
\newcommand{\vandy}{Vanderbilt University, Nashville, Tennessee 37235, USA}
\newcommand{\weizmann}{Weizmann Institute, Rehovot 76100, Israel}
\newcommand{\wigner}{Institute for Particle and Nuclear Physics, Wigner Research Centre for Physics, Hungarian Academy of Sciences (Wigner RCP, RMKI) H-1525 Budapest 114, POBox 49, Budapest, Hungary}
\newcommand{\yonsei}{Yonsei University, IPAP, Seoul 120-749, Korea}
\newcommand{\zagreb}{University of Zagreb, Faculty of Science, Department of Physics, Bijeni\v{c}ka 32, HR-10002 Zagreb, Croatia}
\affiliation{\abilene}
\affiliation{\augie}
\affiliation{\banaras}
\affiliation{\barc}
\affiliation{\baruch}
\affiliation{\bnlcoll}
\affiliation{\bnlphys}
\affiliation{\caucr}
\affiliation{\charlesczech}
\affiliation{\chonbuk}
\affiliation{\ciae}
\affiliation{\cns}
\affiliation{\colorado}
\affiliation{\columbia}
\affiliation{\czechtech}
\affiliation{\debrecen}
\affiliation{\elte}
\affiliation{\ewha}
\affiliation{\fsu}
\affiliation{\gsu}
\affiliation{\hanyang}
\affiliation{\hiroshima}
\affiliation{\howard}
\affiliation{\ihepprot}
\affiliation{\illuiuc}
\affiliation{\inrras}
\affiliation{\instpasczech}
\affiliation{\isu}
\affiliation{\jaea}
\affiliation{\jyvaskyla}
\affiliation{\karoly}
\affiliation{\kek}
\affiliation{\korea}
\affiliation{\kurchatov}
\affiliation{\kyoto}
\affiliation{\labllr}
\affiliation{\lahorelums}
\affiliation{\lawllnl}
\affiliation{\losalamos}
\affiliation{\lpc}
\affiliation{\lund}
\affiliation{\maryland}
\affiliation{\mass}
\affiliation{\michigan}
\affiliation{\muhlenberg}
\affiliation{\myongji}
\affiliation{\nagasaki}
\affiliation{\nara}
\affiliation{\natmephi}
\affiliation{\newmex}
\affiliation{\nmsu}
\affiliation{\ohio}
\affiliation{\ornl}
\affiliation{\orsay}
\affiliation{\peking}
\affiliation{\pnpi}
\affiliation{\riken}
\affiliation{\rikjrbrc}
\affiliation{\rikkyo}
\affiliation{\saispbstu}
\affiliation{\saopaulo}
\affiliation{\seoulnat}
\affiliation{\stonybrkc}
\affiliation{\stonycrkp}
\affiliation{\tenn}
\affiliation{\titech}
\affiliation{\tsukuba}
\affiliation{\vandy}
\affiliation{\weizmann}
\affiliation{\wigner}
\affiliation{\yonsei}
\affiliation{\zagreb}
\author{A.~Adare} \affiliation{\colorado} 
\author{C.~Aidala} \affiliation{\losalamos} \affiliation{\michigan} 
\author{N.N.~Ajitanand} \affiliation{\stonybrkc} 
\author{Y.~Akiba} \affiliation{\riken} \affiliation{\rikjrbrc} 
\author{R.~Akimoto} \affiliation{\cns} 
\author{J.~Alexander} \affiliation{\stonybrkc} 
\author{M.~Alfred} \affiliation{\howard} 
\author{K.~Aoki} \affiliation{\kek} \affiliation{\riken} 
\author{N.~Apadula} \affiliation{\isu} \affiliation{\stonycrkp} 
\author{Y.~Aramaki} \affiliation{\cns} \affiliation{\riken} 
% ...
\collaboration{PHENIX Collaboration} \noaffiliation

\date{\today}

%------------------------------------------------------------------------------|

\begin{abstract}

%\linenumbers

The PHENIX Collaboration at the Relativistic Heavy Ion Collider has 
measured open heavy flavor production in minimum bias Au$+$Au collisions 
at $\sqrt{s_{_{NN}}}=200$~GeV via the yields of electrons from 
semileptonic decays of charm and bottom hadrons. Previous heavy flavor 
electron measurements indicated substantial modification in the momentum 
distribution of the parent heavy quarks due to the quark-gluon plasma 
created in these collisions.  For the first time, using the PHENIX silicon 
vertex detector to measure precision displaced tracking, the relative 
contributions from charm and bottom hadrons to these electrons as a 
function of transverse momentum are measured in Au$+$Au collisions.  We 
compare the fraction of electrons from bottom hadrons to previously 
published results extracted from electron-hadron correlations in $p$$+$$p$ 
collisions at $\sqrt{s_{_{NN}}}=200$~GeV and find the fractions to be 
similar within the large uncertainties on both measurements for 
$p_T>4$~GeV/$c$. We use the bottom electron fractions in Au$+$Au and 
$p$$+$$p$ along with the previously measured heavy flavor electron 
$R_{AA}$ to calculate the $R_{AA}$ for electrons from charm and bottom 
hadron decays separately. We find that electrons from bottom hadron decays 
are less suppressed than those from charm for the region $3<p_T<4$~GeV/$c$.

\end{abstract}

	
\maketitle

%%=====================================================================
	\section{Introduction}
	\label{sec:introduction}

High-energy heavy ion collisions at the Relativistic Heavy Ion Collider 
(RHIC) and the Large Hadron Collider (LHC) create matter that is well 
described as an equilibrated system with initial temperatures in excess of 
340--420 
MeV~\cite{Adcox:2004mh,Adams:2005dq,Adare:2008ab,Romatschke:2009im,Heinz:2013th}.  
In this regime, the matter is understood to be a quark-gluon plasma (QGP) 
with bound hadronic states no longer in existence as the temperatures far 
exceed the transition temperature of approximately 155~MeV calculated by 
lattice quantum chromodynamics (QCD)~\cite{Bazavov:2014pvz}. This QGP 
follows hydrodynamical flow 
behavior with extremely small dissipation, characterized by the shear 
viscosity to entropy density ratio $\eta/s \approx 1/4\pi$ and is thus 
termed a near-perfect 
fluid~\cite{Kovtun:2004de,Gyulassy:2004zy,Heinz:2005zg,Adcox:2004mh}.

Charm and bottom quarks ($m_{c} \approx 1.3$~GeV/$c^{2}$ and $m_{b} 
\approx 4.2$~GeV/$c^{2}$) are too heavy to be significantly produced via 
the interaction of thermal particles in the QGP.  Thus the dominant 
production mechanism is via hard interactions between partons in the 
incoming nuclei, i.e. interactions that involve large momentum transfer, 
$q^2$.  Once produced, these heavy quarks are not destroyed by the strong 
interaction and thus propagate through the QGP and eventually emerge in 
heavy flavor hadrons, for example \D and \B mesons.

Early measurement of heavy flavor electrons from the PHENIX Collaboration 
in \auau collisions at RHIC indicated that although the total heavy flavor 
production scales with the number of binary collisions within 
uncertainties~\cite{Adler:2004ta,Adcox:2002cg}, the momentum distribution 
of these heavy quarks is significantly modified when compared with that in 
\pp collisions~\cite{Adare:2010de,Adare:2006nq}.  These results indicate a 
large suppression for high-\pt $> 5$ \gev electrons and a substantial 
elliptic flow for \pt $= 0.3$--$3.0$ \gev electrons from heavy quark decays. 
Here, and throughout the paper, we use ``electrons'' to refer to both 
electrons and positrons. The suppression of the charm quark has since been 
confirmed through the direct reconstruction of \D mesons by the STAR 
Collaboration~\cite{Adamczyk:2014uip}. In \pbpb collisions at the LHC at 
\sqsn = 2.76~TeV, similar momentum distribution modifications of heavy 
flavor electrons and \D mesons have been 
measured~\cite{Abelev:2014ipa,ALICE:2012ab}. Recently, the CMS experiment 
has reported first measurements of $B \rightarrow J/\psi$ 
~\cite{Chatrchyan:2012np} and b-jets~\cite{Chatrchyan:2013exa} in \pbpb 
collisions. In contrast to this suppression pattern found in \auau 
collisions, \dau and peripheral \cucu collisions at \sqsntwo exhibit an 
enhancement at intermediate electron \pt in the heavy flavor electron 
spectrum~\cite{Adare:2013yxp,Adare:2012yxa} that must be understood in 
terms of a mechanism that enhances the \pt spectrum, e.g. the Cronin 
effect~\cite{Antreasyan:1978cw}. That mechanism potentially moderates the large suppression 
observed in \auau collisions at \sqsntwo. It is notable that in central 
\auau collisions at \sqsn = 62~GeV an enhancement is also observed at 
intermediate \pt~\cite{Adare:2014rly}.

The possibility that charm quarks follow the QGP flow was postulated early 
on~\cite{Batsouli:2002qf}, and more detailed Langevin-type calculations 
with drag and diffusion of these heavy quarks yield a reasonable 
description of the electron 
data~\cite{Moore:2004tg,Cao:2013ita,Rapp:2009my,vanHees:2007me,Gossiaux:2010yx,Adare:2013wca}. 
Many of these theory calculations incorporate radiative and collisional 
energy loss of the heavy quarks in the QGP that are particularly important 
at high-\pt, where QGP flow effects are expected to be sub-dominant.  The 
large suppression of heavy flavor electrons extending up to \pt $\approx 
9$ \gev has been a particular challenge to understand theoretically, in 
part due to an expected suppression of radiation in the direction of the 
heavy quarks propagation -- often referred to as the ``dead-cone'' 
effect~\cite{Dokshitzer:2001zm}.

This observation of the high-\pt 
suppression~\cite{Djordjevic:2014yka,Djordjevic:2006kw} is all the more 
striking because perturbative QCD (pQCD) calculations indicate a 
substantial contribution from bottom quark decays for \pt $> 5$ 
\gev~\cite{Cacciari:2005rk}.  First measurements in \pp collisions at 200 
GeV via electron-hadron correlations confirm this expected bottom 
contribution to the electrons that increases as a function of 
\pt~\cite{Adare:2009ic,Aggarwal:2010xp}. To date, there are no direct 
measurements at RHIC of the contribution of bottom quarks in \auau 
collisions.

For the specific purpose of separating the contributions of charm and 
bottom quarks at midrapidity, the PHENIX Collaboration has added 
micro-vertexing capabilities in the form of a silicon vertex tracker 
(VTX).  The different lifetimes and kinematics for charm and bottom 
hadrons decaying to electrons enables separation of their contributions 
with measurements of displaced tracks (i.e. the decay electron not 
pointing back to the collision vertex).  In this paper, we report on first 
results of separated charm and bottom yields via single electrons in 
minimum bias (MB) \auau collisions at \sqsntwo.

%%=================================================================
	\section{PHENIX Detector}
	\label{sec:phenix}

As detailed in Ref.~\cite{Adcox:2003zm}, the PHENIX detector was originally 
designed with precision charged particle reconstruction 
combined with excellent electron identification.   In 2011, the VTX was 
installed thus enabling micro-vertexing capabilities.   The dataset utilized
in this analysis comprises \auau collisions at \sqsntwo.   

\subsection{Global detectors and MB trigger}

A set of global event-characterization detectors are utilized to select 
\auau events and eliminate background contributions. Two beam-beam 
counters (BBC) covering pseudorapidity $3.0 < |\eta| < 3.9$ and full 
azimuth are located at $\pm$ 1.44 meters along the beam axis and relative 
to the nominal beam-beam collision point. Each of the BBCs comprises 64 
\v{C}erenkov counters.

Based on the coincidence of the BBCs, \auau collisions are selected via an 
online MB trigger, which requires at least two counters on each side of 
the BBC to fire. The MB sample covers $96\pm3$\% of the total inelastic 
\auau cross section as determined by comparison with Monte Carlo Glauber 
models~\cite{Miller:2007ri}.  The BBC detectors also enable a selection on 
the $z$-vertex position of the collision as determined by the 
time-of-flight difference between hits in the two sets of BBC counters.  
The $z$-vertex resolution of the BBC is approximately $\sigma_{z} = 0.6$ 
cm in central \auau collisions.  A selection within approximately $\pm 12$ 
cm of the nominal detector center was implemented and $\sim$ 85\% of all 
\auau collisions within that selection were recorded by the PHENIX 
high-bandwidth data acquisition system.

\subsection{The central arms}

\begin{figure}[!hbt]
	\includegraphics[width=1.0\linewidth]{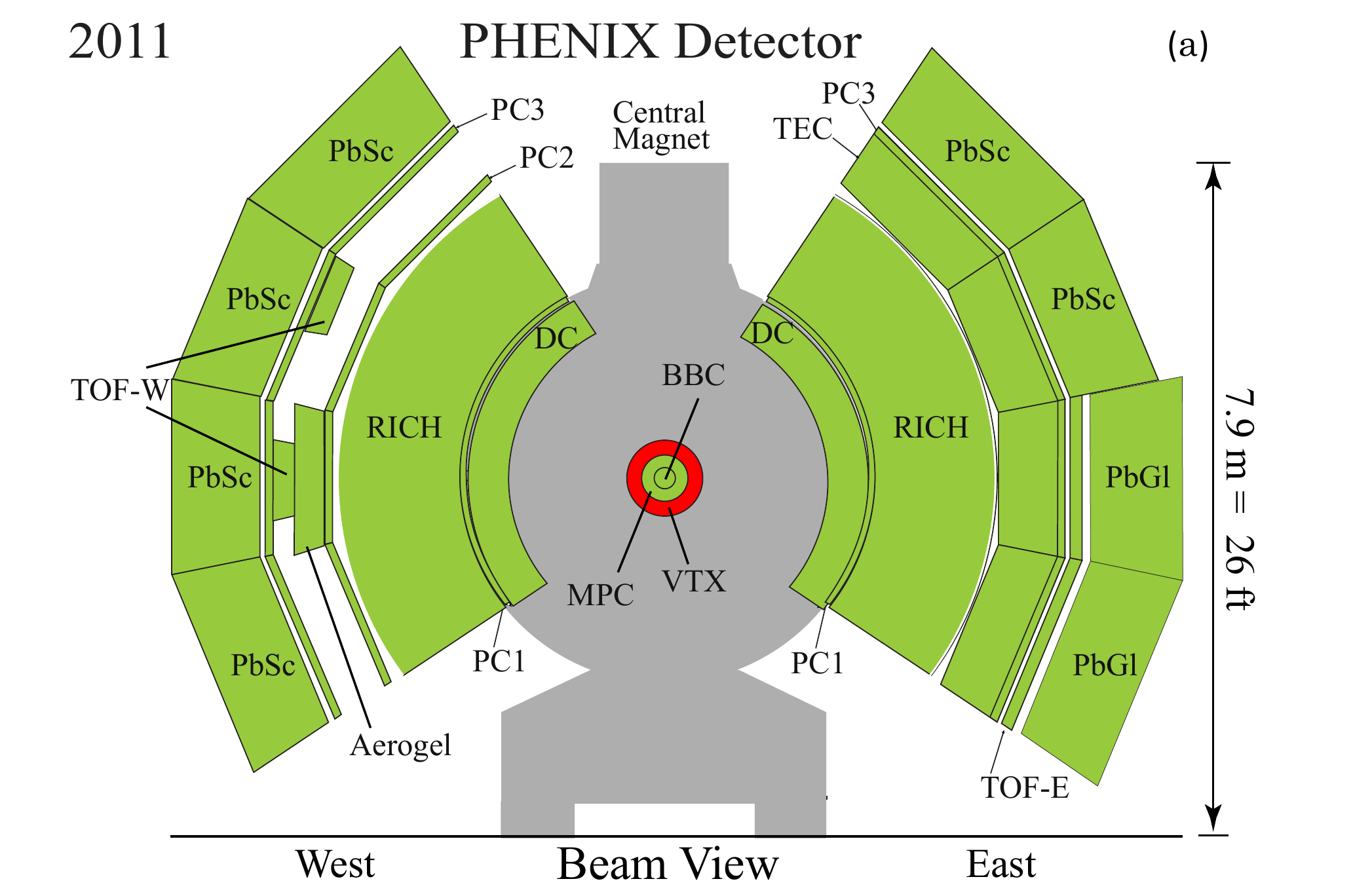}
	\includegraphics[width=0.8\linewidth]{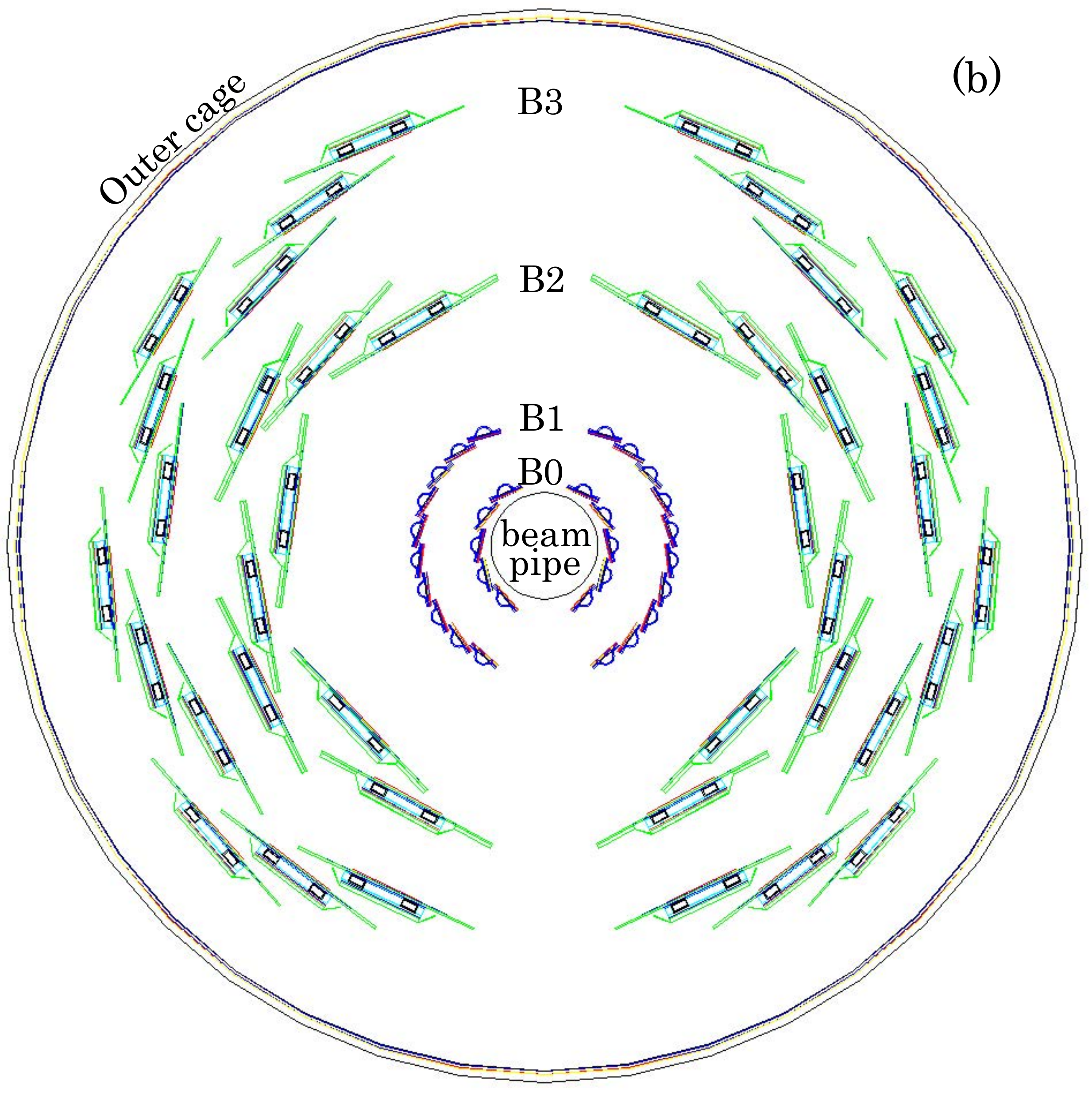}
	\caption{\label{fig:phenix}(Color Online)  
(a) A schematic view of the PHENIX detector configuration for the 2011 run.
(b) A schematic view of the VTX detector with the individual ladders shown.}
\end{figure}

Electrons ($e^+$ and $e^-$) are reconstructed using two central 
spectrometer arms as shown in Fig.~\ref{fig:phenix}(a), each of which 
covers the pseudorapidity range $|\eta|<0.35$ and with azimuthal angle 
$\Delta\phi = \pi/2$. The detector configuration of the central arms is 
the same as in previous PHENIX Collaboration heavy flavor electron 
publications~\cite{Adare:2010de,Adare:2006nq}. Charged particle tracks are 
reconstructed outside of an axial magnetic field using layers of drift 
chamber (DC) and multi-wire proportional pad chambers (PC).  The momentum 
resolution is $\sigma_p/p \simeq$~0.7\% $\oplus$~0.9\%~$p$~(\gev). For 
central arm charged particle reconstructions the trajectory is only 
measured for radial positions $r > 2.02$~meters, and the momentum vector 
is calculated by assuming the track originates at the \auau collision 
point determined by the BBC detectors and assuming 0 radial distance.

Electron identification is performed by hits in a ring imaging 
\v{C}erenkov detector (RICH) and a confirming energy deposit in an 
electromagnetic calorimeter (EMCal).  The RICH uses CO$_{2}$ gas at 
atmospheric pressure as a \v{C}erenkov radiator.  Electrons and pions 
begin to radiate in the RICH at \pt $>$ 20~MeV/$c$ and \pt $>$ 4.9 \gev, 
respectively.  The EMCal is composed of four sectors in each arm. The 
bottom two sectors of the east arm are lead-glass and the other six 
are lead-scintillator. The energy resolution of the EMCal is 
$\sigma_E/E \simeq$ 4.5\% $\oplus$ 8.3$/\sqrt{E{\rm(GeV)}}$ and 
$\sigma_E/E \simeq$ 4.3\% $\oplus$ 7.7$/\sqrt{E{\rm(GeV)}}$ for 
lead-scintillator and lead-glass, respectively.

%%%%%%%%%%%%%%%%%%%%%%%%%%%%%%%%%%%%%%%%%%%%%% Fig_1

\subsection{The VTX detector}

In 2011, the central detector was upgraded with the VTX detector as shown 
in Fig.~\ref{fig:phenix}.  In addition, a new beryllium beam pipe with 
2.16~cm inner diameter and 760 $\mu$m nominal thickness was installed to 
reduce multiple-scattering before the VTX detector.

The VTX detector~\cite{VTX1:2004,Nouicer:2012pr,Kurosawa:2013taa} consists 
of four radial layers of silicon detectors as shown in 
Fig.~\ref{fig:phenix}(b). The detector is separated into two arms, each 
with nominal acceptance $\Delta \phi \approx 0.8 \pi$ centered on the 
acceptance of the outer PHENIX central arm spectrometers. The detector 
covers pseudorapidity $|\eta|<$~1.2 for collisions taking place at $z=0$. 
The VTX can precisely measure the vertex position of a collision within 
$|z|<10$~cm range of the center of the VTX.

The two inner layers, referred to as B0 and B1, of the VTX detector 
comprise silicon pixel detectors, as detailed in Ref.~\cite{Ryo:2009}.  
B0 (B1) comprises 10 (20) ladders with a central radial position of 2.6 
(5.1) cm.  The silicon pixel technology is based on the \mbox{ALICE1LHCb} 
sensor-readout chip~\cite{Snoeys:2001xc}, which was developed at CERN. Each 
ladder is electrically divided into two independent half-ladders. Each 
ladder comprises four sensor modules mounted on a mechanical support 
made from carbon-fiber composite. Each sensor module comprises a silicon 
pixel sensor with a pixel size of 
50~$\mu$m($\phi$)~$\times$~425~$\mu$m($z$) bump-bonded with four pixel 
readout chips.  One pixel readout chip reads 256 $(\phi) \times$ 32 $(z)$= 
8192 pixels and covers approximately 1.3 cm $(\Delta \phi) \times$ 1.4 cm 
$(\Delta z)$ of the active area of the sensor. The position resolution is 
$\sigma_\phi$ = 14.4~$\mu$m~in the azimuthal direction.

The two outer layers of the VTX detector, referred to as B2 and B3, are 
constructed using silicon stripixel sensors, as detailed in 
Ref.~\cite{Ryo:2009}. The B2 (B3) layer comprises 16 (24) silicon 
stripixel ladders at a central radial distance of 11.8 (16.7) cm. The 
stripixel sensor is a novel silicon sensor, and is a single-sided, N-type, 
DC-coupled, two-dimensional (2-D) sensitive 
detector~\cite{Li:2004if,Nouicer:2009}. One sensor has an active area of 
approximately 30~mm $\times$ 60~mm, which is divided into two independent 
sectors of 30~mm $\times$ 30~mm. Each sector is divided into 384 $\times$ 
30 pixels. Each pixel has an effective size of 80 $\mu$m ($\phi$) $\times$ 
1000 $\mu$m ($z$), leading to a position resolution of 
$\sigma_\phi$=23~$\mu$m. A pixel comprises two implants (A and B) 
interleaved such that each of the implants registers half of the charge 
deposited by ionizing particles. There are 30 A implants along the beam 
direction, connected to form a 30 mm long X-strip, and 30 B implants are 
connected with a stereo angle of 80 mrad to form a U-strip. X-strip and 
U-strip are visualized in \cite{Nouicer:2009}. When a charged particle 
hits a pixel, both the X- and the U-strip sharing the pixel register a 
hit. Thus the hit pixel is determined as the intersection of the two 
strips. The stripixel sensor is read out with the SVX4 chip developed by a 
FNAL-LBNL Collaboration \cite{GarciaSciveres:2003cw}.

The total number of channels in the VTX pixel and stripixel layers is 
3.9~million pixels and 0.34~million strips. The compositions of the pixel 
and strip are illustrated in \cite{Ryo:2009,Nouicer:2009}. The main 
characteristics of the VTX detector are summarized in Table~\ref{tab:vtx}.

%%%% TABLE VTX parameters %%%%%%%%

%============================================ Table_II
\begin{table*}[tbh]
\caption{\label{tab:vtx}
A summary of the VTX detector. For each layer (B0 to B3), the detector 
type, the central radius ($r$), ladder length ($l$), sensor thickness 
($t$), sensor active area ($\Delta \phi \times \Delta z$), the number of 
sensors per ladder ($N_S$), the number of ladders ($N_L$), pixel/strip 
size in $\phi$ ($\Delta \phi$) and z ($\Delta z$), the number of read-out 
channels ($N_{ch}$), and the average radiation length including the 
support and on-board electronics ($X_0$) are given.}
\begin{ruledtabular} \begin{tabular}{lcccccccccccc}
&\multicolumn{4}{c}{}
&\multicolumn{2}{c}{sensor active area}
& &\multicolumn{3}{c}{pixel/strip size}
&\multicolumn{2}{c}{}\\
& type & $r$(cm) & $l$(cm) & $t$ ($\mu$m) 
&$\Delta \phi$(cm)&$\Delta z$(cm)
& $N_S$&$N_L$ & $\Delta \phi$ ($\mu$m) & $\Delta z$ ($\mu$m) 
& $N_{ch}$ & $X_{0}$(\%) \\ 
\hline
B0 & pixel     & 2.6  & 22.8 & 200 
& 1.28 & 5.56 
& 4 & 10 & 50 & 425 & $1.3\times 10^6$ & 1.3 \\
B1 & pixel     & 5.1  & 22.8 & 200
& 1.28 & 5.56 
& 4 & 20 & 50 
& 425 & $2.6\times 10^6$ & 1.3 \\
B2 & stripixel & 11.8 & 31.8 & 625 
& 3.07 & 6.00 
& 5 & 16 & 80 
& $3\times 10^4$ & $1.2\times 10^5$ & 5.2 \\
B3 & stripixel & 16.7 & 38.2  & 625
& 3.07 & 6.00 
& 6 & 24 & 80 
& $3\times 10^4$ & $2.2\times 10^5$ & 5.2 \\
\end{tabular} \end{ruledtabular}
\end{table*}

%%========================================================================
	\section{Analysis}
	\label{sec:analysis}

	\subsection{Overview}

The purpose of the analysis is to separate the electrons from charm and 
bottom hadron decays. The life time of $B$ mesons ($c\tau_{B^0}$= 455 
$\mu$m, $c\tau_{B^\pm}$ = 491 $\mu$m \cite{Agashe:2014kda}) is 
substantially longer than that of $D$ mesons ($c\tau_{D^0}$ = 123 $\mu$m, 
$c\tau_{D^\pm}$ = 312 $\mu$m) and the decay kinematics are different. This 
means that the distribution of values for the distance of closest 
approach (DCA) of the track to the primary vertex for electrons from bottom decays 
will be broader than that of electrons from charm decays. There are other 
sources of electrons, namely Dalitz decays of $\pi^0$ and $\eta$, photon 
conversions, $K_{e3}$ decays, and $J/\psi\rightarrow e^+e^-$ decays. With 
the exception of electrons from $K_{e3}$ decays, these background 
components have DCA distributions narrower than those from charm decay 
electrons. Thus we can separate $b\rightarrow e$, $c\rightarrow e$ and 
background electrons via precise measurement of the DCA distribution.

In the first step of the analysis, we select good events where the 
collision vertex is within the acceptance of the VTX detector, and its 
function is normal (Sec.~\ref{sec:data}). We then reconstruct electrons in 
the PHENIX central arms (Sec.~\ref{sec:track_reconstruction}). The 
electron tracks are then associated with hits in the VTX detector and 
their DCA is measured (Sec.~\ref{sec:vtx_tracking}). At this point we 
have the DCA distribution of inclusive electrons that has contributions 
from heavy flavor ($b\rightarrow e$ and $c\rightarrow e$) and several 
background components.

The next step is to determine the DCA shape and normalization of all 
background components~(Sec.~\ref{sec:dca_background}). They include 
mis-identified hadrons, background electrons with large DCA caused by 
high-multiplicity effects, photonic electrons (Dalitz decay electrons, 
photon conversions), and electrons from $K_{e3}$ and quarkonia decays. The 
shapes of the DCA distributions of the various background electrons are 
determined via data driven methods or Monte Carlo simulation. We then 
determine the normalization of those background electron components in the 
data (Sec.~\ref{sec:norm}).

Because the amount of the VTX detector material is substantial (13\% of one 
radiation length) the largest source of background electrons is photon 
conversion within the VTX. We suppress this background by a conversion 
veto cut (Sec.~\ref{sec:photonic_electrons})

Once the shape and the normalization of all background components are 
determined and subtracted, we arrive at the DCA distribution of heavy 
flavor decay electrons that can be described as a sum of $b \rightarrow e$ 
and $c \rightarrow e$ DCA distributions. The heavy flavor DCA 
distribution is decomposed by an unfolding method 
(Sec.~\ref{sec:unfolding}).

	\subsection{Event selection}
	\label{sec:data}

The data set presented in this analysis is from \auau collisions at 
\sqsntwo recorded in 2011 after the successful commissioning of the VTX 
detector. As detailed earlier, the MB \auau data sample was recorded using 
the BBC trigger sampling $96 \pm 3$\% of the inelastic \auau cross 
section. A number of offline cuts were applied for optimizing the detector 
acceptance uniformity and data quality as described below. After all cuts, 
a data sample of 2.4$\times 10^9$ \auau events was analyzed.

	\subsubsection{z-vertex selection}

The acceptance of the PHENIX central arm spectrometers covers collisions 
with $z$-vertex within $\pm$ 30~cm of the nominal interaction point. The 
VTX detector is more restricted in $|z|$ acceptance, as the B0 and B1 
layers cover only $|z|<11.4$ cm. Thus the BBC trigger selected only events 
within the narrower vertex range of $|z_{\rm BBC}|< 12$ cm. In the offline 
reconstruction, the tracks reconstructed from VTX information alone are 
used to reconstruct the \auau collision vertex with resolution 
${\sigma_z}=75$~$\mu$m. All \auau events in the analysis are required to have a 
$z$-vertex within $\pm$10~cm as reconstructed by the VTX.

	\subsubsection{Data quality assurance}

Due to a number of detector commissioning issues in this first data taking 
period for the VTX, the data quality varies substantially. Therefore we 
divide the entire 2011 \auau data taking period into four periods. The 
acceptance of the detector changes significantly between these periods.

In addition, several cuts are applied to ensure the quality and the 
stability of the data. Applying electron identification cuts described in 
Sec.~\ref{sec:eid}, the electron to hadron ratios were checked for each 
run, a continuous data taking period typically lasting of order one hour, 
and three runs out of 547 with ratios outside of 5$\sigma$ from the mean 
were discarded. The B2 and B3 stripixel layers had an issue in stability 
of read-out electronics where some of the sensor modules would drop out, 
resulting in a reduced acceptance within a given run. Additional 
instabilities also existed in the B0 and B1 pixel layers. Detailed channel 
by channel maps characterizing dead, hot, and unstable channels were 
generated for all layers within a given run. These maps were used to mask 
dead, hot, and unstable channels from the analysis, as well as to define 
the fiducial area of the VTX in simulations.

During this first year of data taking, the instability of the read-out 
electronics discussed above caused significant run-to-run variations in 
the acceptance and efficiency of the detector. It is therefore not 
possible to reliably calculate the absolute acceptance and efficiency 
correction while maintaining a large fraction of the total data set 
statistics. Instead, we report on the relative yields of charm and bottom 
to total heavy flavor. We have checked that the DCA distributions are 
consistent between running periods and are not impacted by the changing 
acceptance. Thus we can measure the shape of the DCA distribution using 
the entire data set. In the following, we use the shape of the 
measured DCA distribution only to separate $b\rightarrow e$ 
and $c\rightarrow e$ components.

	\subsection{Electron reconstruction in central arms}
	\label{sec:track_reconstruction}

	\subsubsection{Track reconstruction}

Charged particle tracks are reconstructed using the outer central arm 
detectors, DC and PC, as detailed in Ref.~\cite{Adare:2006nq}. The DC has 
six types of wire modules stacked radially, named X1, U1, V1, X2, U2, and 
V2.  The X wires run parallel to the beam axis in order to measure the 
$\phi$-coordinate of the track and the U and V wires have stereo angles 
varying from 5.4 to 6.0 degrees. Tracks are required to have hits in both 
the X1 and X2 sections along with uniquely associated hits in the U or V 
stereo wires and at least one matching PC hit, to reduce mis-reconstructed 
tracks. The track momentum vector is determined assuming the particle 
originated at the \auau collision vertex as reconstructed by the BBC.

\subsubsection{Electron identification}
\label{sec:eid}

Electron candidates are selected by matching tracks with hits in the RICH 
and energy clusters in the EMCal. The details on the electron selection 
cuts are given in Ref.~\cite{Adare:2010de}. In this analysis we select 
electron candidates within $1.5<\pt\ [\gev]<5.0$, and we briefly describe 
the cuts in the RICH and EMCal below.

\begin{figure}[!hbt]
\includegraphics[width=1.0\linewidth]{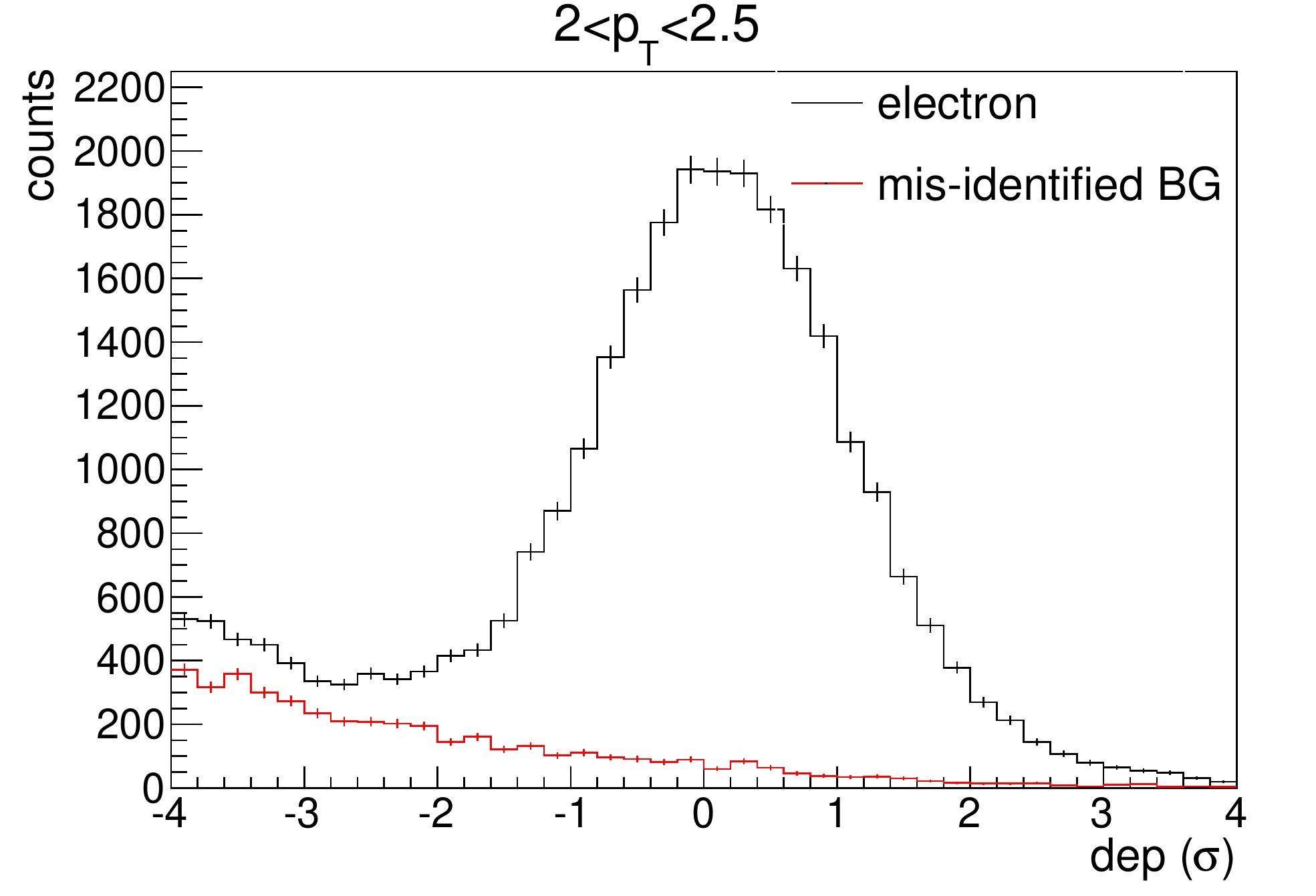}
\caption{\label{fig:dep} (Color Online) 
Matching variable between the reconstructed track momentum ($p$) and the 
energy measured in the EMCal ($E$):  $dep = (E/p - \mu_{E/p}) 
/\sigma_{E/p}$.  The black distribution is for identified electrons with 
\pt = 2.0--2.5~\gev, and the red distribution is the estimated 
contribution from mis-identified electrons via the RICH swap-method.}
\end{figure}

\v{C}erenkov photons from an electron track produce a ring-shaped cluster 
in the RICH. At least three associated PMT hits are required in the RICH 
and a ring-shape cut is applied. The center of the ring is required to be 
within 5 cm of the track projection. The probability that the associated 
cluster in the EMCal comes from an electromagnetic shower is calculated 
based on the shower shape.  Based on that probability, tracks are selected 
in a way that maintains high efficiency for electrons while rejecting 
hadrons. Further, the energy ($E$) in the EMCal is required to match the 
track determined momentum ($p$).  This match is calculated as $dep = (E/p 
- \mu_{E/p}) /\sigma_{E/p}$, where $\mu_{E/p}$ and $\sigma_{E/p}$ are the 
mean and standard deviation respectively of a Gaussian fit to the $E/p$ 
distribution, determined as a function of momentum (see 
Fig.~\ref{fig:dep}).  A cut of $dep >-2$ is used to further reject hadrons 
that have an $E/p$ ratio $<1$, because they do not deposit their full 
energy in the EMCal.

In high-multiplicity \auau events there is a significant probability for a 
random association between the track and hits in the RICH and EMCal.  
This mis-identified hadron probability is estimated as follows.  The $z<0$ 
and $z>0$ sides of the RICH have their hits swapped in software, and the 
tracks are re-associated with RICH hits. Because the two longitudinal 
sides of the RICH are identical, this gives a good estimate of the random 
hadron background in the electron sample.

The distribution of electron candidates at \pt=2.0--2.5 \gev for the 
normalized EMCal energy to track momentum ratio, $dep$ defined above, is 
shown in Fig.~\ref{fig:dep}. There is a large peak near zero from true 
electrons as expected and a clear low-side tail from mis-identified 
hadron.  Also shown is the result of the above swap method. The difference 
between the data and the ``swap" distribution (red) is explained as 
contributions from off-vertex electrons caused by conversions from the 
outer layer of the VTX and weak decay. In the final accounting for all 
contributions to the identified-electron DCA distribution, we utilize 
this swap method to statistically estimate the contribution of 
mis-identified hadron in each \pt selection as detailed in 
Section~\ref{sec:hadron_contamination}.

%%%%%%%%%%%%%%%%%%%%%%%%%%%%%%%%%%%%%%%%%%%%%% Fig_2

\subsection{DCA measurement with the VTX}
\label{sec:vtx_tracking}

Charged particle tracks reconstructed in the central arms must be 
associated with VTX hits in order to calculate their DCA.  
Three-dimensional (3-D) hit positions in the 4 layers of VTX are 
reconstructed. For each collision, the primary vertex is reconstructed by 
the VTX. Then central arm tracks are associated with hits in the VTX, and 
VTX-associated tracks are formed.  Finally, the DCA between the primary 
vertex and the VTX-associated tracks are measured.

\subsubsection{VTX alignment}

In order to achieve good DCA resolution to separate $b\rightarrow e$
and $c\rightarrow e$, alignment of the detector ladders to high
precision is required.
The detector alignment is accomplished via an iterative 
procedure of matching outer central arm tracks from the DC and PC to the VTX 
hits.  The procedure is convergent for the position of each ladder.
The alignment was repeated each time the detector was repositioned following 
a service access.   The final alignment contribution to the
DCA resolution in both $\phi$ and $z$ is a few tens of microns.

\subsubsection{VTX hit reconstruction}

For layers B0 and B1, clusters of hit pixels are formed by connecting 
contiguous hit pixels by a recursive clustering algorithm. An average 
cluster size is 2.6 (6.7) pixels for the pixel (stripixel). The center of 
the cluster in the local 2-D coordinate system of the sensor 
is calculated as the hit position.

For B2 and B3 layers, 2D hit points on the sensor are reconstructed from 
the X-view and the U-view. Hit lines in the X-view (U-view) are formed by 
clustering contiguous hit X-strips (U-strips) weighted by deposited 
charges, and then 2D hit points are formed as the intersections of all hit 
lines in X- and U- views. When one hit line in U-view crosses more than 
two hit lines in X-view, ghost hits can be formed, because which crossing 
point is the true hit is ambiguous. These ghost hits increase the number 
of reconstructed 2D hits approximately by 50\% (30\%) in B2 (B3) in 
central \auau collisions. The ghost hit rate was studied using a full 
\geant~\cite{GEANT} simulation with the HIJING~\cite{Wang:1991hta} generator as input. 
However, because the occupancy of the detector at the reconstructed 2D hit 
point level is low, less than 0.1\%, these ghost hits do not cause any 
significant issue in the analysis.

The positions of all 2-D hits in the VTX are then transferred into the 
global PHENIX 3-D coordinate system.  Correction of the sensor position 
and orientation, determined by the alignment procedure described in the 
previous section, is applied in the coordinate transformation. The 
resulting 3-D hit positions in the global coordinate system are then used 
in the subsequent analysis.

\subsubsection{The primary vertex reconstruction}

With the VTX hit information alone, charged particle tracks can be 
reconstructed only with modest momentum resolution $\delta p / p \approx$ 
10\% due to the limited magnetic field integrated over the VTX volume
and the multiple scattering within the VTX. These tracks can be
utilized to determine the collision
vertex in three-dimensions 
($z_{0}$ along the beam axis, and $x_{0}$,$y_{0}$ in the transverse plane)
for each \auau event under the safe
assumption that the majority of particles originate at the collision vertex.
This vertex position is called the primary vertex position.

The position resolution of the primary vertex for each direction 
depends on the sensor pixel and strip sizes, the precision of the 
detector alignment, and the number of particles used for
the primary vertex calculation and their 
momentum in each event.
For MB \auau collisions, the resolution values are $\sigma_x = 96$ 
$\mu$m, ${\sigma_y}=43$~$\mu$m, and ${\sigma_z}=75$~$\mu$m.
The worse resolution in $x$ compared to $y$ is due to the 
orientation of the two VTX arms. For comparison, the beam profile in
the transverse plane is 
$\sigma^{\rm lumi}_x \approx \sigma_y^{\rm lumi} \approx 90$~$\mu$m in
the 2011 \auau run.

\subsubsection{Association of a central arm track with VTX}

Each central arm track is projected from the DC through the magnetic field 
to the VTX detector. Hits in VTX are then associated with the track using 
a recursive windowing algorithm as follows.

The association starts from layer B3. VTX hits in that layer that are 
within a certain ($\Delta \phi \times \Delta z$) window around the track 
projection are searched. If hits are found in this window, the track is 
connected to each of the found hits, and then projected inward to the next 
layer. In this case the search window in the next layer is decreased, 
because there is much less uncertainty in projection to the next layer. If 
no hit is found, the layer is skipped, and the track is projected inward 
to the next layer, keeping the size of the projection window. This process 
continues until the track reaches layer B0, and a chain of VTX hits that 
can be associated with the track is formed. The window sizes are momentum 
dependent and determined from a full \geant simulation of the 
detector so that the inefficiency of track reconstruction due to the 
window size is negligible.

After all possible chains of VTX hits that can be associated with a given 
central arm track are found by the recursive algorithm, a track model fit 
is performed for each of these possible chains, and the $\chi^2$ of the 
fit, $\chi^2_{\rm vtx}$, is calculated. The effect of multiple scattering 
in each VTX layer is taken into account in calculation of $\chi^2_{\rm 
vtx}$. Then the best chain is chosen based on the value of $\chi^2_{\rm 
vtx}$ and the number of associated hits. This best chain and its track 
model are called a VTX-associated track. Note that at most one 
VTX-associated track is formed from each central arm track.

In this analysis we require that VTX-associated tracks have associated 
hits in at least the first three layers, i.e. B0, B1, and B2. An 
additional track requirement is $\chi^2_{\rm vtx}/{\rm NDF}<2$ for \pt$<2$ 
\gev and $\chi^2_{\rm vtx}/{\rm NDF}<3$ for \pt$>2$ \gev, where NDF is the 
number of degrees of freedom in the track fit.

	\subsubsection{\DCAR and \DCAZ}
	\label{sec:dca_def}

\begin{figure}[!hbt]
\includegraphics[width=1.0\linewidth]{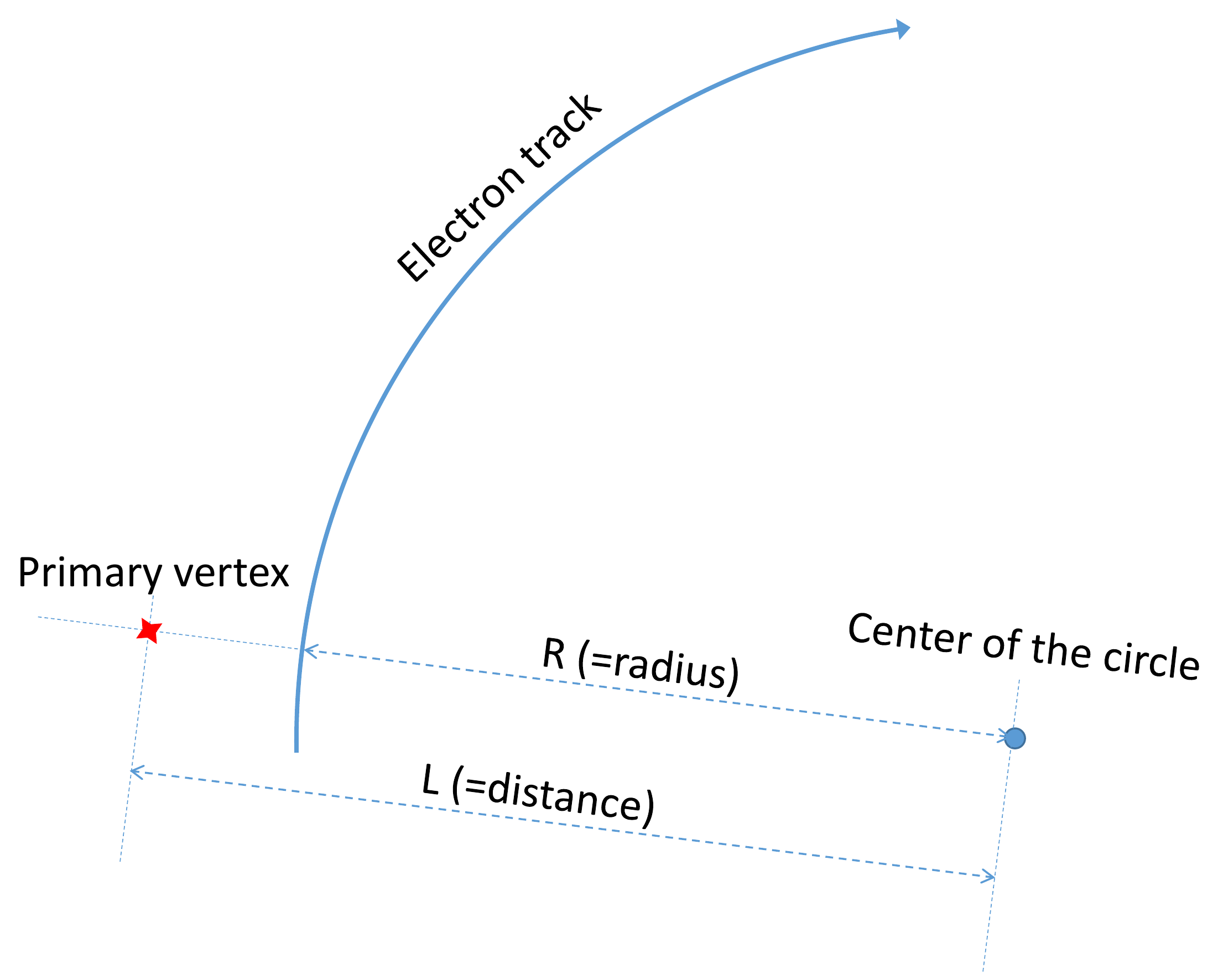}
\caption{(Color Online) Illustration of the definition of \DCAR $\equiv$ $L$ - $R$ 
in the transverse plane.
}
\label{fig:dca_definition}
\end{figure}

Using the primary vertex position determined above, the DCA of a track 
is calculated separately in the transverse plane (\DCAR) and along the 
beam axis (\DCAZ). Because by design the \DCAR has a better resolution 
than \DCAZ, we first find \DCAR with a track model of a circle trajectory 
assuming the uniform magnetic field over the VTX.  We define \DCAR as
\begin{equation}
\DCAR \equiv L - R,
\end{equation}
where $L$ is the distance from the collision vertex to the center of the 
circle defining the particle trajectory, and $R$ is the radius of the 
circle as shown in Fig.~\ref{fig:dca_definition}. \DCAZ is the distance 
between the z-coordinate of the point \DCAR found and z-coordinate of the 
primary vertex.

It is notable that \DCAR has a sign in this definition. The distinction 
between positive and negative values of \DCAR---whether the trajectory is 
bending towards or away from the primary vertex---is useful since certain 
background contributions have asymmetric distributions in positive and 
negative \DCAR, as discussed in section~\ref{sec:dca_background}. For 
electrons, the positive side of \DCAR distribution has less background 
contribution. There is no such positive/negative asymmetry in \DCAZ.

%%%%%%%%%%%%%%%%%%%%%%%%%%%%%%%%%%%%%%%%%%%%%% Fig_3

	\subsubsection{{\rm DCA} measurement}
	\label{sec:dca_measurement}

\begin{figure}[!hbt]
\includegraphics[width=1.0\linewidth]{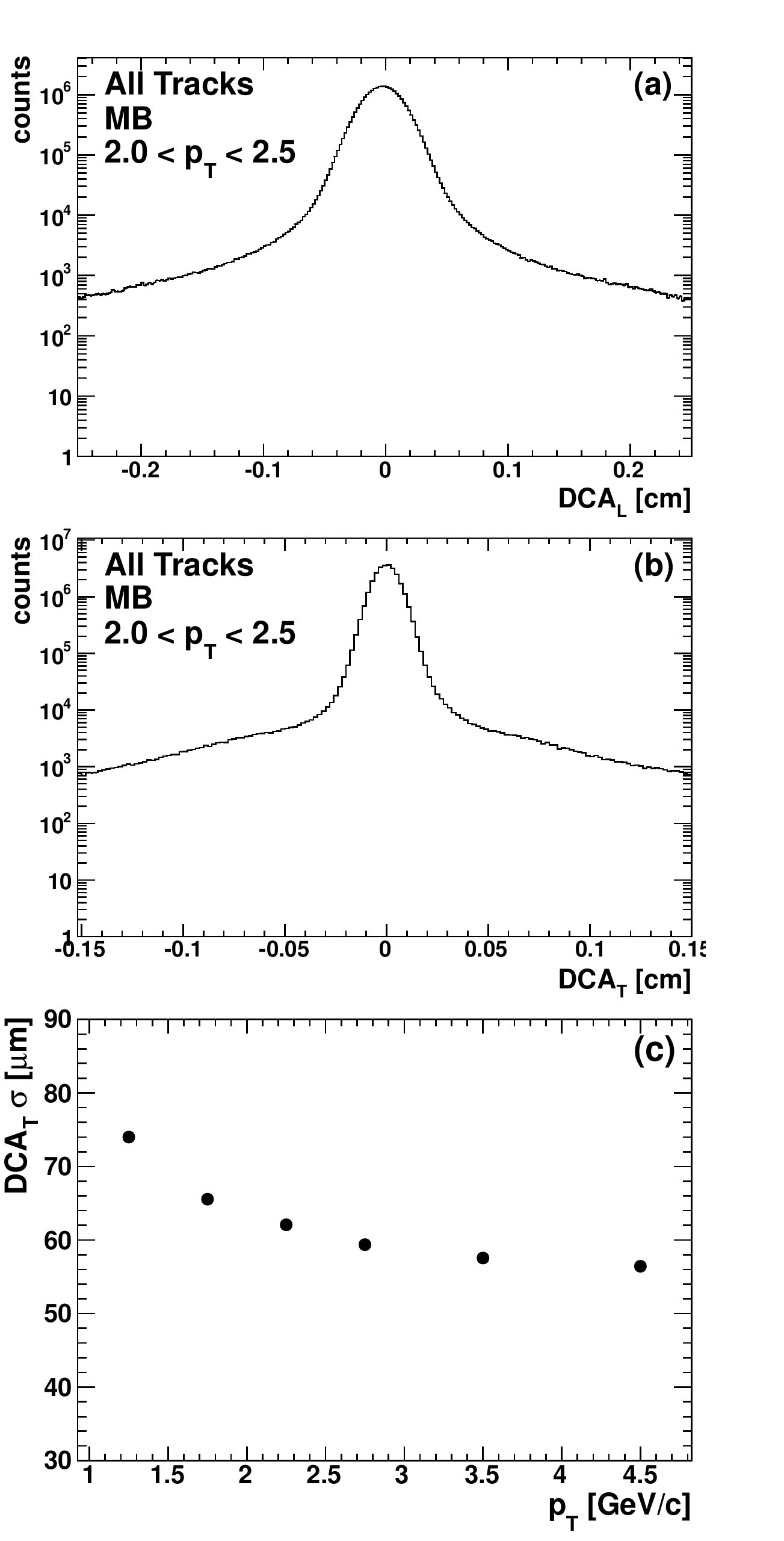}
\caption{\label{fig:hadron_dca} 
Distance-of-closest-approach distributions for 
(a) along the beam axis \DCAZ and (b) transverse plane \DCAR for all 
VTX-associated tracks in \auau at \sqsntwo in the range $2.0<\pt\,[{\rm~GeV}/c]<2.5$. 
(c) The \DCAR resolution as a function of \pt for all tracks.
}
\end{figure}

For each VTX-associated track, the DCA is calculated separately in the 
radial and longitudinal direction (\DCAR and \DCAZ) from the track model 
and the primary vertex position. Shown in Fig.~\ref{fig:hadron_dca} is the 
resulting \DCAR and \DCAZ distributions for all VTX-associated tracks with 
\pt = 2.0--2.5~\gev. Since the vast majority of charged tracks are hadrons 
originating at the primary vertex, we observe a large peak around \DCAR, 
\DCAZ = 0 that is well fit to a Gaussian distribution where the $\sigma$ 
represents the \DCAR, \DCAZ resolution. A selection of $|$\DCAZ$| < 
0.1$~cm is applied to reduce background.

There are broad tails for $|$\DCAR$|>$ 0.03 cm. Monte Carlo simulation 
shows that the main source of the broad tails is the decay of long lived 
light hadrons such as $\Lambda$ and $K_S^0$.

The \DCAR resolution as a function of the track \pt is extracted using a 
Gaussian fit to the peak and is shown in Fig.~\ref{fig:hadron_dca}~c). The 
\DCAR resolution is approximately 75~$\mu$m for the 1.0--1.5 \gev bin and 
decreases with increasing \pt as the effect of multiple scattering becomes 
smaller for higher \pt. The \DCAR resolution becomes less than 60 $\mu$m 
for \pt $>$ 4 \gev, where it is limited by the position resolution of the 
primary vertex.

%%%%%%%%%%%%%%%%%%%%%%%%%%%%%%%%%%%%%%%%%%%%%% Fig_4

\begin{figure*}[!hbt]
 \includegraphics[width=0.4\linewidth]{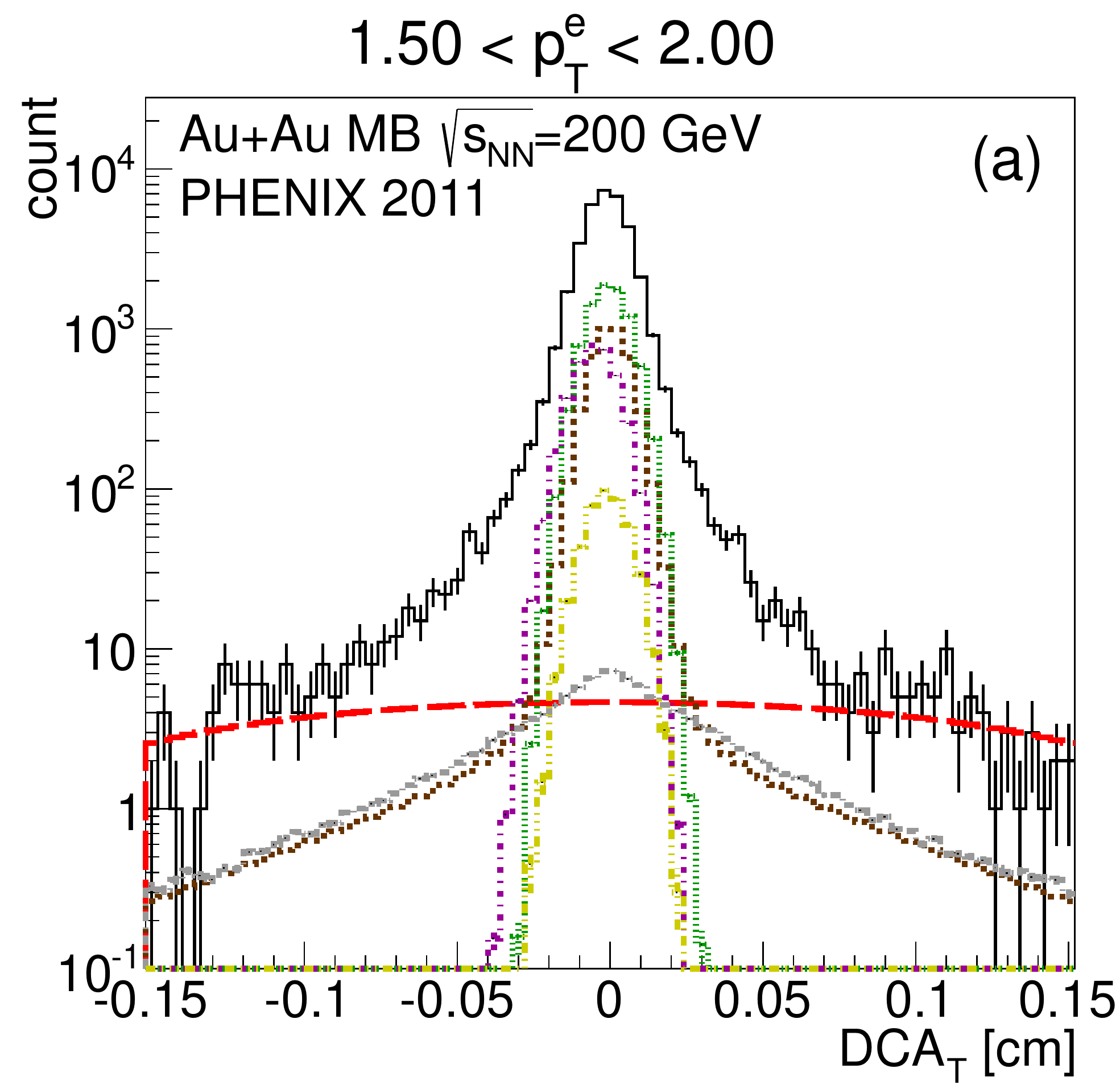}
 \includegraphics[width=0.4\linewidth]{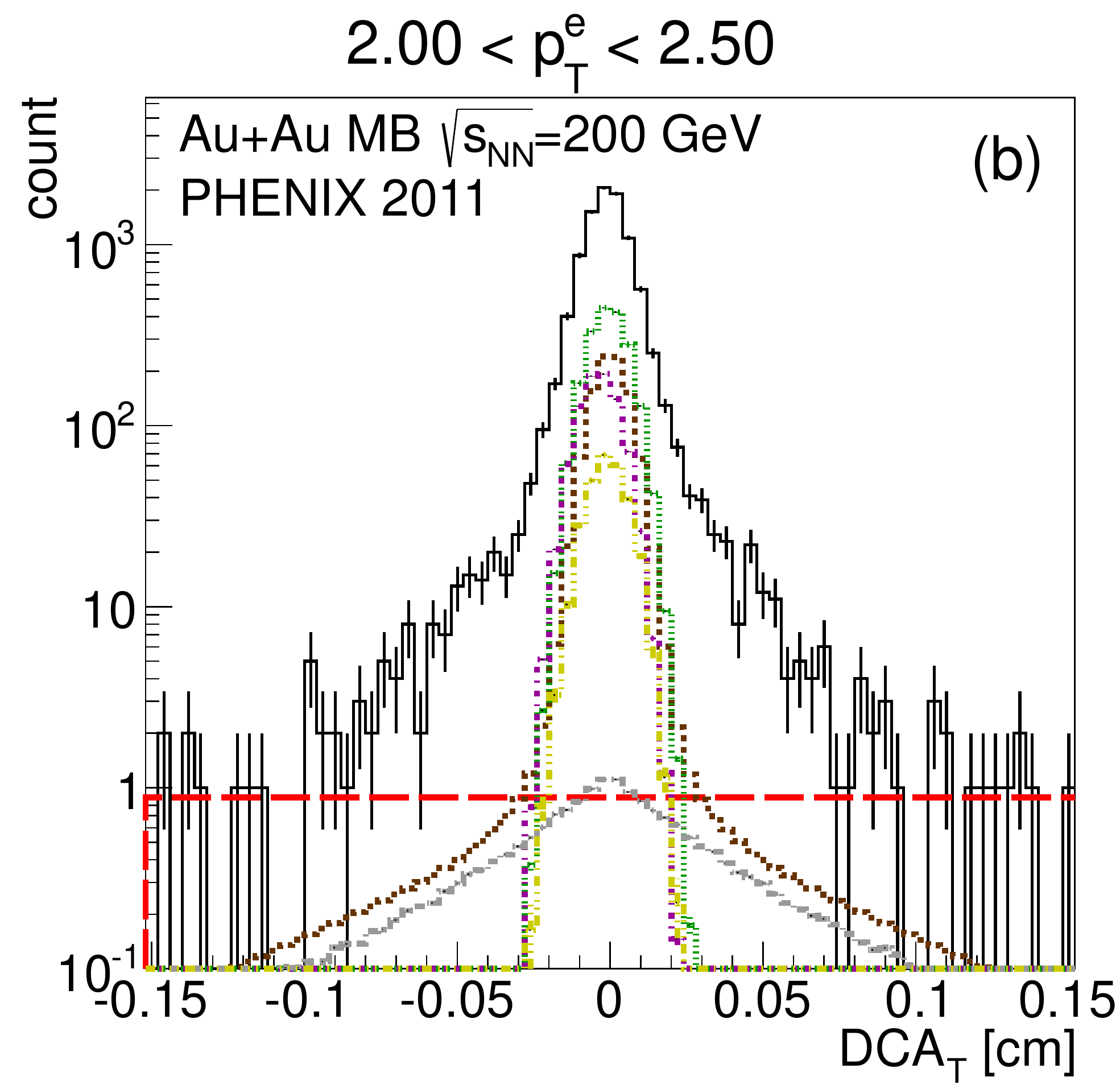}\\
 \includegraphics[width=0.4\linewidth]{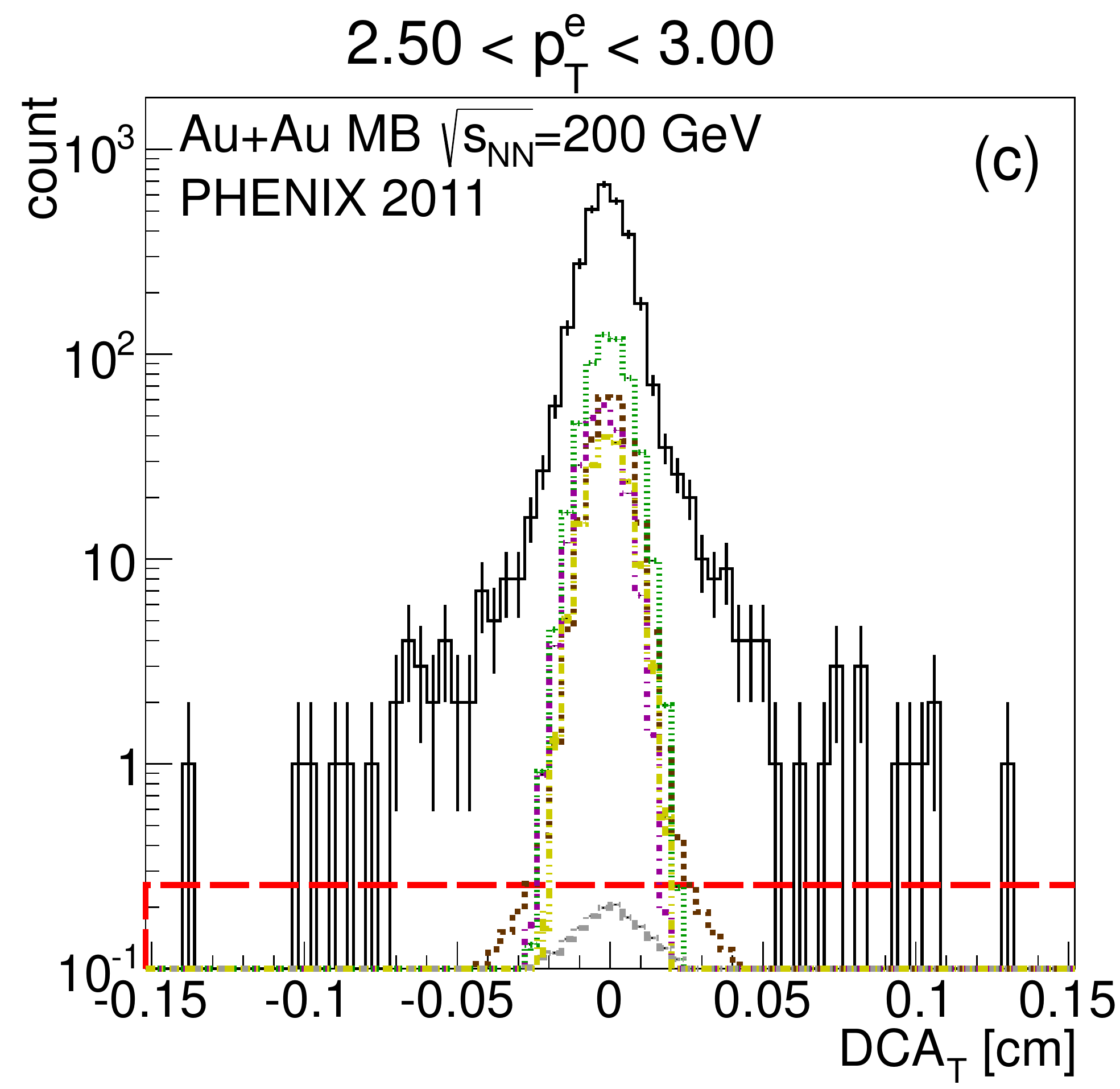}
 \includegraphics[width=0.4\linewidth]{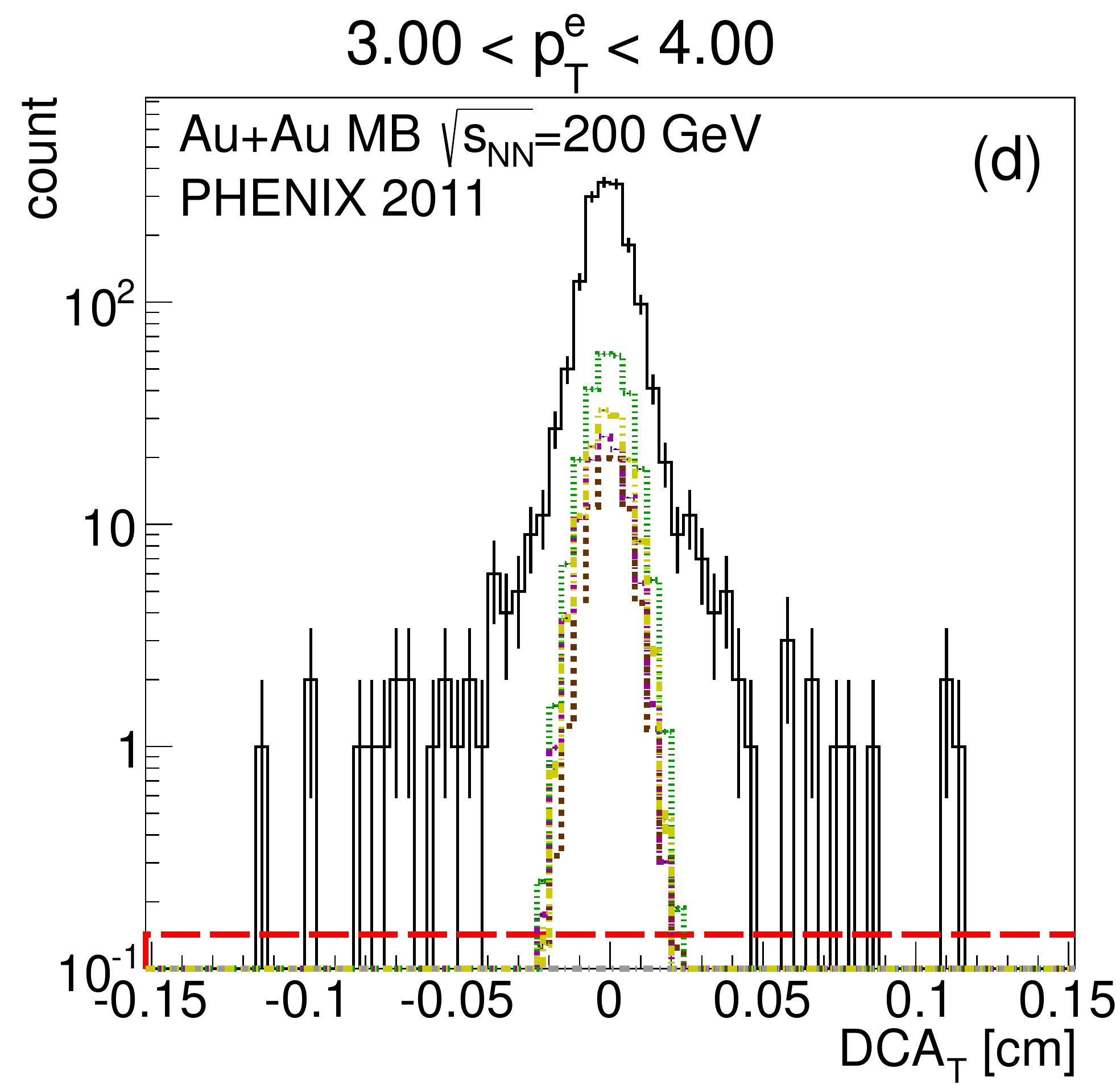}\\
 \includegraphics[width=0.4\linewidth]{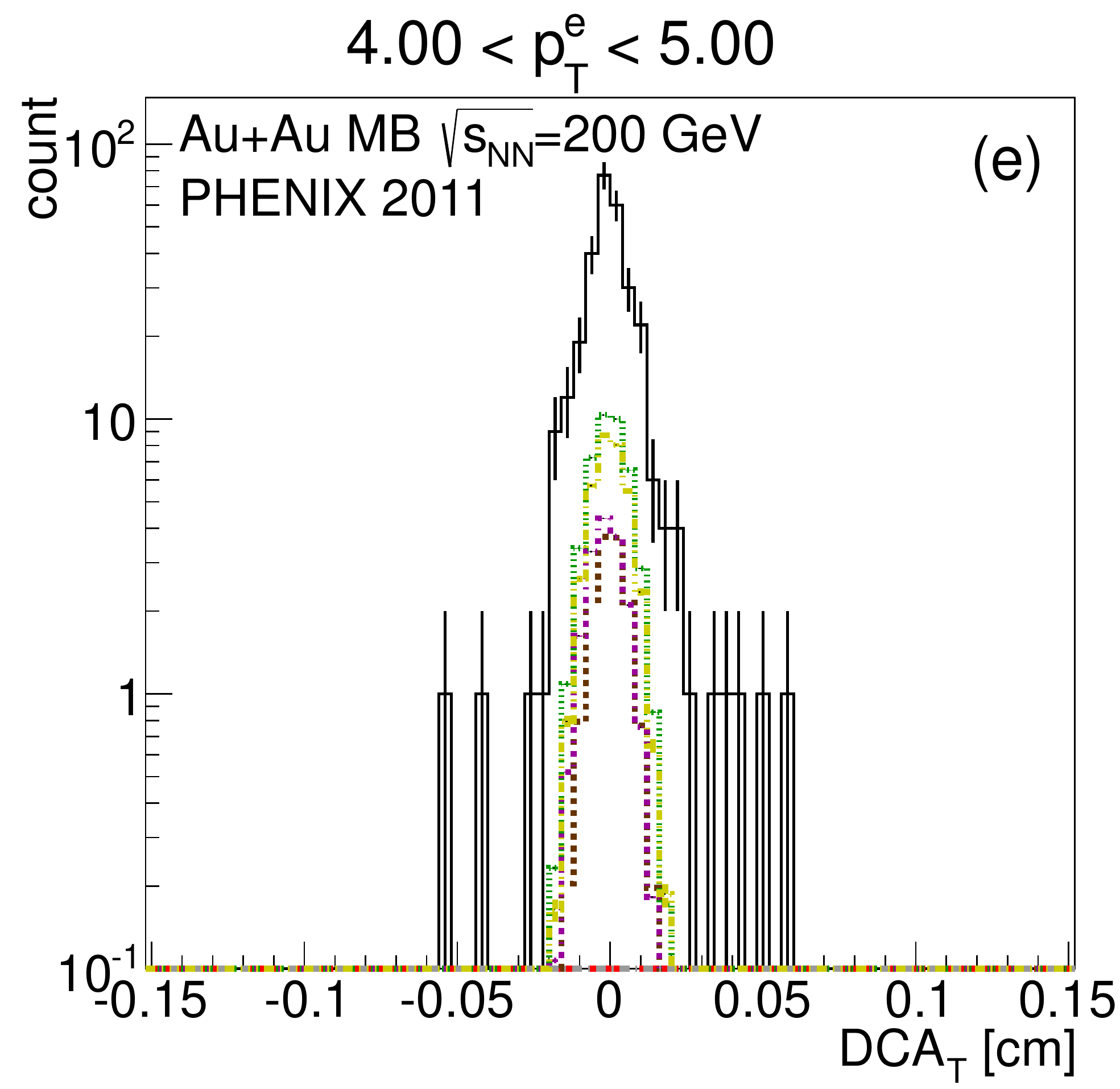}
 \includegraphics[width=0.4\linewidth]{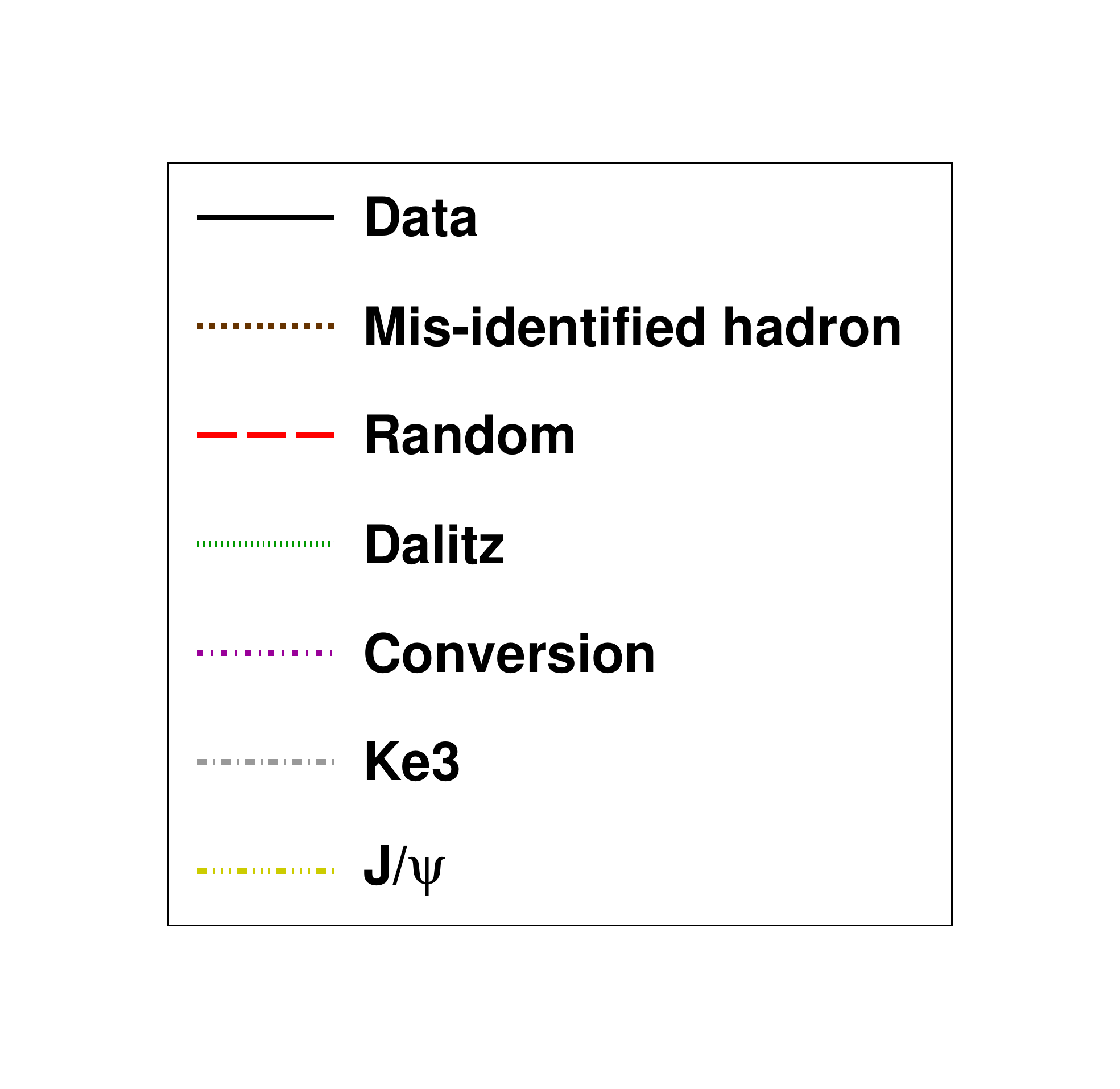}
\caption{\label{fig:DCA0} (Color Online) 
\DCAR distributions for electrons in MB Au$+$Au at \sqsntwo that pass the 
reconstruction and conversion veto cut in the indicated five electron-\pt 
selections.  Also shown are the normalized contributions for the various 
background components detailed in Section~\ref{sec:dca_background}.
}
\end{figure*}

We divide the electrons into five \pt bins and show the \DCAR 
distributions for each in Fig.~\ref{fig:DCA0}. These distributions are in 
integer-value counts and are not corrected for acceptance and efficiency. 
The DCA distributions include various background components other than 
heavy flavor contributions. The background components are also shown in 
the figure and are discussed in the next section 
(Section~\ref{sec:dca_background}).

%%%%%%%%%%%%%%%%%%%%%%%%%%%%%%%%%%%%%%%%%%%%%% Fig_5

While the \DCAR distributions in Fig.~\ref{fig:DCA0} are plotted within 
$|\DCAR|<0.15$~cm, only a $|\DCAR|<0.1$~cm is used in the analysis to 
extract the charm and bottom yield described later. At large \DCAR, the 
distribution is dominated by high-multiplicity background 
(Sec.~\ref{sec:dca_sideband}) and therefore provides little constraint in 
the extraction of the charm and bottom contributions.

	\subsection{DCA distribution of Background Components}
	\label{sec:dca_background}

The sample of candidate electron tracks that pass all the analysis cuts 
described above contains contributions from a number of sources other than 
the desired electrons from semi-leptonic decays of charm and bottom 
hadrons. In order to extract the heavy flavor contributions, all 
background components must be fully accounted for and their \DCAR shapes 
as a function of \pt incorporated. These background components are listed 
in the order presented below.

\begin{enumerate}
\item Misidentified hadrons
\item High-multiplicity background
\item Photonic electrons
\item Kaon decay electrons
\item Heavy-quarkonia decay electrons
\end{enumerate}
 
As described in this and the following section, all background components 
are constrained by PHENIX measurements in \auau and are fully simulated 
through a \geant description of the detector. This method is similar to 
the cocktail method of background subtraction used in the previous 
analysis of inclusive heavy flavor electrons~\cite{Adare:2010de}.

Next, we describe these background sources and their DCA distributions. 
The first two components are caused by detector and multiplicity effects. 
DCA distributions and normalization of these two components are 
determined by data driven methods, as detailed in this section. The last 
three components are background electrons that are not the result of 
semi-leptonic decays of heavy flavor hadrons. Their DCA distributions are 
determined by Monte Carlo simulation, and their normalization is 
determined by a bootstrap method described in section~\ref{sec:norm}. Of 
those background electrons, photonic electrons are the dominant 
contribution. We developed a conversion veto cut to suppress this 
background (\ref{sec:photonic_electrons}).

	\subsubsection{Mis-identified hadron}
	\label{sec:hadron_contamination}

As detailed in the discussion on electron identification, there is a 
nonzero contribution from mis-identified electrons. This contribution is 
modeled via the RICH swap-method described in Section~\ref{sec:eid}. From 
this swap method, we obtain the probability that a charged hadron is 
mis-identified as an electron as a function of \pt. This probability is 
then applied to the DCA distribution of charged hadrons to obtain the 
DCA distribution of mis-identified hadrons.

The resulting \DCAR distribution is shown in each panel of 
Fig.~\ref{fig:DCA0}. Note that this component is properly normalized 
automatically. For each \pt bin, the DCA distribution of mis-identified 
prompt hadrons has a narrow Gaussian peak at $\DCAR=0$. The broad tails 
for large $|$\DCAR$|$ are mainly caused by decays of $\Lambda$ and 
$K_S^0$. In all \pt bins the magnitude of this background is no more than 
10\% of the data for all \DCAR

\subsubsection{High-multiplicity background}
\label{sec:dca_sideband}

Due to the high multiplicity in \auau collisions, an electron candidate 
track in the central arms can be associated with random VTX hits. Such 
random associations can cause a background that has a very broad \DCAR 
distribution. Although the total yield of this background is only $\simeq$ 
0.1\% of the data, its contribution is significant at large \DCAR where we 
separate $b\rightarrow e$ and $c \rightarrow e$.

To evaluate the effect of event multiplicity on the reconstruction 
performance, we embed simulated single electrons---{\it i.e.} the response 
of the PHENIX detector to single electrons that is obtained from a \geant 
simulation---into data events containing VTX detector hits from real \auau 
collisions.  The events are then processed through the standard 
reconstruction software to evaluate the reconstruction performance in MB 
\auau collisions.

%%%%%%%%%%%%%%%%%%%%%%%%%%%%%%%%%%%%%%%%%%%%%% Fig_6
\begin{figure}[!hbt]
\includegraphics[width=1.0\linewidth]{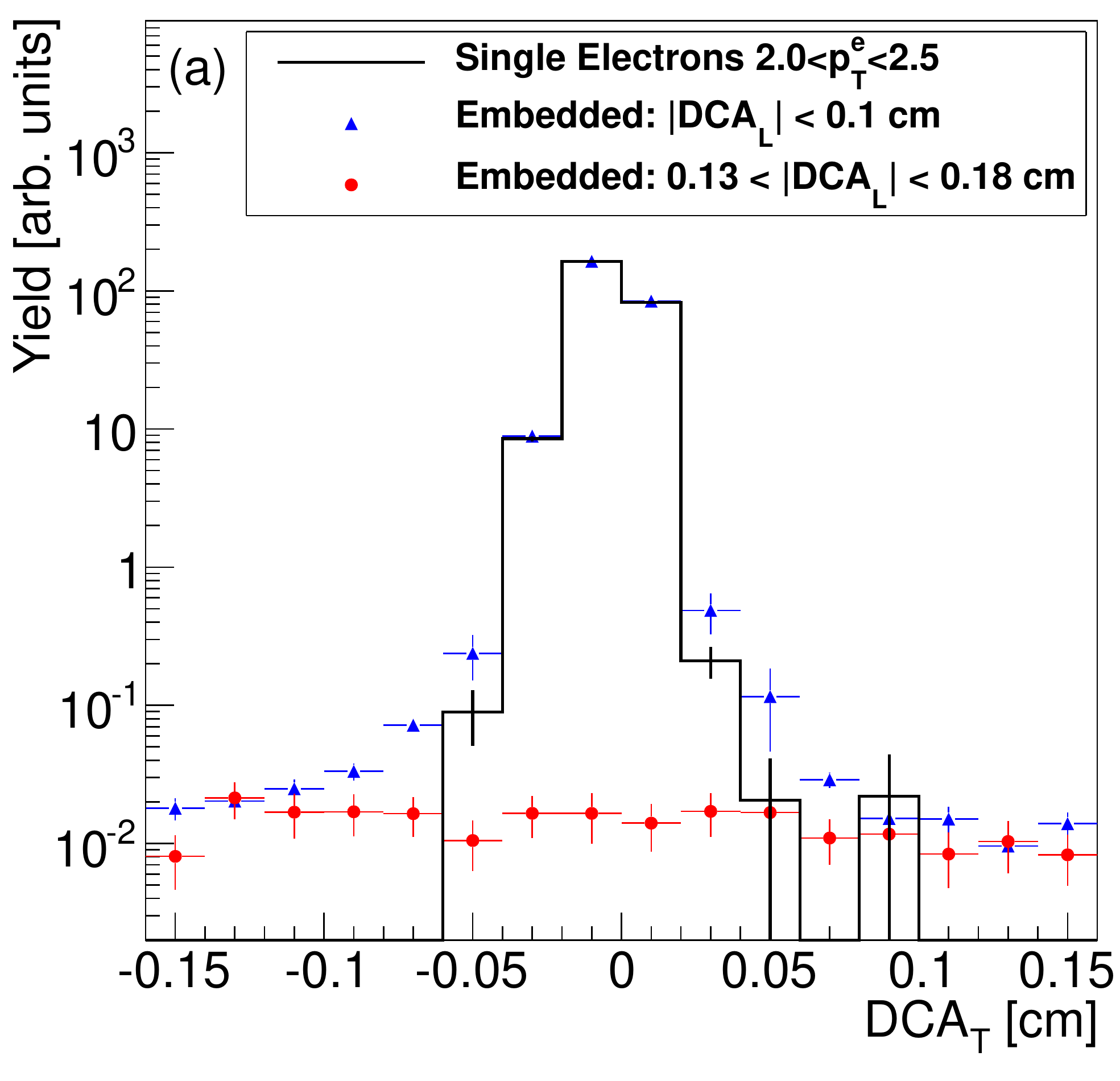}
\includegraphics[width=1.0\linewidth]{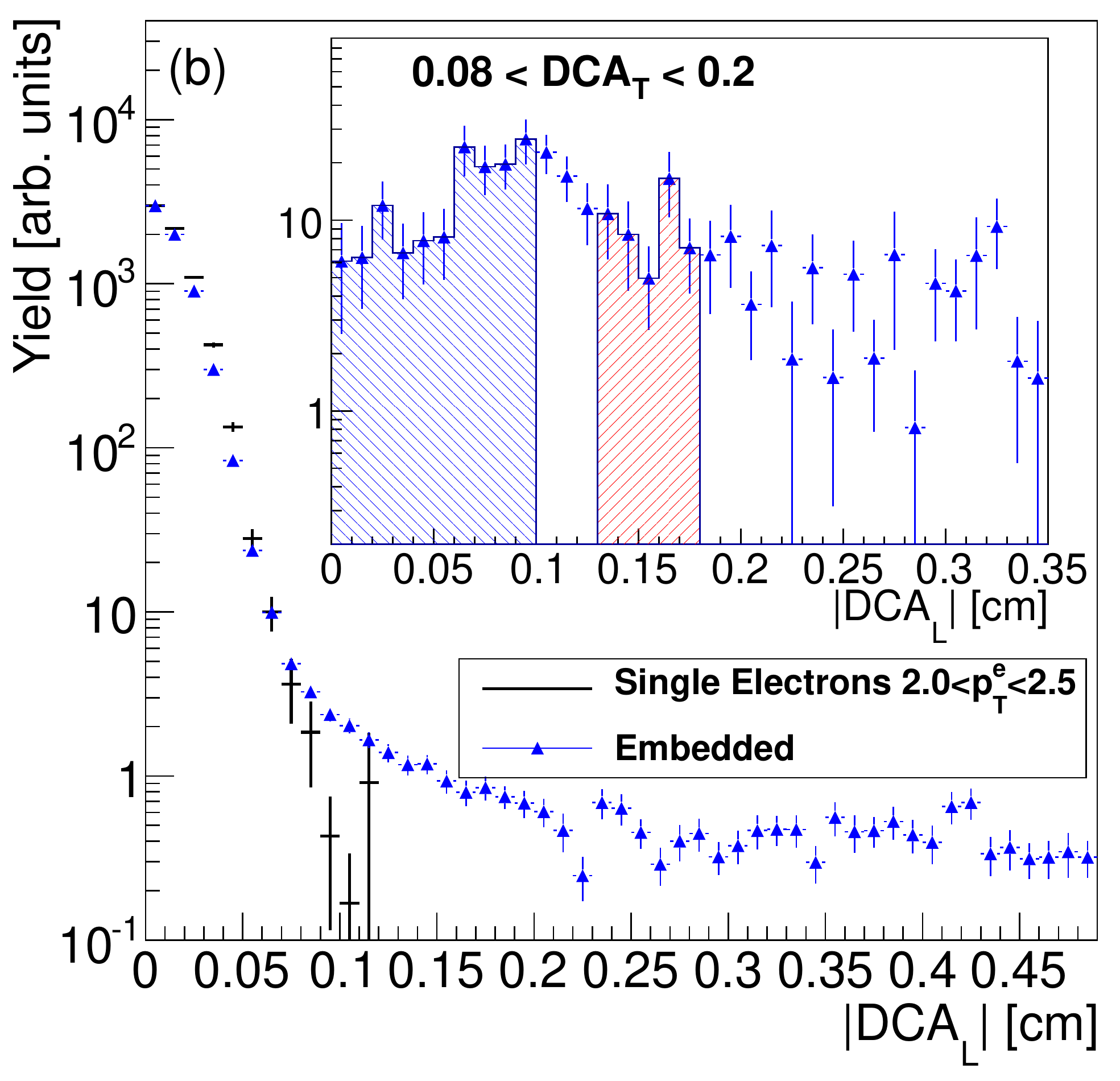}
\caption{\label{fig:embedSideband} (Color Online) 
Simulated primary electron (a) \DCAR and (b) \DCAZ distribution before and 
after embedding in real \auau data.
}
\end{figure}

The reconstructed \DCAR and \DCAZ for embedded primary electrons in MB 
\auau collisions is shown in Fig.~\ref{fig:embedSideband}. Here the 
histograms, labeled as ``Single Electrons", show the reconstructed \DCAR 
and \DCAZ distributions of primary electrons before embedding. The \DCAR 
distribution comprises a narrow Gaussian with no large \DCAR tail and 
the \DCAZ distribution comprises a similar, but slightly broader, 
Gaussian with no large tail. The blue filled triangles show the \DCAR and 
\DCAZ distributions after embedding. The \DCAR and \DCAZ distributions 
comprise a Gaussian peaked at $\DCAR(\DCAZ)\sim0$ which is consistent 
with the distribution before embedding. This demonstrates that the DCA 
resolution of the VTX is not affected by the high multiplicity 
environment. However, the embedded distributions have broad tails at large 
$|\DCAR|$ and $|\DCAZ|$.

As shown in Fig.~\ref{fig:embedSideband}(b), tracks with $|\DCAZ|>0.13$~cm 
are dominated by random associations, as they are not present in the 
``Single Electron'' sample. We therefore use the \DCAR distribution for 
tracks with large $|\DCAZ|$ as an estimate of this random 
high-multiplicity background. We choose the region $0.13<|\DCAZ|\ {\rm 
cm}<0.18$ to represent this background, and restrict our signal to 
$|\DCAZ|<0.1$~cm. The \DCAR distribution of tracks with $0.13<|\DCAZ|\ 
{\rm cm}<0.18$ must be normalized in order to be used as an estimate of 
the high-multiplicity background for tracks within $|\DCAZ|<0.1$~cm. This 
normalization is determined by matching the integrated yield of embedded 
primary electrons in each $|\DCAZ|$ region for $0.08<\DCAR\ {\rm cm}<0.2$, 
as shown in the inlay of Fig.~\ref{fig:embedSideband}(b). The region 
$0.08<\DCAR\ {\rm cm}<0.2$ is dominated by random associations, as shown 
in Fig.~\ref{fig:embedSideband}(a), and is therefore safe to use for 
determining the normalization. The normalization of the high-multiplicity 
background is determined to be $2.89\pm0.29$. The red filled circles in 
Fig.~\ref{fig:embedSideband}(a) show the embedded \DCAR distribution with 
large \DCAZ ($0.13<|\DCAZ|\ {\rm cm}<0.18$). This distribution agrees with 
the embedded \DCAR distribution (blue filled triangles in 
Fig.~\ref{fig:embedSideband}) for large \DCAR. This demonstrates that the 
tails for large \DCAR are well normalized by the distribution of electrons 
with large \DCAZ. However, there is a small excess in the region 
$0.05<|\DCAR|\ {\rm cm}<0.10$ that is not accounted for by the 
distribution with large \DCAZ. We address this excess in the systematic 
uncertainties, as described in Sec.~\ref{sec:unfolding_systematics}, where 
it is found to have only a small effect on the extraction of $b\rightarrow 
e$ and $c\rightarrow e$.

In each panel of Fig.~\ref{fig:DCA0} the high-multiplicity background is 
shown as a red line. It is determined from the \DCAR distribution of the 
data within $0.13<|\DCAZ|\ {\rm cm}<0.18$, as described above. The number 
of electron tracks in the large \DCAZ region is small. We therefore fit 
the resulting \DCAR data in each \pt bin with a smooth function to obtain 
the shape of the red curves shown in Fig.~\ref{fig:DCA0}. A second order 
polynomial is used in the lowest \pt bin, where there are enough 
statistics to constrain it. The higher \pt bins are fit with a constant 
value. All curves are multiplied by the same normalization factor, 
determined from embedded simulations as described above.

	\subsubsection{Photonic electrons and conversion veto cut}
	\label{sec:photonic_electrons}

%-Conversion and Dalitz components

Photon conversions and Dalitz decays of light neutral mesons ($\pi^{0}$ 
and $\eta$) are the largest electron background.  We refer to this 
background as photonic electron background as it is produced by external 
or internal conversion of photons.

The PHENIX Collaboration has previously published the yields of $\pi^{0}$ 
and $\eta$ mesons in \auau collisions at 
\sqsntwo~\cite{Adare:2008qa,Adare:2010dc}. In addition to the electrons 
from Dalitz decays of these mesons, the decay photons may convert to an 
$e^{+}e^{-}$ pair in the detector material in the beam pipe or each layer 
of the VTX.  The PHENIX Collaboration has also published the yields of 
direct photons in \auau collisions at 
\sqsntwo~\cite{Afanasiev:2012dg,Adare:2008ab}, that can also be a source 
for conversions.

In principle with these measured yields, combined with simple decay 
kinematics and a detailed \geant description of the detector material and 
reconstruction algorithm, one could fully account for these photonic 
electron contributions as a function of \DCAR and \pt. However, systematic 
uncertainties on the measured yields for the $\pi^{0}$, $\eta$, and direct 
photons would then dominate the uncertainty of the heavy flavor electron 
extraction. Therefore, we utilize the VTX detector itself to help reject 
these contributions in a controlled manner.

We require that at least the first three layers of the VTX have hits 
associated with the electron track. Conversions in B1 and subsequent 
layers are rejected by the requirement of a B0 hit, leaving only 
conversions in B0 and the beam pipe. The requirement of B1 and B2 hits 
enables us to impose a conversion veto cut, described below, that 
suppresses conversions from the beam pipe and B0.

%%%%%%%%%%%%%%%%%%%%%%%%%%%%%%%%%%%%%%%%%%%%%% Fig_7
\begin{figure}[!hbt]
\includegraphics[width=1.0\linewidth]{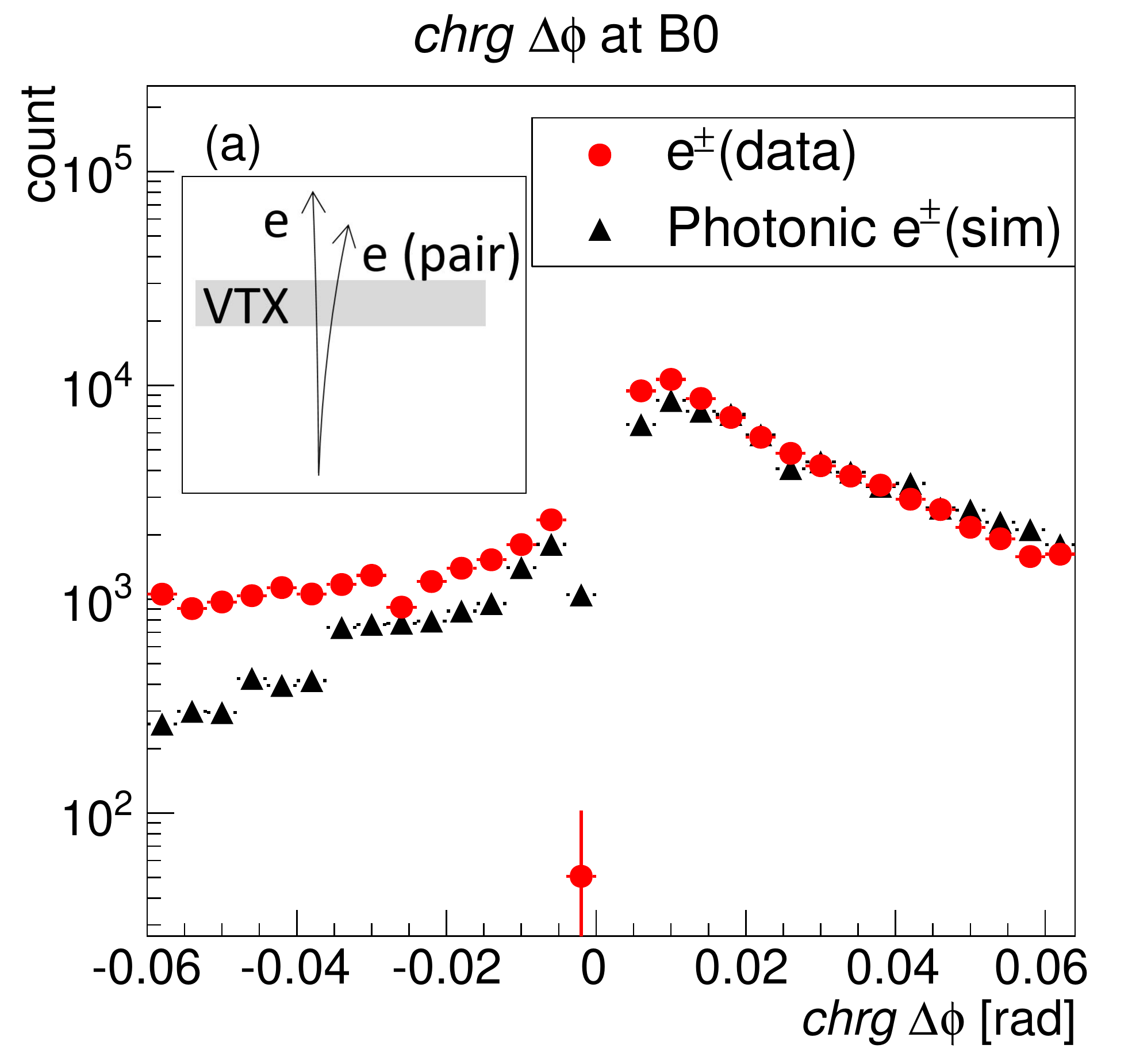}
\includegraphics[width=1.0\linewidth]{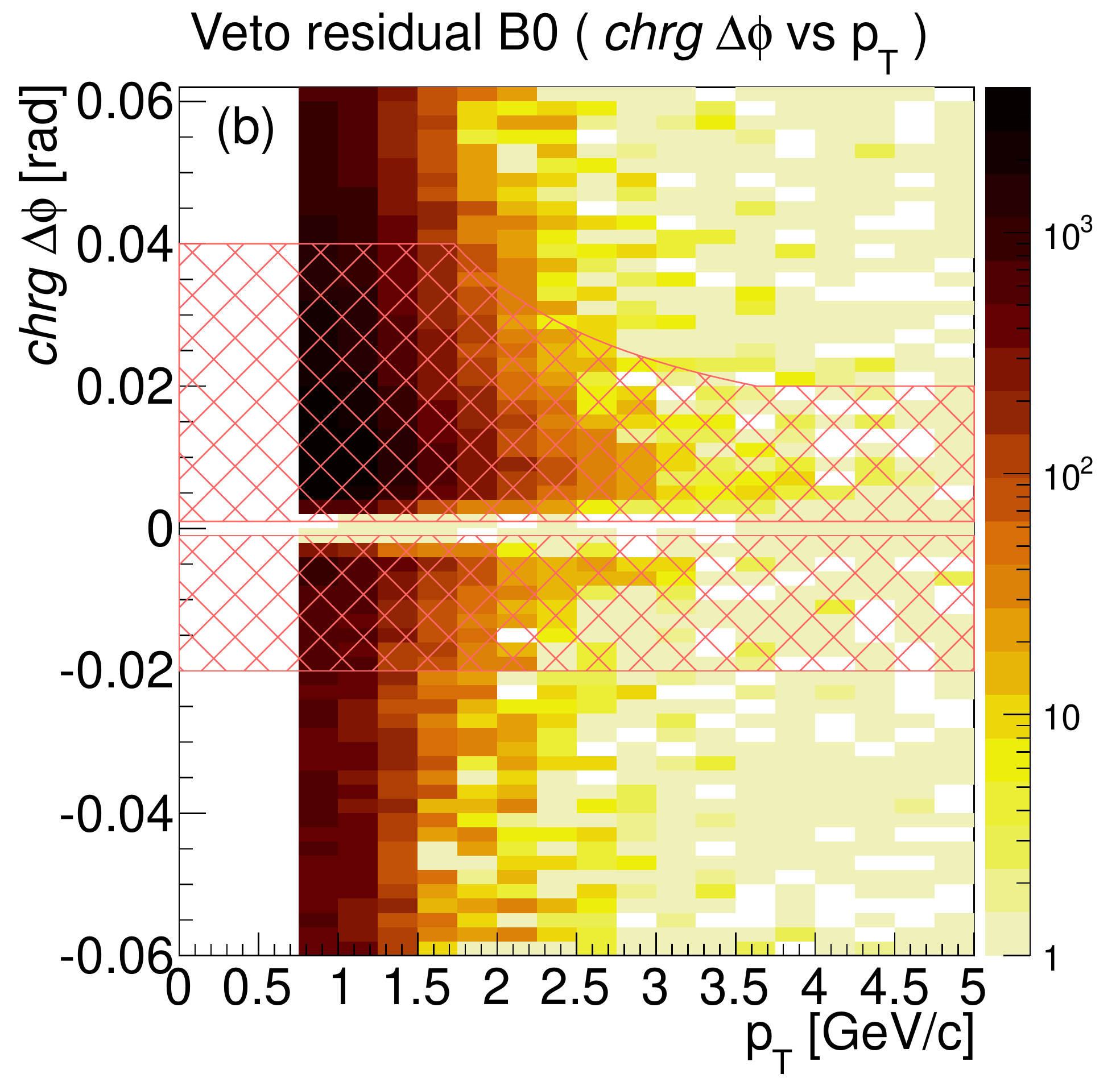}
\caption{\label{fig:convveto} (Color Online) (a) Distribution of correlated hits in B0 near 
electron tracks for $1<p_T<2$ \gev. The red (circle) points are from \auau data and 
the black (triangle) points are from Monte Carlo simulation. The insert in (a) 
illustrates the electron pairs from Dalitz decays. (b) The window of the 
conversion veto cut for B0 layer (hatched) and the hit distribution near 
electron track in 2D space of $chrg\ \Delta\phi$ vs \pt of electrons in 
\auau collisions. (See the text for details).}
\end{figure}

The conversion veto cut rejects tracks with another VTX hit within a 
certain window in $\Delta \phi$ and $\Delta z$ around hits associated with 
a VTX-associated track. Photons that convert to an $e^{+}e^{-}$ pair in 
the beam pipe will leave two nearby hits in the first layer (B0) and/or 
subsequent layers of the VTX, and thus be rejected by the conversion veto 
cut. Similarly, conversions in B0 will result in two nearby hits in the 
second layer (B1) and/or subsequent outer layers. The same is true for 
$e^{+}e^{-}$ from a Dalitz decay, though with a larger separation due to a 
larger opening angle of the pair.

Figure~\ref{fig:convveto}(a) shows distribution of $chrg\ \Delta 
\phi$ of hits in B0 relative to the electron track, where $chrg$ is the 
charge of the track. The red (circle) histogram shows the data in MB \auau 
collisions. If the track at the origin is not an electron, we have a flat 
distribution due to random hits in the detector. These random hits have 
been subtracted in Fig.~\ref{fig:convveto}(a). The transverse momentum of 
the electron track is in the interval $1<p_T\ \gev<2$.

As mentioned above, these correlated hits around electron tracks are 
caused by the partner $e^+$ or $e^-$ of Dalitz decays or photon 
conversions. The left-right asymmetry of the distribution is caused by the 
fact that the partner $e^\pm$ track is separated from the electron track 
by the magnetic field and the direction of the separation is determined by 
the charge of the electron track. In the distribution of $chrg\ 
\Delta \phi$, the partner track is bent towards the positive direction.

The black (triangle) histogram in Fig.~\ref{fig:convveto}(a) shows the distribution 
from Monte Carlo simulations. In the simulation, the response of the 
PHENIX detector to single $\pi^0$s is modeled by \geant, and the resulting 
hits in the VTX and the central arms are then reconstructed by the same 
reconstruction code as the data. The correlated hits in the simulation are 
caused by the Dalitz decay of $\pi^0$ and photon conversion in the 
material of the beam pipe and the VTX itself. The simulation reproduces 
the data well for $chrg\ \Delta \phi > 0$. There is a difference 
between the data and the simulation for $chrg\ \Delta \phi < 0$. 
This is caused by a subtle interplay between the conversions and high 
multiplicity effects. The difference disappears for peripheral collisions. 
Similar correlated hits are observed in B1 to B3 layers in the data and 
they are also well explained by the simulation.

We define a ``window" of the conversion veto cut around an electron track 
in each layer B0 to B3 and require that there is no hit other than the hit 
associated with the electron track in the window. Since a photonic 
electron (Dalitz and conversion) tends to have a correlated hit in the 
window, as one can see in Fig.~\ref{fig:convveto}, this conversion veto 
cut rejects photonic background. A larger window size can reject photonic 
background more effectively, but this can also reduce the efficiency for 
the heavy flavor electron signal due to random hits in the window. The 
window for the conversion veto cut is a compromise in terms of the 
rejection factor on photonic backgrounds and efficiency for heavy flavor 
electrons. We optimized the size of the window of the conversion veto cut 
based on a full \geant simulation.

The red hatched area shown in Fig.~\ref{fig:convveto}(b) shows the window 
of the conversion veto cut in layer B0. The window size is asymmetric 
since correlated hits are mainly in the positive side of $chrg\ 
\Delta \phi$. The window size is reduced for higher electron \pt since the 
distribution of correlated hits becomes narrower for higher \pt. The 
windows for B1-B3 are similarly determined based on \geant simulation.
 
\begin{figure}[!hbt]
\includegraphics[width=1.0\linewidth]{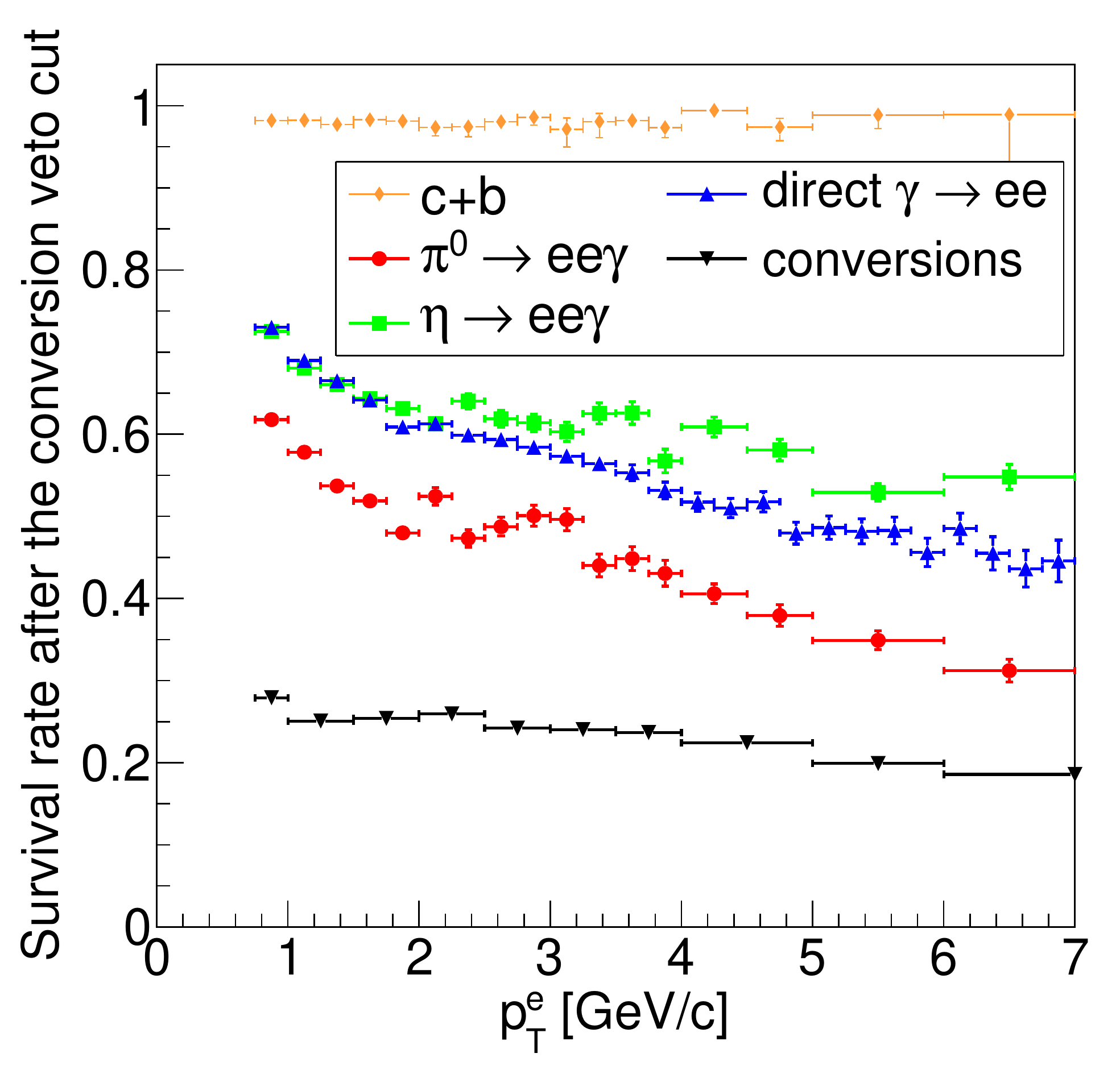}
\caption{\label{fig:survival_rate} (Color Online) 
The survival rate as a function of electron \pt (\pte) for electrons from 
photon conversion (black), Dalitz decay of $\pi^0$ (red), $\eta$ (green), 
electrons from direct photon (blue) and heavy flavor decay electrons (dark 
orange). %The black
}
\end{figure}

Figure~\ref{fig:survival_rate} shows the survival fraction of the 
conversion veto cut for electrons from photon conversions and Dalitz 
decays as a function of electron \pt from a full \geant simulation of the 
detector with hits run through the reconstruction software. The survival 
probability for conversions is less than 30\% at $\pt=1$ \gev and 
decreases further at higher \pt. The survival probability for Dalitz 
decays is higher since a Dalitz decay partner is more likely to fall 
outside of the window of the conversion veto cut due to the larger opening 
angle. Also shown in Fig.~\ref{fig:survival_rate} is the survival fraction 
of electrons from heavy flavor decays which pass the conversion veto cut 
($S_{\rm HF}$). As expected, their efficiency for passing the conversion 
veto cut is quite high and \pt independent.

The efficiencies shown in Fig.~\ref{fig:survival_rate} are calculated 
without the \auau high-multiplicity that may randomly provide a hit 
satisfying the conversion veto cut. Since these are random coincidences, 
they are a common reduction for all sources including the desired signal 
--- heavy flavor electrons. This common reduction factor, 
$\delta_{random}$, is measured from the reduction of the hadron track 
yield by the conversion veto cut to be $\simeq$ 35\% at $\pt=1$ \gev to 
$\simeq$ 25\% at $\pt=5$ \gev for MB \auau collisions. Note that 
when we determine the \DCAR distribution of the various background 
components using a full \geant simulation we apply the same conversion 
veto cuts.

%%%%%%%%%%%%%%%%%%%%%%%%%%%%%%%%%%%%%%%%%%%%%% Fig_8

The \DCAR distributions from photonic background processes that survive 
the conversion veto cut are shown in Fig.~\ref{fig:DCA0}. The means of the 
\DCAR distributions from Dalitz decays and conversions are shifted to 
negative \DCAR values due to the mis-reconstruction of the momentum caused 
by the assumption that the tracks originate at the primary vertex, as 
explained in the next paragraph. The shift is largest at the lowest \pt 
bin and decreases with increasing \pt.

For Dalitz electrons, the shift is due to the energy loss via induced 
radiation (bremsstrahlung). The total radiation length of the VTX is 
approximately 13\% as shown in Table~\ref{tab:vtx}. Thus a Dalitz electron 
coming from the primary vertex loses approximately 
$1-e^{-0.13}\approx{12}$\% of its energy on average when it passes through 
the VTX. The momentum measured by the DC is close to the one after the 
energy loss due to the reconstruction algorithm. Since the momentum 
determined by the DC is used when projecting inward from the hit in B0 to 
the primary vertex and in calculation of \DCAR, this results in a slight 
shift in the \DCAR distribution. This effect is fully accounted for in the 
\DCAR template of Dalitz electrons since it is generated through the full 
\geant and reconstruction simulation.

In the case of conversions, the effect is even larger, as one can clearly 
see in Fig.~\ref{fig:DCA0}. While a photon goes straight from the primary 
vertex to the beam pipe or B0 layer where it converts, \DCAR is calculated 
assuming that the electron track is bent by the magnetic field. Thus the 
\DCAR distribution is shifted by the difference of the actual straight 
line trajectory and the calculated bent trajectory. Again, this is fully 
accounted for with the full \geant simulation. The effect is verified by 
selecting conversion electrons with a reversed conversion veto cut.

\subsubsection{$K_{e3}$}

The background from $K_{e3}$ decays ($K^0_S$, $K^\pm$ $\rightarrow e \nu \pi$) 
contributes electrons over a broad range of \DCAR due to the long lifetime 
of the kaons. Both contributions are determined using \pythia and a full 
\geant simulation, taking into account the exact track reconstruction, 
electron identification cuts, and conversion veto cut. The resulting \DCAR 
distribution for these kaon decays is shown in Fig.~\ref{fig:DCA0}.  As 
expected, though the overall yield is small, this contributes at large 
\DCAR in the lower \pt bins and is negligible at higher \pt.

\subsubsection{Quarkonia}

Quarkonia ($J/\psi$ and $\Upsilon$) decay into electron pairs.  Due to the 
short lifetime, these decays contribute to electrons emanating from the 
primary vertex. The $J/\psi$ yields in \auau collisions at \sqsntwo have 
been measured by the PHENIX Collaboration~\cite{Adare:2006ns}.  The 
detailed modeling of these contributions out to high \pt is detailed in 
Ref.~\cite{Adare:2010de}. While these measurements include a small 
fraction of $B\rightarrow J/\psi$ decays, all $J/\psi$'s are considered 
prompt when modeling the \DCAR distribution.  The $J/\psi$ contribution is 
shown in Fig.~\ref{fig:DCA0}, and is quite small and peaked about \DCAR = 
0 as expected.  Thus, the systematic uncertainty from the quarkonium yields 
in \auau collisions is negligible in all electron \pt bins.

\subsection{Normalization of electron background components}
\label{sec:norm}

If the detector performance were stable, we could convert the \DCAR 
distributions from counts into absolutely normalized yields. Then one 
could straightforwardly subtract the similarly absolutely normalized 
background contributions described above---with the normalization 
constrained by the previously published PHENIX yields for $\pi^{0}$, 
$\eta$, etc. However, due to detector instability during the 2011 run, 
such absolute normalization of background contributions can have a large 
systematic uncertainty. Thus we bootstrap the relative normalization of 
these background contributions utilizing our published \auau 
results~\cite{Adare:2010de} from data taken in 2004.

The idea of the method is the following. PHENIX measured the invariant 
yield of open heavy flavor decay electrons from the 2004 dataset. In this 
2004 analysis we first measured inclusive electrons ({\it i.e.} the sum of 
background electrons and heavy flavor electrons). We then determined and 
subtracted the background electron components from the inclusive electron 
yields to obtain the heavy flavor contribution. Thus the ratio of the 
background components to the heavy flavor contribution were determined and 
published in~\cite{Adare:2010de}. We use these ratios to determine the 
normalization of background components in the 2011 data, as described in 
the next paragraph. Some backgrounds have the same ratio to signal 
regardless of the year the data was collected, while others will differ 
due to the additional detector material added by the VTX.

The invariant yield in \auau collisions at \sqsntwo of heavy flavor 
electrons and background electrons from Dalitz decays is a physical 
observable independent of the year the data was taken. Thus we can use the 
ratio of heavy flavor/Dalitz that is determined in the 2004 analysis in 
the 2011 data. On the other hand, the invariant yield of conversion 
electrons depends on the detector material present and is thus different 
in the 2011 data taking period with the VTX installed compared with the 
2004 data.  We account for this difference by calculating the fraction of 
nonphotonic electrons in the 2011 data. A detailed description of the 
normalization procedure is given in Appendix~\ref{app:norm}.

With this bootstrapped normalization completed, the correctly normalized 
background components are shown for all five \pt bins vs \DCAR in 
Fig.~\ref{fig:DCA0}. Note that the normalization of mis-identified hadron 
and random background is determined from the data as explained in 
sections~\ref{sec:hadron_contamination} and ~\ref{sec:dca_sideband}, 
respectively. The electron yield beyond the sum of these background 
components is from the combination of charm and bottom heavy flavor 
electrons.

	\subsection{Unfolding}
	\label{sec:unfolding}

	\subsubsection{Introduction}

With the \DCAR distributions as a function of electron \pt and the various 
background components in hand, we proceed to extract the remaining charm 
and bottom components.  If one knew the shape of the parent charm and 
bottom hadron \pt and rapidity distributions, one could calculate in 
advance the \DCAR shape for electrons from each heavy flavor via a model 
of the decay kinematics.  Since the decay lengths of charm and bottom 
hadrons are significantly different, they will yield different \DCAR 
distributions.  In this case, one could simultaneously fit the \DCAR 
distribution for each \pt bin with all background components fixed across 
\pt bins, and extract the one free parameter: the ratio of charm to bottom 
contributions.  However, the \pt distribution of charm hadrons is known to 
be significantly modified in \auau collisions --- see for example 
Ref.~\cite{Adamczyk:2014uip}.  For bottom hadrons this is also likely to 
be the case.  Therefore one does not know \textit{a priori} the heavy 
flavor \DCAR distribution since it depends on the parent \pt distribution.

Since the \DCAR distributions for all electron \pt result from the same 
parent charm and bottom hadron \pt spectrum, one can perform a 
simultaneous fit to all the electron \pt and \DCAR data in order to find 
the most likely heavy flavor parent hadron \pt distributions. The 
estimation of a set of most likely model parameters using a simultaneous 
fit to data is often referred to as unfolding. Statistical inference 
techniques are often employed to solve such problems; see for example the 
extraction of reconstructed jet cross sections~\cite{Cowan:2002in}.

The \DCAR distributions are in counts and have not been corrected for the 
\pt-dependent reconstruction efficiency in \auau collisions, and therefore 
hold no yield information. To further constrain the extraction of the 
charm and bottom components, we include the total heavy flavor electron 
invariant yield as measured by PHENIX~\cite{Adare:2010de} in \auau 
collisions at \sqsntwo.  This measurement is more accurate than currently 
available with the 2011 data set, where the VTX acceptance changes with 
time.

The unfolding procedure, using a particular sampling method (described in 
Section~\ref{sec:unfoldtechnique}), chooses a set of trial charm and 
bottom parent hadron yields. The trial set of yields is multiplied by a 
decay matrix (described in Section~\ref{sec:decay_matrix}), which encodes 
the probability for a hadron in a given \pt interval to decay to an 
electron at midrapidity as a function of electron \pt and \DCAR. The 
resulting distributions of electron \pt and \DCAR are compared with the 
measured data using a likelihood function (described in 
Section~\ref{sec:likelihood}). In order to dampen discontinuities and 
oscillatory behavior, a penalty upon the likelihood (described in 
Section~\ref{sec:prior}) is added to enforce smoothness in the resulting 
hadron \pt distributions.

\subsubsection{Unfolding method}
\label{sec:unfoldtechnique}

Here we apply Bayesian inference techniques to the unfolding problem. A 
detailed pedagogical introduction to these techniques is given in 
Ref.~\cite{Choudalakis:2012hz}. Techniques involving maximum likelihood 
estimation or maximum \textit{a posteriori} estimation, often used in 
frequentist statistics, can at best compute only a point estimate and 
confidence interval associated with individual model parameters. In 
contrast, Bayesian unfolding techniques have the important advantage of 
providing a joint probability density over the full set of model 
parameters. In this analysis, the vector of model parameters, $\thetavec$, 
is the vector of parent charm and bottom hadron yields binned in \pt.

Given a vector of measured data, $\xvec$, and our vector of model 
parameters, $\thetavec$, we use Bayes' theorem
\begin{equation}\label{eq:bayesrule}
  \post = \frac{\like \prior}{P(\xvec)},
\end{equation}

to compute the posterior probability density $\post$ from the likelihood 
$\like$ and prior information $\prior$. The function $\like$, quantifies 
the likelihood of observing the data given a vector of model parameters. 
In frequentist statistics, the $\like$ is often used alone to determine 
the best set of model parameters. Bayesian inference, on the other hand, 
allows for the inclusion of the analyzer's \textit{a priori} knowledge 
about the model parameters, as encoded in $\prior$. The implementation of 
$\prior$ used in this analysis is discussed in Sec.~\ref{sec:prior}. The 
denominator $P(\xvec)$ serves as an overall normalization of the combined 
likelihood $\like\prior$ such that $\post$ can be interpreted as a 
probability density. In this analysis, $\post$ gives the probability for a 
set of charm and bottom hadron yields,

\begin{equation}
\thetavec = (\thetavec_c; \thetavec_b), 
\end{equation}
given the values of the measured electron data points $\xvec$. Since we 
are only interested in the parameters which maximize $\post$, we can 
dispense with the calculation of $P(\xvec)$, as it serves only as an 
overall normalization.

Here $\thetavec$ comprises 17 bins of both charm and bottom hadron \pt, 
yielding a 34-dimensional space which must be sampled from in order to 
evaluate $\post$. To accomplish this we employ a Markov Chain Monte Carlo 
(MCMC) algorithm to draw samples of $\thetavec$ in proportion to $\post$. 
This makes accurate sampling of multidimensional distributions far more 
efficient than uniform sampling. In implementation, it is in fact the 
right hand side of Eq.~\ref{eq:bayesrule} that is sampled. The MCMC 
variant used here is an affine-invariant ensemble sampler described in 
Ref.~\cite{goodman2010} and implemented as described in 
Ref.~\cite{ForemanMackey:2012ig}. It is well suited to distributions that are highly 
anisotropic such as spectra which often vary over many orders of 
magnitude.

\subsubsection{Modeling the likelihood function}
\label{sec:likelihood}

This analysis is based on 21 data points of total heavy flavor electron 
invariant yield, $\eptdata$, in the range 1.0--9.0 \gev from the 2004 data 
set~\cite{Adare:2010de}, and five electron \DCAR distributions 
$\dcadata{j}$, where $j$ indexes each electron \pt interval within the 
range 1.5--5.0 \gev from the 2011 data set. Therefore,
\begin{equation}
\xvec=(\eptdata,\dcadata{0},\dcadata{1},\dcadata{2},\dcadata{3},\dcadata{4})
\end{equation}
in Eq.~\ref{eq:bayesrule}.

Our ultimate goal is to accurately approximate the posterior distribution 
over the parent hadron invariant yields $\thetavec$ by sampling from it. 
For each trial set of hadron yields, the prediction in electron \pt, 
$\epttheta$, and \DCAR, $\dcatheta{j}$, is calculated by
\begin{eqnarray}
\epttheta &= \My\thetavec_c + \My\thetavec_b \\
\dcatheta{j} &= \Md\thetavec_c + \Md\thetavec_b,
\end{eqnarray}
where $\My$ and $\Md$ are decay matrices discussed in 
Section~\ref{sec:decay_matrix}. We then evaluate the likelihood between 
the prediction and each measurement in the data sets $\eptdata$ and 
$\{\dcadata{j}\}_{j=0}^4$. As is customary, the logarithm of the 
likelihood function is used in practice. The combined (log) likelihood for 
the data is explicitly
\begin{equation} \label{eq:logl} \small
  \ln \like =  
  \ln P(\eptdata|\epttheta) + \sum_{j=0}^4 \ln P(\dcadata{j}|\dcatheta{j}).
\end{equation}

The $\eptdata$ dataset is assigned statistical uncertainties that are 
assumed to be normally distributed and uncorrelated. Thus, the likelihood 
$\ln P(\eptdata|\epttheta)$ is modeled as a multivariate Gaussian with 
diagonal covariance. The systematic uncertainties on the $\eptdata$ 
dataset and their effect on the unfolding result are discussed in 
Sec.~\ref{sec:unfolding_systematics}.

The \DCAR data sets, in contrast, each comprise a histogrammed 
distribution of integer-valued entries, and the likelihood $\ln 
P(\dcadata{j}|\dcatheta{j})$ is thus more appropriately described by a 
multivariate Poisson distribution. However, the likelihood calculation for 
the \DCAR data sets requires three additional considerations. First, there 
are significant background contributions from a variety of sources, as 
discussed in Section~\ref{sec:dca_background}. Secondly, detector 
acceptance and efficiency effects are not explicitly accounted for in the 
\DCAR distributions. This implies that the total measured yield of signal 
electrons in each \DCAR histogram is below what was actually produced, and 
consequently the measured $\dcadata{j}$ distributions do not match the 
predictions in normalization.  Lastly, because of the high number of 
counts in the region near \DCAR$=0$, this region will dominate the 
likelihood and be very sensitive to systematic uncertainties in the \DCAR 
shape there, even though the main source of discrimination between charm 
and bottom electrons is at larger \DCAR.

To deal with the first issue, the relatively normalized background 
described in Sec.~\ref{sec:dca_background} is added to each prediction of 
the \DCAR distribution for summed electrons from charm and bottom hadrons 
so that the shape and relative normalization of the background component 
of the measurement is accounted for.

To handle the second, each prediction plus the background is scaled to 
exactly match the normalization of $\dcadata{j}$. In this way, only the 
shape of the prediction is a constraining factor.

To deal with the third, a 5\% uncertainty is added in quadrature to the 
statistical uncertainty when the number of counts in a given \DCAR bin is 
greater than a reasonable threshold (which we set at 100 counts). This 
accounts for the systematic uncertainty in the detailed \DCAR shape by 
effectively de-weighting the importance of the region $\DCAR\approx$0 
while maintaining the overall electron yield normalization (as opposed to 
removing the data entirely). This additional uncertainty also necessitates 
changing the modeling of $\ln P(\dcadata{j}|\dcatheta{j})$ from a Poisson 
to a Gaussian distribution. We have checked that varying both the 
additional uncertainty and the threshold at which it is added has little 
effect on the results.

	\subsubsection{Decay model and matrix normalization}
	\label{sec:decay_matrix}

The \pythia-6~\cite{Sjostrand:2006za} generator with heavy flavor 
production process included, via the parameter MSEL=4(5), is used to 
generate parent charm (bottom) hadrons and their decays to electrons. 
Electrons within $|\eta|<0.35$ decayed from the ground state charm hadrons 
($D^{\pm}$, $D^0$, $D_s$, and $\Lambda_c$) or bottom hadrons ($B^{\pm}$, 
$B^0$, $B_s$, and $\Lambda_b$) are used to create a decay matrix between 
hadron \pt (\pth, representing charm hadron \pt, \ptc, or bottom hadron 
\pt, \ptb) and electron \pt (\pte) and \DCAR. Here we treat the feed down 
decay $B\rightarrow D\rightarrow e$ as a bottom hadron decay and exclude 
it from charm hadron decays.

%%%%%%%%%%%%%%%%%%%%%%%%%%%%%%%%%%%%%%%%%%%%%% Fig_9
\begin{figure}[!hbt]
	\includegraphics[width=1.0\linewidth]{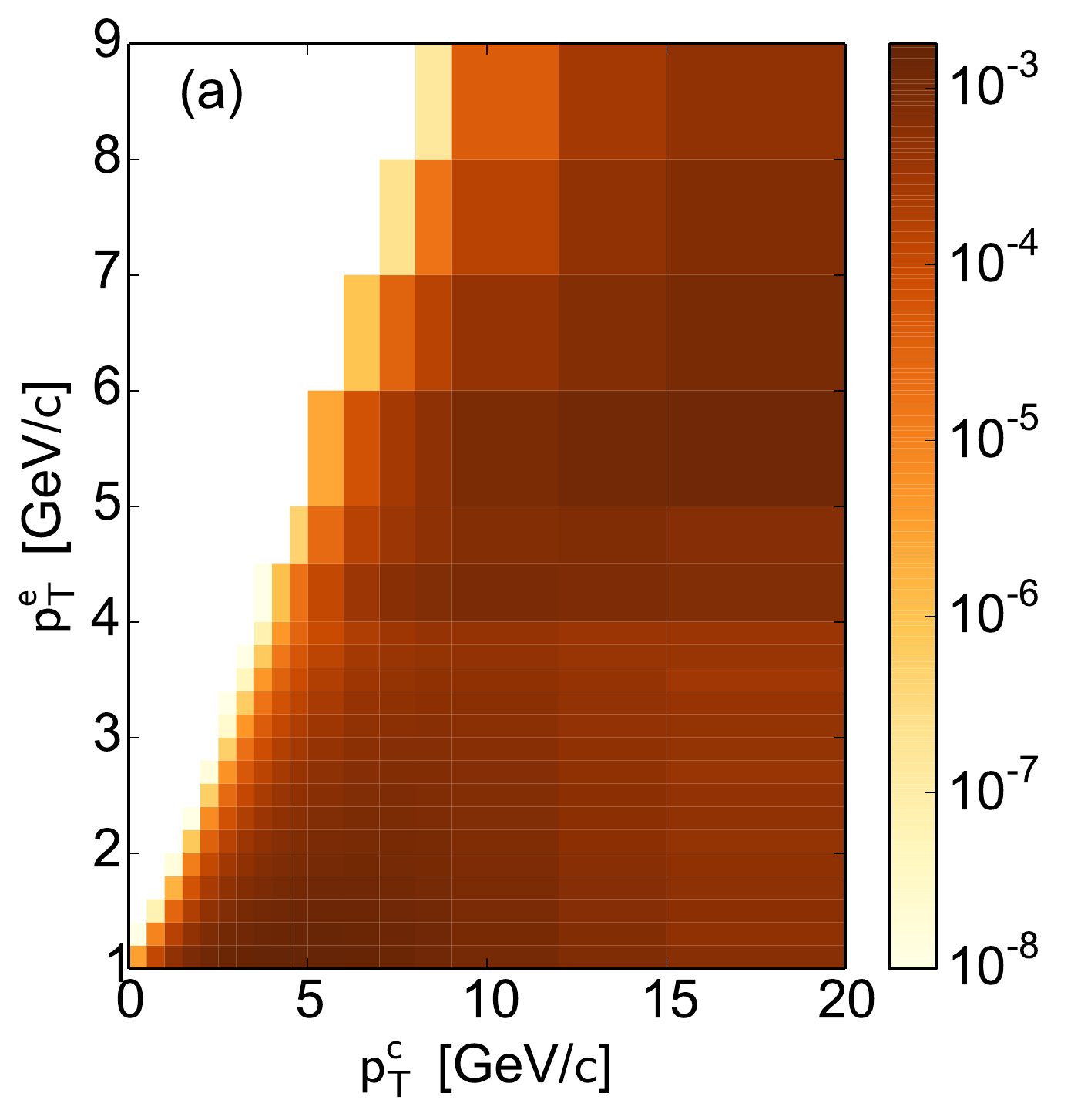}
	\includegraphics[width=1.0\linewidth]{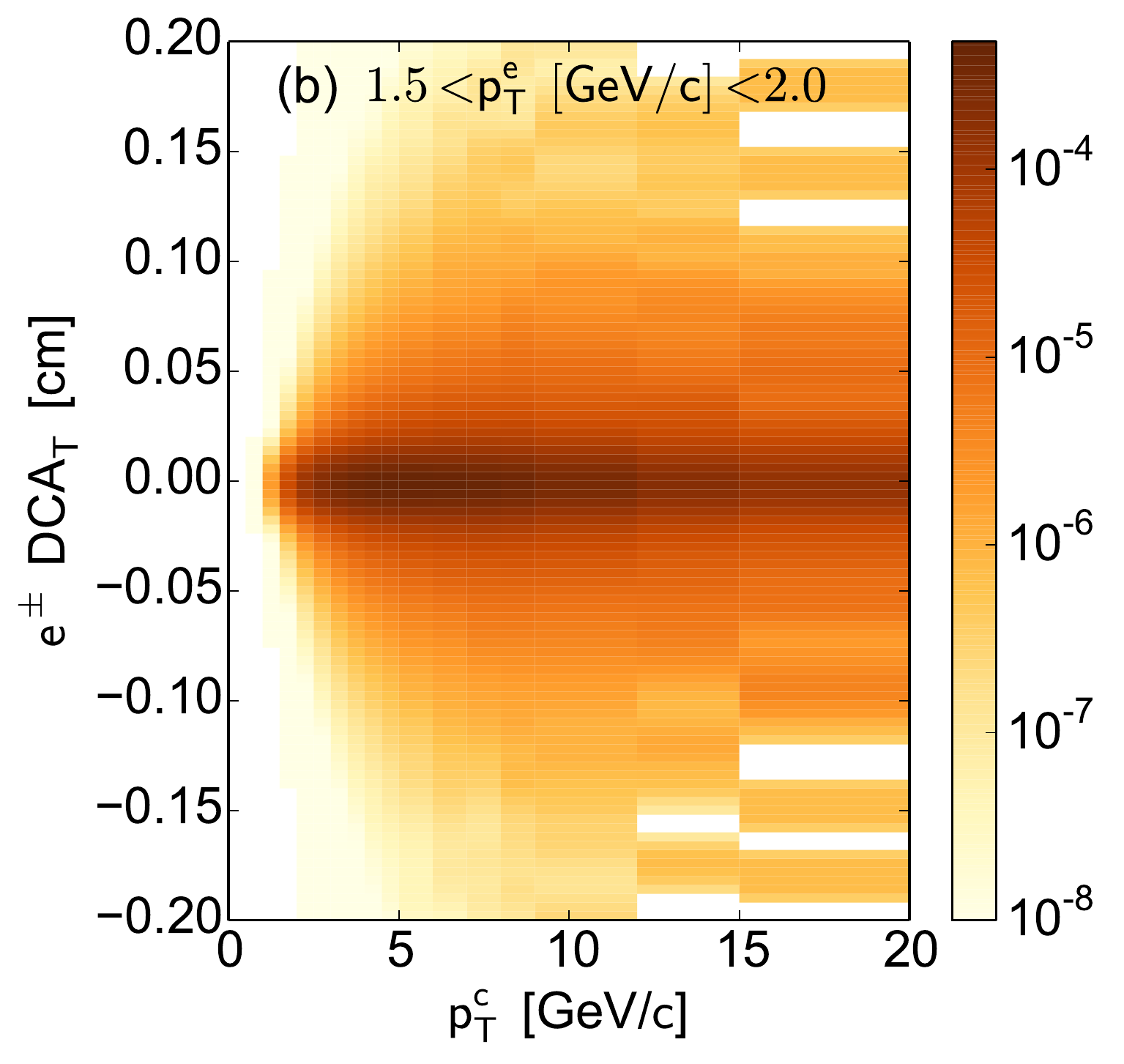}
\caption{(Color Online) 
(a) The decay matrix, $\My$, encoding the probability for charmed hadrons 
decaying to electrons within $|\eta|<0.35$ as a function of both electron 
\pt (\pte) and charm hadron \pt (\ptc). (b) An example decay matrix, 
$\Md$, encoding the probability for charmed hadrons decaying to electrons 
within $|\eta|<0.35$ and $1.5<\pte\ [{\rm~GeV}/c]<2.0$ as a function of both 
electron \DCAR and charm hadron \pt (\ptc). In both cases the color 
intensity represents the probability of decay in the given bin.}
\label{fig:decaymat}
\end{figure}

\begin{figure}[!hbt]
    \includegraphics[width=0.98\linewidth]{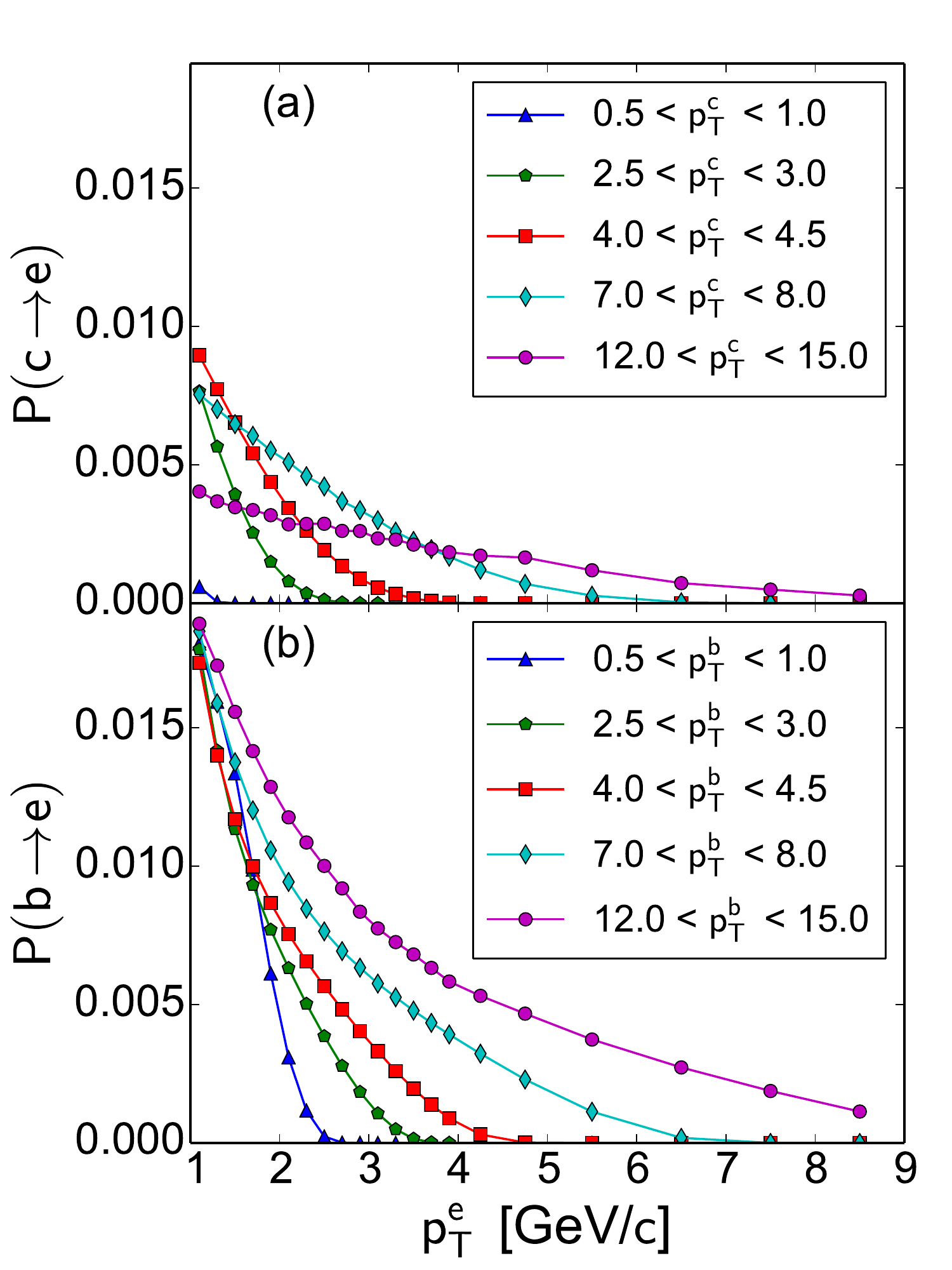}
  \caption{ (Color Online) 
The probability for (a) charm and (b) bottom hadrons in a given range of 
hadron \pt (\ptc and \ptb for charm and bottom hadrons respectively) to decay 
to electrons at midrapidity as a function of electron \pt (\pte).
}
  \label{fig:decayprob}
\end{figure}

The probability for a charm or bottom hadron at a given \pth to decay to an 
electron at a given \pte and \DCAR is encoded in the multidimensional 
matrices $\My$ and $\Md$. An example decay matrix for charmed hadrons is 
shown in Fig.~\ref{fig:decaymat}. Note that the 17 bins in \ptc correspond 
to the same bins shown along the $x$-axis in Fig.~\ref{fig:hadron-pt}, and 
that the binning in \pte and \DCAR seen in Fig.~\ref{fig:decaymat} is the 
same as that shown in Fig.~\ref{fig:ept-refold} and 
Fig.~\ref{fig:dca-refold} respectively. Furthermore, note that the marginal 
probabilities do not integrate to unity in these matrices. This is because 
the decay probabilities are normalized to the number of hadrons that are 
generated at all momenta, in all directions, and over all decay channels. 
The probability distribution for a hadron integrated over all rapidities and 
decay channels within a given \pth range to decay to an electron at \midy 
with a given \pte (integrated over \DCAR) is shown in 
Fig.~\ref{fig:decayprob} for an example set of \pth bins.

%%%%%%%%%%%%%%%%%%%%%%%%%%%%%%%%%%%%%%%%%%%%%% Fig_10

In principle, this decay matrix introduces a model dependence to the 
result. In the creation of the decay matrix we are integrating over all 
hadron rapidities as well as combining a number of hadron species and 
their decay kinematics to electrons. This involves two assumptions. The 
first is that the rapidity distributions of the hadrons are unmodified. 
BRAHMS found that the pion and proton $R_{AA}$ did not depend strongly on 
rapidity up to $y\approx3$~\cite{Staszel:2005aw}, justifying the 
assumption. This assumption will further lead us to quote charm and bottom 
hadron yields as a function of \pt integrated over all rapidity. The 
second assumption is that all ground state charm hadrons experience the 
same modification as a function of \ptc. While different than the charm 
suppression, all bottom hadrons are assumed to experience the same 
modification. 

An enhancement in the baryon to meson production ratios in both 
nonstrange and strange hadrons has been measured at 
RHIC~\cite{Abelev:2006jr}, which may carry over into the heavy quark 
sector, invalidating the second assumption. While there are some 
models~\cite{MartinezGarcia:2007hf} that attempt to incorporate this 
anomalous enhancement into the charm hadrons to help explain the measured 
heavy flavor electron \raa, there are few measurements to help constrain 
this proposed enhancement. Following Ref.~\cite{Sorensen:2005sm}, we have 
tested the effect of this assumption by applying the observed baryon/meson 
enhancement to both the $\Lambda_c/D$ and $\Lambda_b/B$ ratios. As in 
Ref.~\cite{Sorensen:2005sm}, we assume that the modification 
asymptotically approaches 1 for hadron $\pt>8$~GeV/$c$. We find that 
including the enhancement gives a lower charm hadron yield at high-\pt and 
a larger bottom hadron yield at high-\pt, but the modifications are within 
the systematic uncertainties discussed in 
Sec.~\ref{sec:unfolding_systematics} and shown in 
Fig.~\ref{fig:hadron-pt}. We also find a larger bottom electron fraction, 
which is again within the systematic uncertainties shown in 
Fig.~\ref{fig:bfrac}. While we have not used other particle generators to 
create alternate decay matrices, we find that the $D^0$ and $D^{\pm}$ 
meson \pt and rapidity distributions from \pythia are similar to those 
given by Fixed Order + Next-to-Leading Log (\fonll) 
calculations~\cite{Cacciari:2005rk}. We have not included any systematic 
uncertainty due to this model dependence in the final result.

	\subsubsection{Regularization/prior}
	\label{sec:prior}

To penalize discontinuities in the unfolded distributions of charm and 
bottom hadrons, we include a regularization term to the right hand side of 
equation~\ref{eq:logl}. In this analysis we included a squared-exponential 
function
\begin{equation} \label{eq:l2reg}
  \ln \prior = -\alpha^2 \left(|\mathbf{L} \rvec_c|^2 + |\mathbf{L} \rvec_b|^2\right)
\end{equation}
where $\rvec_c$ and $\rvec_b$ are ratios of the charm and bottom 
components of the parent hadron $p_T$ vector to the corresponding 17 
components of the prior, $\thetavec_{{\rm prior}}$, and $\mathbf{L}$ is a 
17-by-17 second-order finite-difference matrix of the form
\begin{equation} \label{eq:lmatrix}
\mathbf{L} = \frac{17}{2}
\begin{pmatrix}
-1 &  1 &    &   & & & &\\
 1 & -2 &  1 &   & & & &\\
   &  1 & -2 & 1 & & & &\\
& & \ddots & \ddots & \ddots & & & &\\
& & & \ddots & \ddots & \ddots & & & \\
& & & & 1 & -2 & 1 &\\
& & & & & 1 & -2 & 1\\
& & & & & & 1 & -1
\end{pmatrix}.
\end{equation}

Thus the addition of this term encodes the assumption that departures from 
$\thetavec_{{\rm prior}}$ should be smooth by penalizing total 
curvature as measured by the second derivative.

Here, $\alpha$ is a regularization parameter set to $\alpha=1.0$ in this 
analysis. We determine $\alpha$ by repeating the unfolding procedure, 
scanning over $\alpha$ and choosing the value of $\alpha$ which maximizes 
the resulting sum of Eq.~\ref{eq:logl} and $-\left(|\mathbf{L} \rvec_c|^2 
+ |\mathbf{L} \rvec_b|^2\right)$ (Eq.~\ref{eq:l2reg} dropping $\alpha^2$). 
In this way we can directly compare log likelihood values for unfolding 
results with different $\alpha$ values. We include variations on $\alpha$ 
in the systematic uncertainty as described in 
Section~\ref{sec:unfolding_systematics}.

We set $\thetavec_{{\rm prior}}$ to \pythia charm and bottom hadron \pt 
distributions scaled by a modified blast wave 
calculation~\cite{Adare:2013wca} which asymptotically approaches \raa 
values of 0.2(0.3) for $D$($B$) mesons at high-\pt. We have tested the 
sensitivity of the result to $\thetavec_{{\rm prior}}$ by alternatively 
using unmodified \pythia charm and bottom hadron \pt distributions. We 
find that the result is sensitive to the choice of $\thetavec_{\rm prior}$ 
dominantly in the lowest charm hadron \pt bins, where there is minimal 
constraint from the data. We have included this sensitivity in the 
systematic uncertainty as discussed in 
Section~\ref{sec:unfolding_systematics}.

	\subsubsection{Parent charm and bottom hadron yield and their 
                       statistical uncertainty} 
	\label{sec:marg}

%%%%%%%%%%%%%%%%%%%%%%%%%%%%%%%%%%%%%%%%%%%%%% Fig_11
\begin{figure*}[!hbt]
\includegraphics[width=0.98\linewidth]{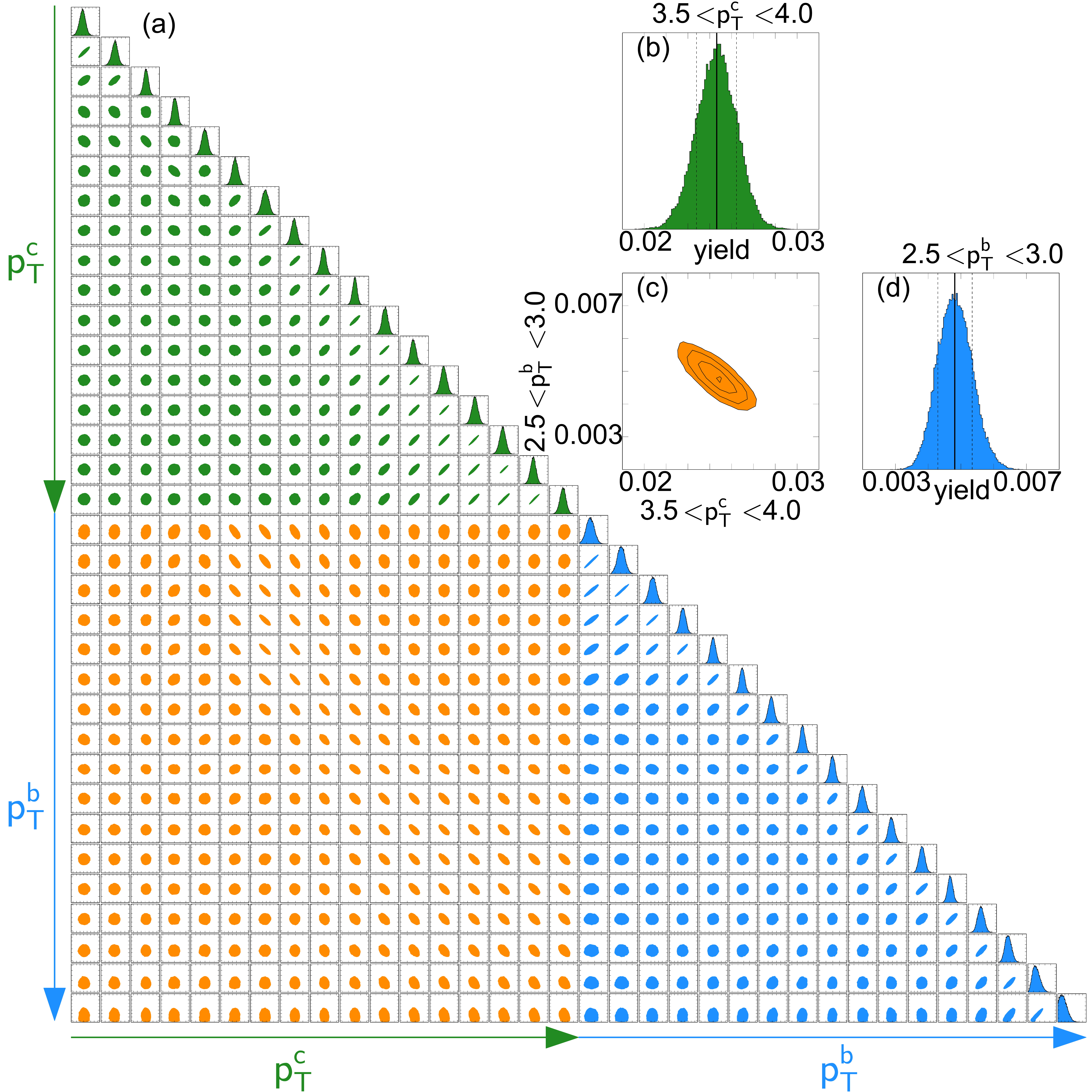}
\caption{(Color Online) 
The joint probability distributions for the vector of hadron yields, 
$\thetavec$, showing the 2-D correlations between 
parameters. The diagonal plots show the marginalized probability 
distributions for each hadron \pt bin (i.e. the 1-dimensional projection 
over all other parameters). Along the Y-axis the plots are organized from 
top to bottom as the 17 charm hadron \pt (\ptc) bins from low to high \ptc 
followed by the 17 bottom hadron \pt (\ptb) bins from low to high \ptb. 
The X-axis is organized similarly from left to right. The \ptc and \ptb 
binning follows that shown in Fig.~\ref{fig:hadron-pt}. The region of 
green plots (top left quadrant) shows the charm hadron yields and the correlations between 
charm hadron yields. The region of blue plots (bottom right quadrant) shows the bottom hadron 
yields and correlations between bottom hadron yields. The region of orange 
plots (bottom left quadrant) shows the correlations between charm and bottom hadron yields. 
Sub-panels (b)-(d) show a set of example distributions. (b) The 1-D 
probability distribution of charm hadron yield in $3.5<\ptc\ \gev<4.0$. 
(d) The 1-D probability distribution of bottom hadron yield in $2.5<\ptb\ 
\gev<3.0$. (c) The correlation between (b) and (d).
}
\label{fig:posttriangle}
\end{figure*}

The outcome of the sampling process is a distribution of $\thetavec$ 
vectors, which is 34-dimensional in this case. In principle, the 
distribution of $\thetavec$ vectors contains the full probability, 
including correlations between the different parameters. The 2-D 
correlations are shown in Fig.~\ref{fig:posttriangle}. While it is 
difficult to distinguish fine details in the 34$\times$34-dimensional grid 
of correlation plots, we can see a few gross features. A circular contour 
in the 2-D panels represents no correlation between the corresponding 
hadron \pt bins. An oval shape with a positive slope indicates a positive 
correlation between corresponding bins, and an oval shape with a negative 
slope represents an anti-correlation between corresponding bins. A large 
positive correlation is seen for adjacent bins for high-\pt charm hadrons 
and low-\pt bottom hadrons. This is a consequence of the regularization, 
which requires a smooth \pt distribution, and is stronger at the higher 
and lower \pt regions where there is less constraint from the data. We 
also see that, while there is little correlation between the majority of 
nonadjacent \pt bins, there does seem to be a region of negative 
correlation between the mid to high \pt charm hadrons and the low to mid 
\pt bottom hadrons. Charm and bottom hadrons in these regions contribute 
decay electrons in the same \pt region, and appear to compensate for each 
other to some extent. An example of this is shown between $3.5<\ptc\ 
\gev<4.0$ and $2.5<\ptb\ \gev<3.0$ in Fig.~\ref{fig:posttriangle}(b)-(d).

To summarize $\post$, we take the mean of the marginalized posterior 
distributions (the diagonal plots in Fig.~\ref{fig:posttriangle}) for each 
hadron \pt bin as the most likely values, and the 16$^{\rm th}$ and 
84$^{\rm th}$ quantiles to represent the $\pm1\sigma$ uncertainty in those 
values due to the statistical uncertainty in the data modified by the 
regularization constraint.

	\subsubsection{Re-folded comparisons to data}

\begin{figure}[!hbt]
\includegraphics[width=0.98\linewidth]{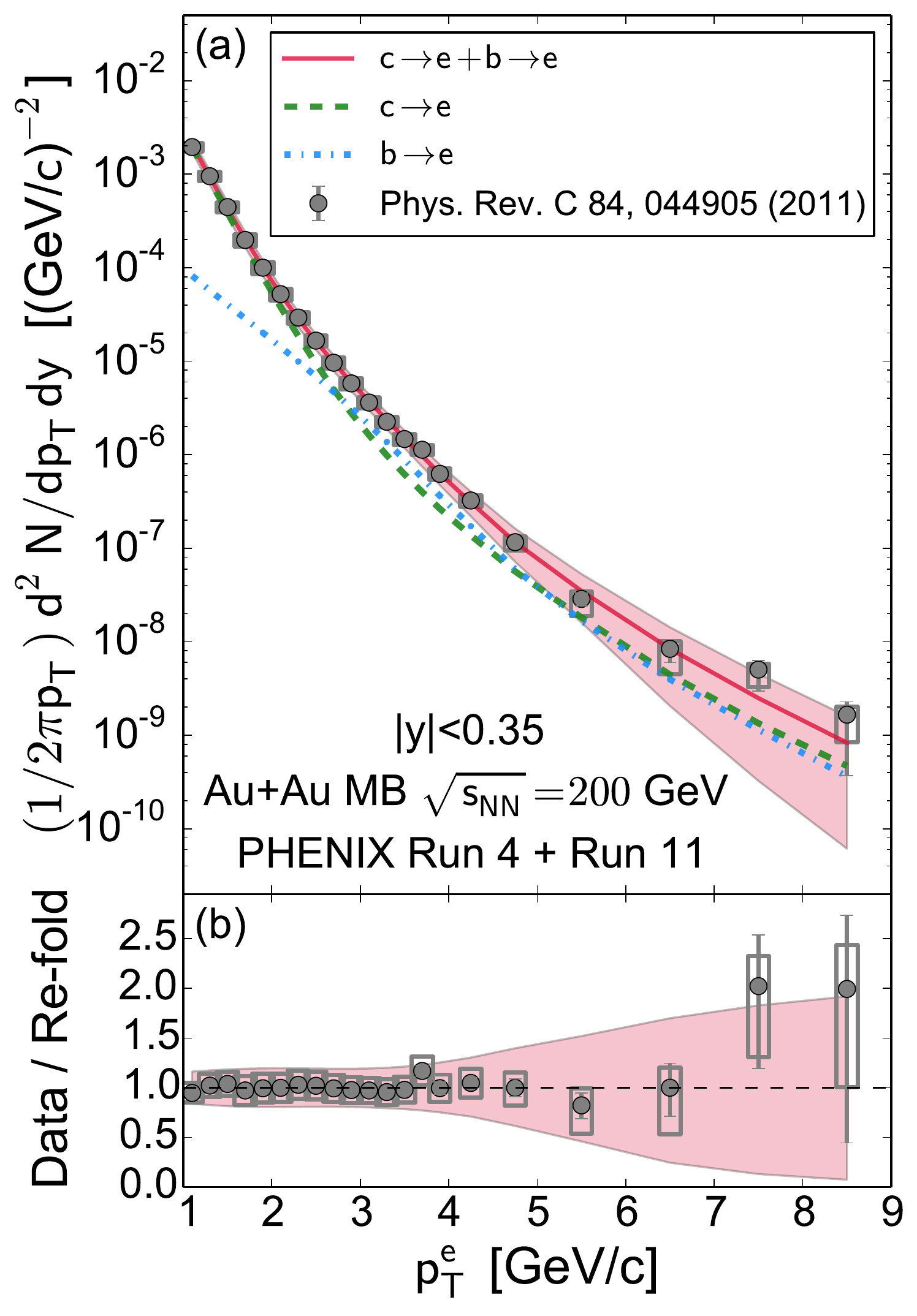}
\caption{ (Color Online) 
The heavy flavor electron invariant yield as a function of \pt from 
measured data~\cite{Adare:2010de} compared to electrons from the re-folded 
charm and bottom hadron yields. The boxes represent the point-to-point 
correlated uncertainties on the measured heavy flavor electron invariant 
yield, while the error bars on the points represent the point-to-point 
uncorrelated uncertainties. The label ``PHENIX Run 4 + Run 11'' on this 
and all subsequent plots indicates that the unfolding result uses the 
heavy flavor electron invariant yield as a function of \pt from data taken 
in 2004 (Run 4) combined with \DCAR measurements from data taken in 2011 
(Run 11).
}
  \label{fig:ept-refold}
\end{figure}

\begin{figure*}[!hbt]
    \includegraphics[width=0.4\linewidth]{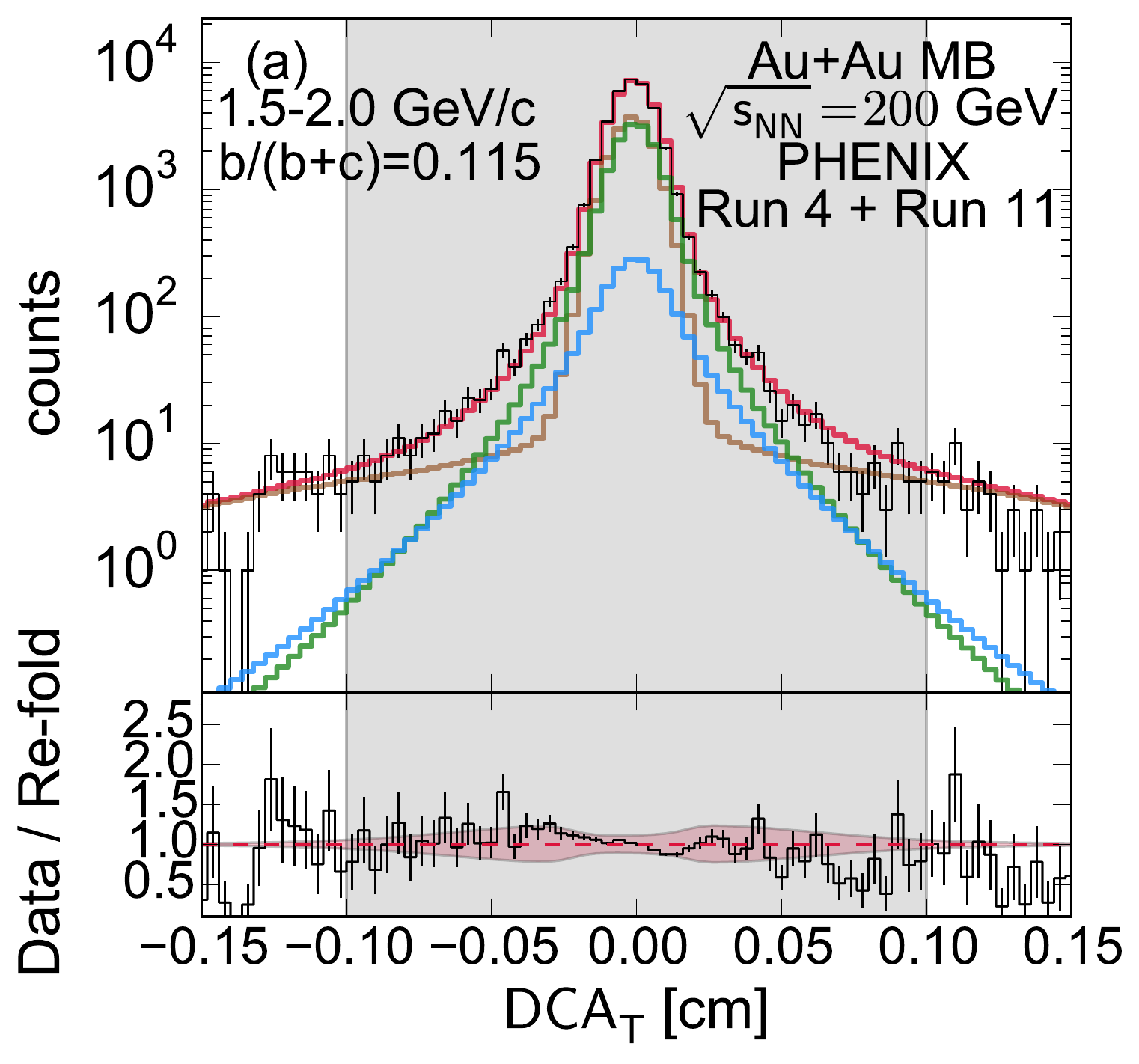}
    \includegraphics[width=0.4\linewidth]{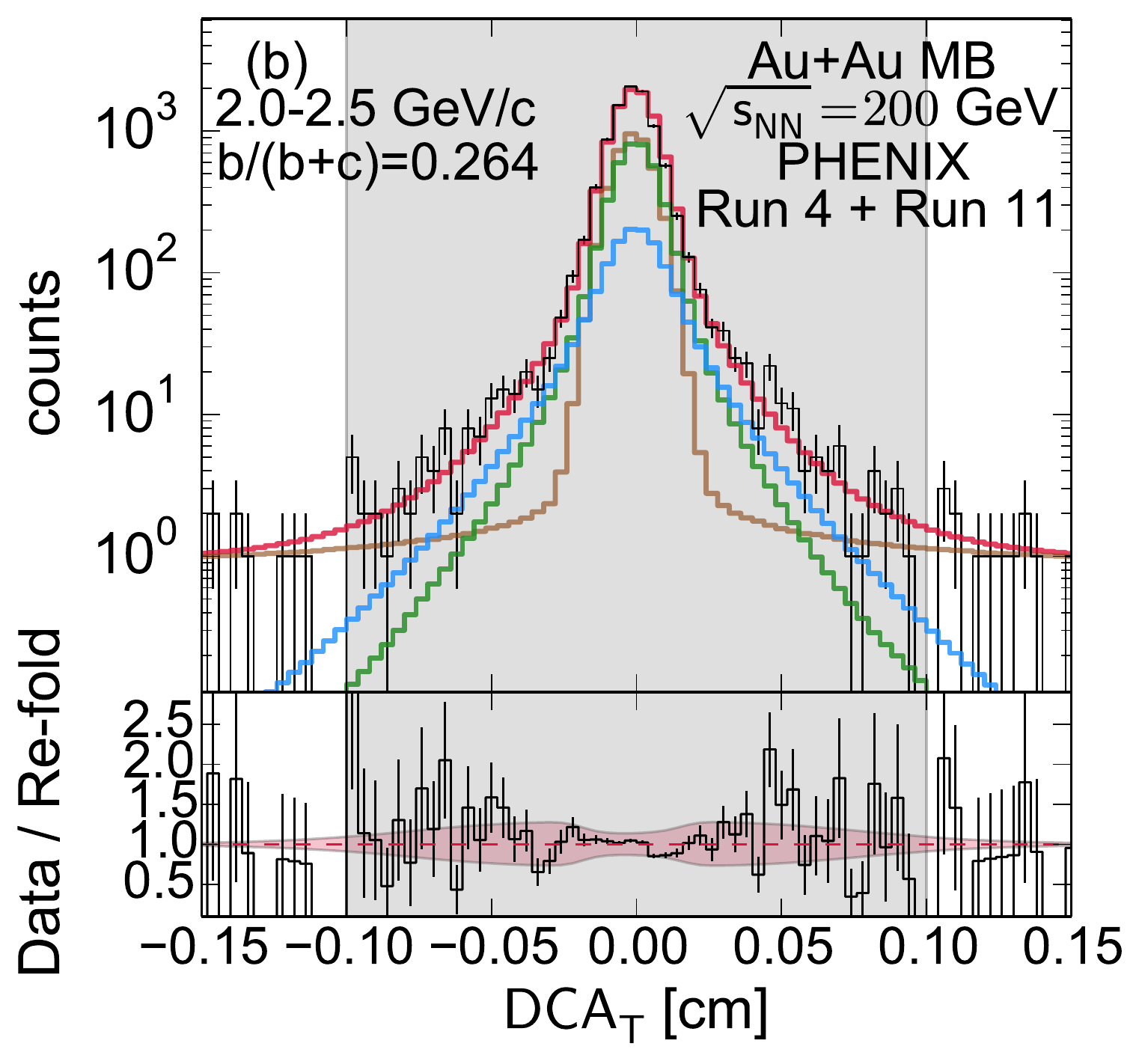}\\
    \includegraphics[width=0.4\linewidth]{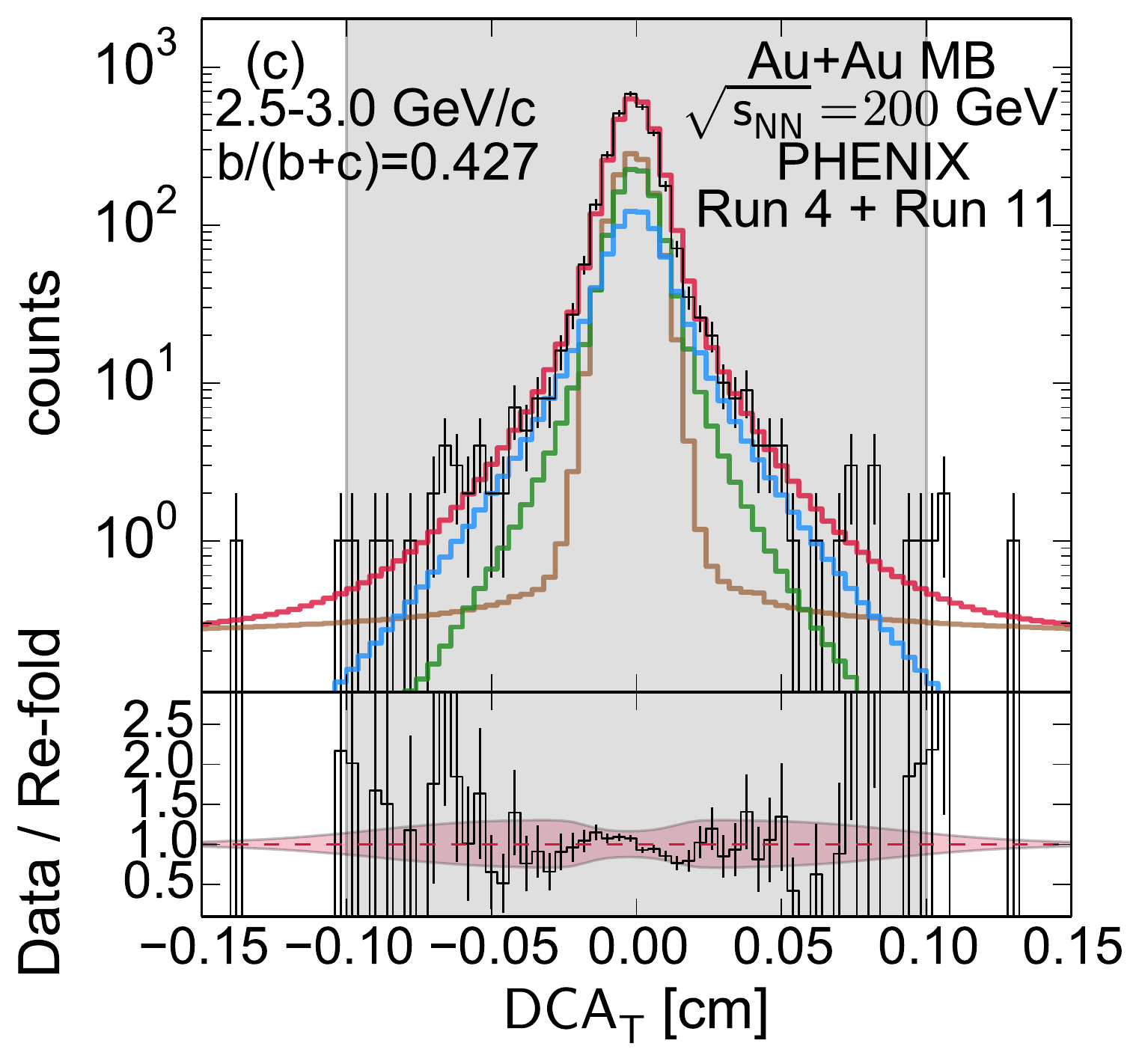}
    \includegraphics[width=0.4\linewidth]{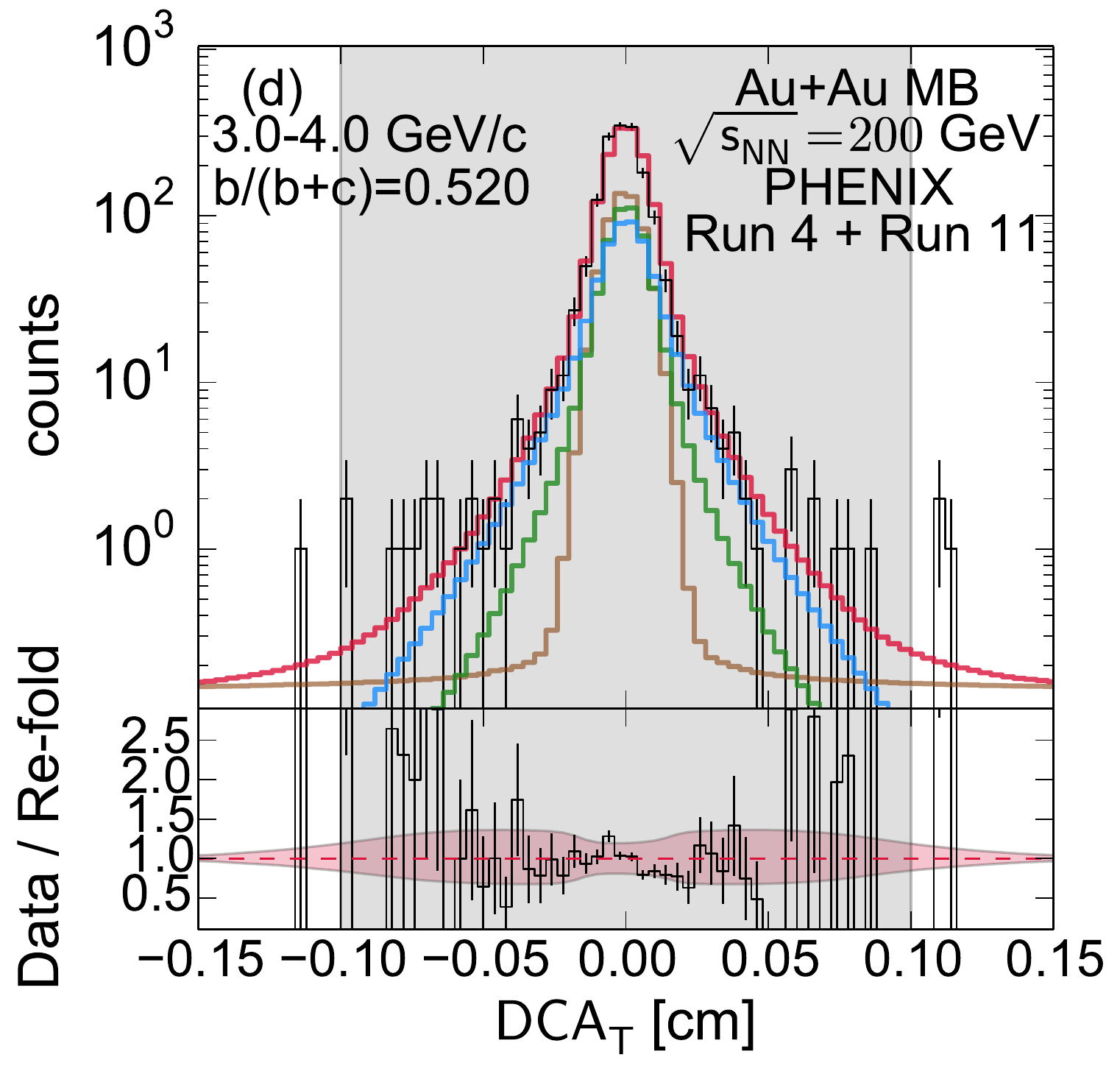}\\
    \includegraphics[width=0.4\linewidth]{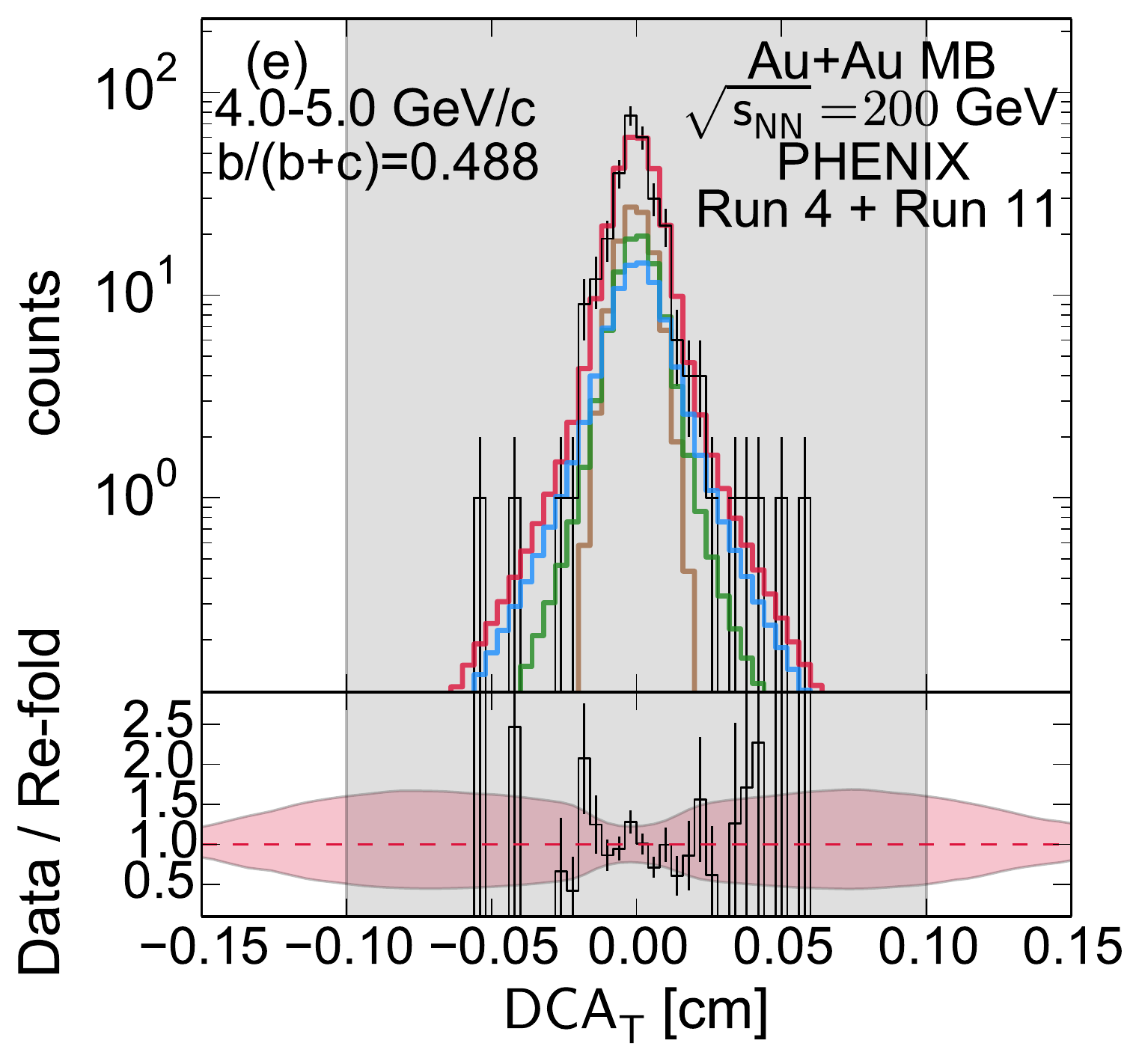}
    \includegraphics[width=0.4\linewidth]{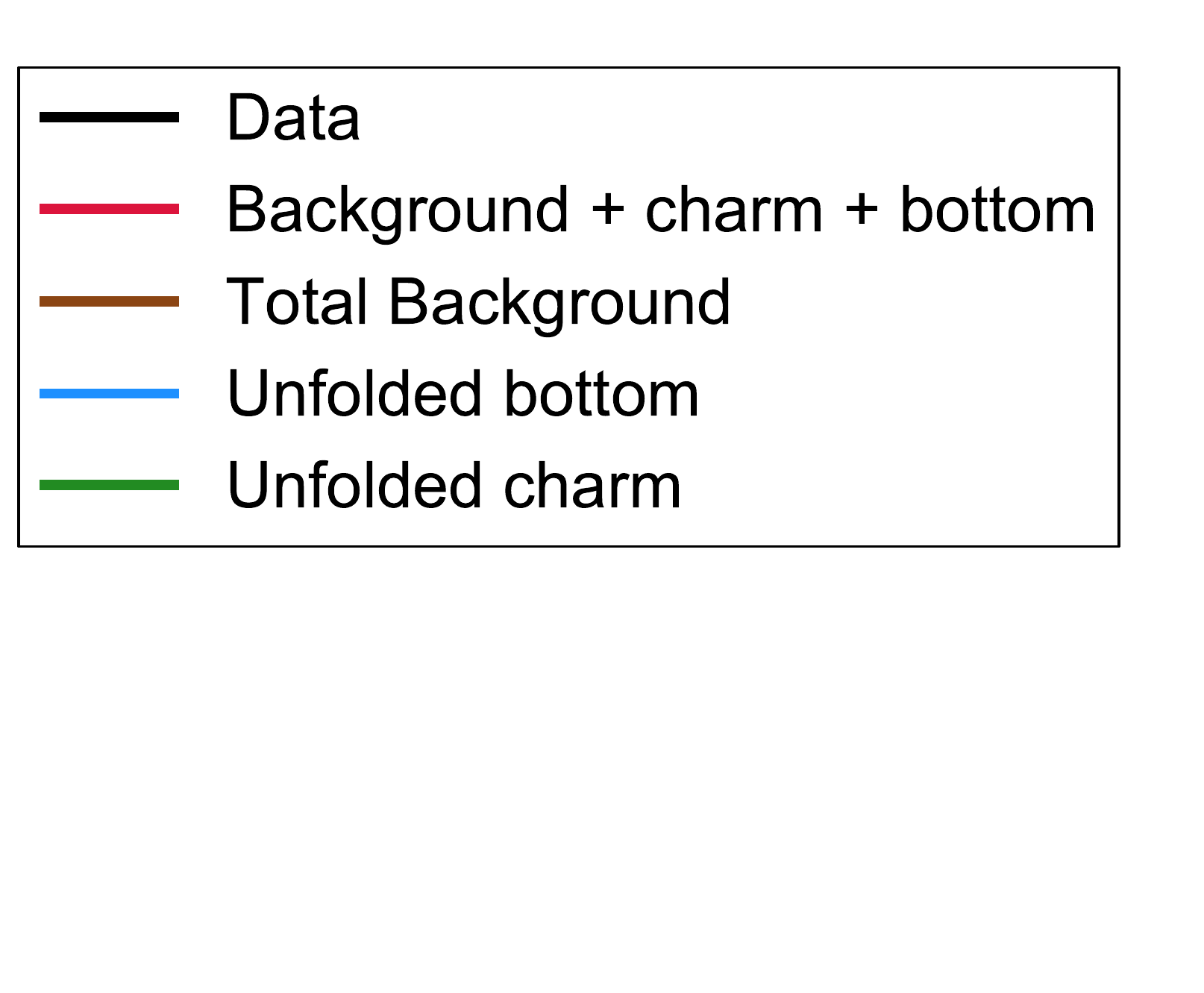}
  \caption{(Color Online) 
The \DCAR distribution for measured electrons compared to the decomposed 
\DCAR distributions for background components, electrons from charm 
decays, and electrons from bottom decays. The sum of the background 
components, electrons from charm and bottom decays is shown as the red (upper)
curve for direct comparison to the data. The gray band indicates the 
region in \DCAR considered in the unfolding procedure. Also quoted in the 
figure is the bottom electron fraction for $|\DCAR|<0.1$~cm integrated 
over the given \pt range. The legend follows the same order from top to bottom as panel (b) at $\DCAR=-0.1$ cm.
}
  \label{fig:dca-refold}
\end{figure*}

The vector of most likely hadron yields, with uncertainties, can be 
multiplied by the decay matrix to check the consistency of the result with 
the measured data (here referred to as re-folding). 
Figure~\ref{fig:ept-refold} shows the measured heavy flavor electron 
invariant yield in \auau collisions~\cite{Adare:2010de} compared with the 
re-folded electron spectra from charm and bottom hadrons. We find good 
agreement between the measured data and the electron spectrum from the 
re-folded charm and bottom hadron yields. Figure~\ref{fig:dca-refold} 
shows the comparison in electron \DCAR space for each bin in electron \pt. 
Shown in each panel is the measured \DCAR distribution for electrons, the 
sum of the background contributions discussed in 
Section~\ref{sec:dca_background}, the \DCAR distribution of electrons from 
charm hadron decays, and the \DCAR distribution of electrons from bottom 
hadron decays. Note that the sum of the background contributions is fixed 
in the unfolding procedure, and only the relative contribution of charm 
and bottom electrons within $|\DCAR|<0.1$~cm, as well as their \DCAR 
shape, vary. For convenience, the region of the \DCAR distribution 
considered in the unfolding procedure is also shown, as discussed in 
Section~\ref{sec:dca_measurement}. The sum of the background 
contributions, charm, and bottom electrons is shown for a direct 
comparison with the data. 

The summed log likelihood values for each of the \DCAR distributions and 
the electron invariant yield are given in Table~\ref{tab:llrefold}. To aid 
in the interpretation of the likelihood values, we use a Monte-Carlo 
method to calculate the expected likelihood from statistical fluctuations 
around the re-folded result. We draw samples from the re-folded result 
based on the data statistics and calculate the distribution of resulting 
likelihood values. The number of standard deviations from the expected 
value is also shown in Table~\ref{tab:llrefold}. We find that the log 
likelihood values are large compared to expectations in the heavy flavor 
electron invariant yield as well as the lowest two \DCAR \pt bins. We note 
that the likelihood values do not incorporate the systematic uncertainties 
on the data, which are handled separately as described in 
Sec.~\ref{sec:unfolding_systematics}. In particular the statistical 
uncertainties on the heavy flavor electron invariant yield are much 
smaller than the systematics at low-\pt, making the likelihood value not 
surprising. We find reasonable agreement within uncertainties between the 
remaining \DCAR \pt bins.

%%%%%%%%%%%%%%%%%%%%%%%%%%%%%%%%%%%%%%%%%%%%%% Fig_12

%%%%%%%%%%%%%%%%%%%%%%%%%%%%%%%%%%%%%%%%%%%%%% Fig_13

%============================================ Table_I
\begin{table}
\caption{\label{tab:llrefold} 
The log likelihood values ($LL$) summed over each \DCAR distribution and 
for the comparison to the heavy flavor electron invariant yield. Also 
quoted is the number of data points ($N_p$) and the deviation from the log 
likelihood value expected from statistical fluctuations ($\Delta LL$), as 
discussed in the text, for each comparison.
}
\begin{ruledtabular} \begin{tabular}{crrc}
	Data set & $N_p$ & LL & $\Delta LL$ [$\sigma$] \\
	\hline
	$e$ \DCAR $1.5<\pte<2.0$ & 50 & -195.5 & -3.8 \\
	$e$ \DCAR $2.0<\pte<2.5$ & 50 & -156.5 & -2.9 \\
	$e$ \DCAR $2.5<\pte<3.0$ & 50 & -115.8 & -0.6 \\
	$e$ \DCAR $3.0<\pte<4.0$ & 50 & -104.1 & -1.8 \\
	$e$ \DCAR $4.0<\pte<5.0$ & 50 &  -53.2 & 0.0 \\
	$e$ Inv. Yield. $1.0<\pte<9.0$ & 21 & -45.9 & -3.5 \\
	Total Sum & 271 & -673.8 &  \\
\end{tabular} \end{ruledtabular}
\end{table}

	\subsection{Systematic uncertainties}
	\label{sec:unfolding_systematics}

When performing the unfolding procedure, only the statistical 
uncertainties on the electron \DCAR and \pt spectra are included. In this 
section we describe how we consider the systematic uncertainties on both 
the measured data and the unfolding procedure. We take the following 
uncertainties into account as uncorrelated uncertainties:

\begin{enumerate}
	\item Systematic uncertainty in the heavy flavor electron \pt invariant yield 
	\item Uncertainty in the high-multiplicity background
	\item Uncertainty in the fraction of nonphotonic electrons ($F_{\rm NP}$) 
	\item Uncertainty in $K_{e3}$ normalization
	\item Regularization hyperparameter $\alpha$
	\item Uncertainty in the form of $\thetavec_{{\rm prior}}$
\end{enumerate}

The uncertainty in $F_{\rm NP}$ (See Sec.~\ref{sec:FNP}), and $K_{e3}$ are 
propagated to the unfolded hadron yields by varying each independently by 
$\pm1\sigma$, and performing the unfolding procedure with the modified 
background template. The difference between the resulting hadron yields 
and the central values is taken as the systematic uncertainty. The same 
procedure is used to determine the uncertainty in the result due to the 
regularization parameter, which is varied by $^{+0.60}_{-0.25}$ based on 
where the summed likelihood from both the data and regularization drops by 
1 from the maximum value.

The uncertainty in the high-multiplicity background includes two 
components. The first is the uncertainty on the normalization of the 
high-multiplicity background \DCAR distribution, as determined in 
Sec.~\ref{sec:dca_sideband} and shown in Fig.~\ref{fig:DCA0}. This is 
propagated to the unfolded hadron yields by varying the normalization by 
$\pm1\sigma$ and performing the unfolding procedure with the modified 
background template, as with the $F_{\rm NP}$ and $K_{e3}$ uncertainties. The 
second component addresses the small excess in the embedded primary 
electron distribution observed in Fig.~\ref{fig:embedSideband} and not 
accounted for by using the \DCAR distribution for large \DCAZ. We 
parametrize the excess, which is more than two orders of magnitude below 
the peak, and apply it to the background components, re-performing the 
unfolding procedure to find its effect on the hadron yield. Both effects 
combined are small relative to the dominant uncertainties.

Incorporating the \pt correlated systematic uncertainty on the heavy 
flavor electron invariant yield is more difficult. Ideally one would 
include a full covariance matrix encoding the \pt correlations into the 
unfolding procedure. In practice, the methodology employed 
in~\cite{Adare:2010de} does not provide a convenient description of the 
$p_T$ correlations needed to shape the covariance matrix. Instead we take 
a conservative approach by considering the cases which we believe 
represent the maximum \pt correlations. We modify the heavy flavor 
electron invariant yield by either tilting or kinking the spectrum about a 
given point. Tilting simply pivots the spectra about the given point so 
that, for instance, the first point goes up by a fraction of the 
systematic uncertainty while the last point goes down by the same fraction 
of its systematic uncertainty, with a linear interpolation in between. 
Kinking simply folds the spectra about the given point so that that the 
spectrum is deformed in the form of a \textbf{V}. We implement the 
following modifications and re-perform the unfolding procedure:

\begin{enumerate}
\item Tilt the spectra about $\pt=1.8$ \gev by $\pm1\sigma$ of the 
systematic uncertainty.
\item Tilt the spectra about $\pt=5$ \gev by $\pm1\sigma$ of the 
systematic uncertainty.
\item Kink the spectra about $\pt=1.8$ \gev by $\pm1\sigma$ of the 
systematic uncertainty.
\item Kink the spectra about $\pt=5$ \gev by $\pm1\sigma$ of the 
systematic uncertainty.
\end{enumerate}

The \pt points about which the spectra were modified were motivated by the 
points in \pt at which analysis methods and details changed, as discussed 
in~\cite{Adare:2010de}. We then take the RMS of the resulting deviations 
on the hadron yield from the central value as the propagated systematic 
uncertainty due to the systematic uncertainty on the heavy flavor electron 
invariant yield.

The effect of our choice of $\thetavec_{{\rm prior}}$ on the charm and 
bottom hadron yields is taken into account by varying 
$\thetavec_{{\rm prior}}$, as discussed in Section~\ref{sec:prior}. The 
differences between each case and the central value are added in 
quadrature to account for the bias introduced by 
$\thetavec_{{\rm prior}}$.

The uncertainties on the unfolded hadron yields due to the six components 
described above and the uncertainty determined from the posterior 
probability distributions are added in quadrature to give the uncertainty 
shown in Fig.~\ref{fig:hadron-pt}.

%%%%%%%%%%%%%%%%%%%%%%%%%%%%%%%%%%%%%%%%%%%%%% Fig_14
\begin{figure}[!hbt]
\includegraphics[width=0.98\linewidth]{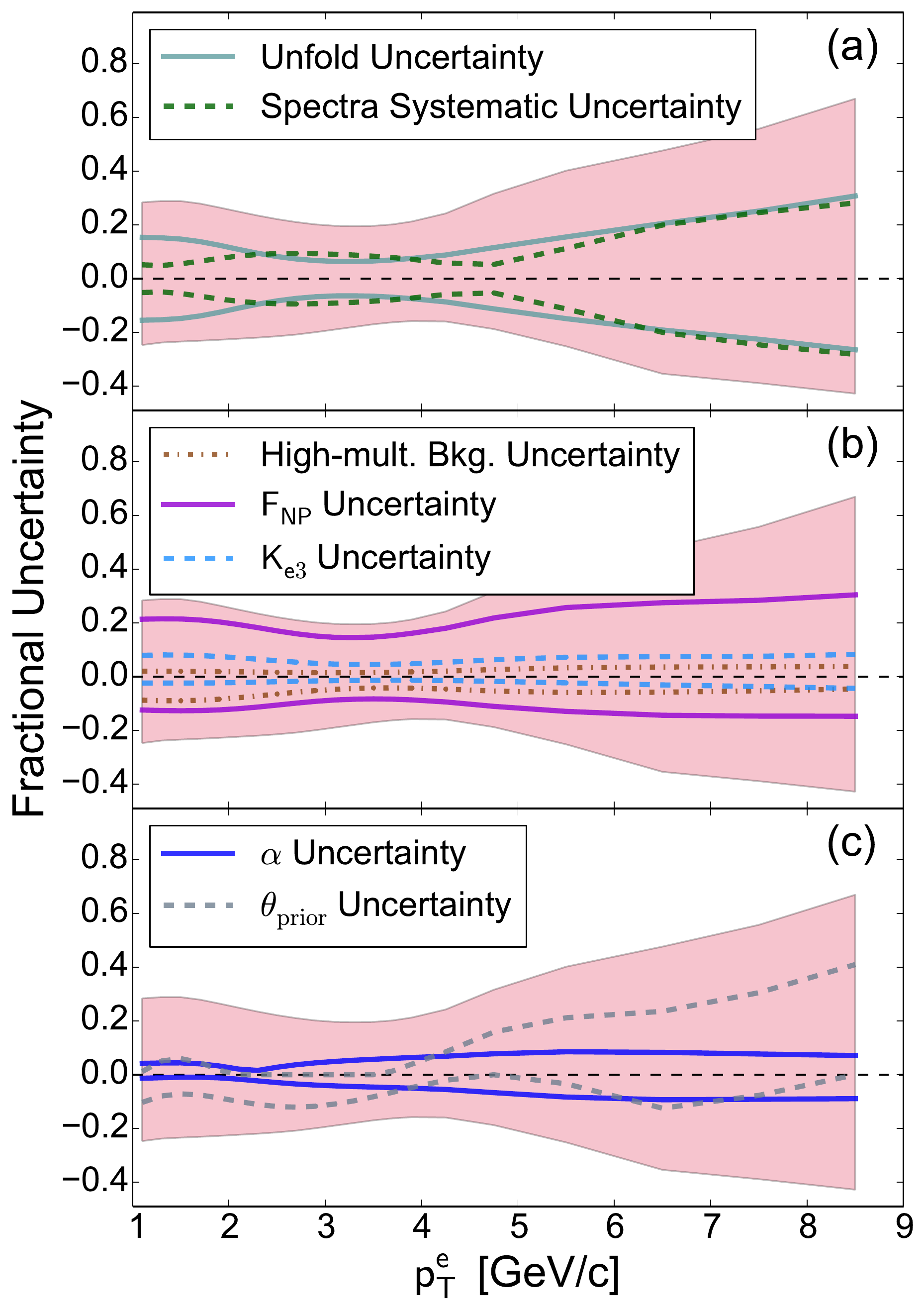}
\caption{\label{fig:bfrac-sys}(Color Online) 
The relative contributions from the different components to the 
uncertainty on the fraction of electrons from bottom hadron decays as a 
function of \pt. The shaded red band in each panel is the total 
uncertainty.
}
\end{figure}

Due to the correlations between charm and bottom yields, the relative 
contributions from the different uncertainties depend on the variable 
being plotted. To give some intuition for this, we have plotted the 
relative contributions from the different uncertainties to the fraction of 
electrons from bottom hadron decays as a function of \pt (discussed in 
Sec.~\ref{sec:bfraction}) in Fig.~\ref{fig:bfrac-sys}. One can see that 
the dominant uncertainties come from the statistical uncertainty on the 
\DCAR and heavy flavor electron invariant yield, the systematic 
uncertainty on the heavy flavor electron invariant yield, and $F_{\rm NP}$. 
We remind the reader that for $\pt>5$ \gev we no longer have 
\DCAR information to directly constrain the unfolding, and all information comes 
dominantly from the heavy flavor electron invariant yield, leading to the 
growth in the uncertainty band in this region.

%%========================================================================
	\section{Results}
	\label{sec:results}

\begin{figure*}[!hbt]
\includegraphics[width=0.75\linewidth]{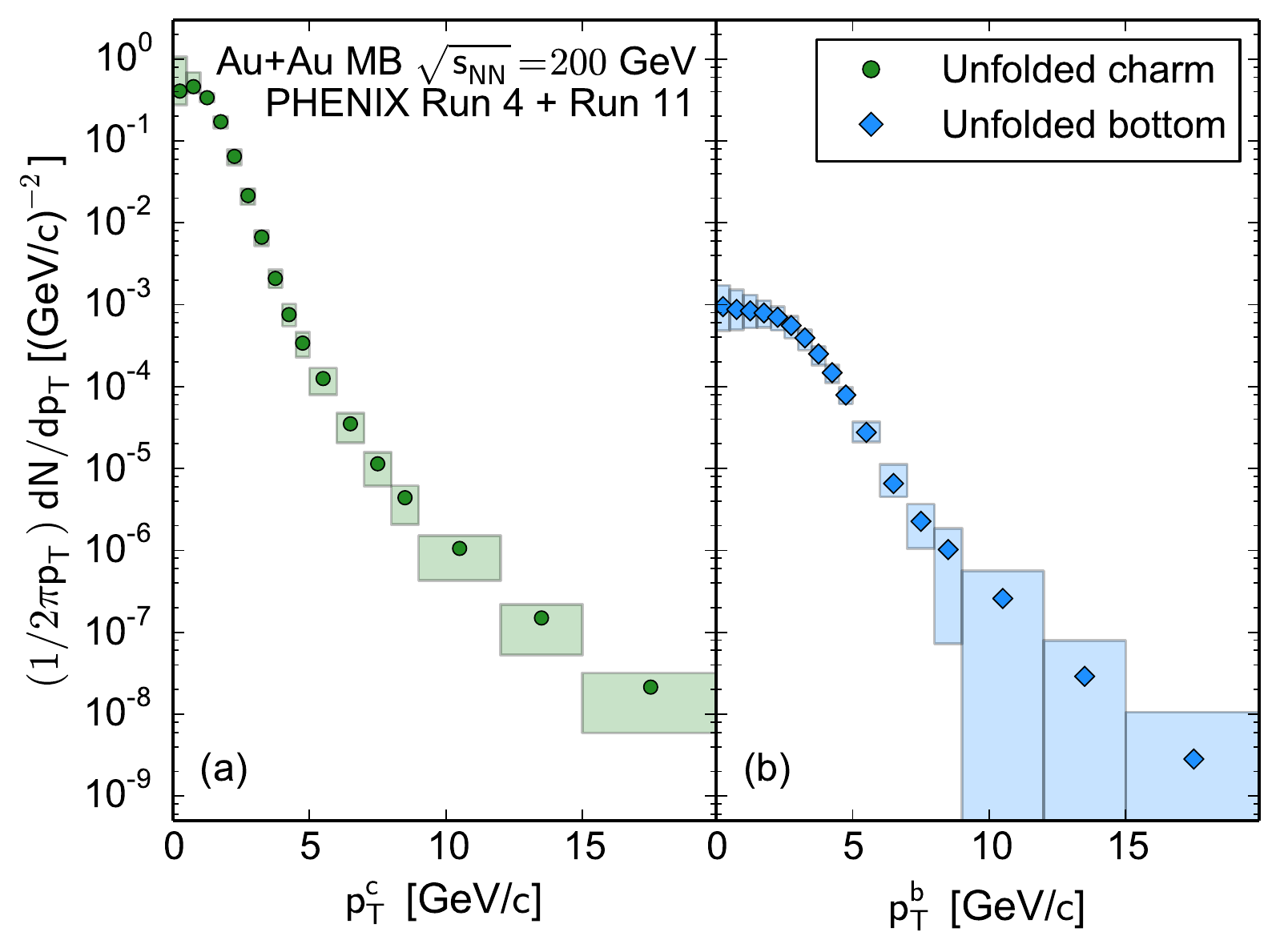}
\caption{\label{fig:hadron-pt}(Color Online) 
Unfolded (a) charm and (b) bottom hadron invariant yield as a 
function of \pt, integrated over all rapidities, as constrained by 
electron yield vs \DCAR in 5 \pte bins and previously published heavy 
flavor electron invariant yield vs \pte~\cite{Adare:2010de}.
}
\end{figure*}

The final result of the unfolding procedure applied simultaneously to the 
heavy flavor electron invariant yield vs \pt (shown in 
Fig.~\ref{fig:ept-refold}) and the five electron \DCAR distributions 
(shown in Fig.~\ref{fig:dca-refold}) is the invariant yield of charm and 
bottom hadrons, integrated over all rapidity, as a function of \pt. As a 
reminder, the hadron yields are integrated over all rapidity by assuming 
the rapidity distribution within \pythia is accurate and that it is 
unmodified in \auau, as detailed in Sec.~\ref{sec:decay_matrix}. The 
unfolded results for MB (0\%--96\%) \auau collisions at \sqsn=200~GeV are 
shown in Fig.~\ref{fig:hadron-pt}. The central point represents the most 
likely value and the shaded band represents the $1\sigma$ limits on the 
combination of the uncertainty in the unfolding procedure and the 
systematic uncertainties on the data, as described in 
Sec.~\ref{sec:unfolding_systematics}. The uncertainty band represents 
point-to-point correlated uncertainties, typically termed Type B in PHENIX 
publications. There are no point-to-point uncorrelated (Type A), or global 
scale uncertainties (Type C), from this procedure.

The uncertainties on the hadron invariant yields shown in 
Fig.~\ref{fig:hadron-pt} grow rapidly for charm and bottom hadrons with 
$\pt>6$ \gev. This is due to the lack of \DCAR information for $\pte>5$ 
\gev. Above $\pte>5$ \gev, the unfolding is constrained by the heavy 
flavor electron invariant yield only. This provides an important 
constraint on the shape of the hadron \pt distributions, but the \DCAR 
distributions provide the dominant source of discriminating power 
between the charm and bottom. However, due to the decay kinematics, even 
high \pt hadrons contribute electrons in the range $1.5<\pte\ 
[{\rm~GeV}/c]<5.0$. We find that charm(bottom) hadrons in the range 
$7<\pth\ [{\rm~GeV}/c]<20$ contribute 18.2\%(0.3\%) of the total 
electron yield in the region $1.5<\pte\ [{\rm~GeV}/c]<5.0$. This 
explains the larger uncertainties in the bottom hadron yield compared to 
the charm hadron yield at high \pth.

%%%%%%%%%%%%%%%%%%%%%%%%%%%%%%%%%%%%%%%%%%%%%% Fig_15

\begin{figure}[!hbt]
\includegraphics[width=0.98\linewidth]{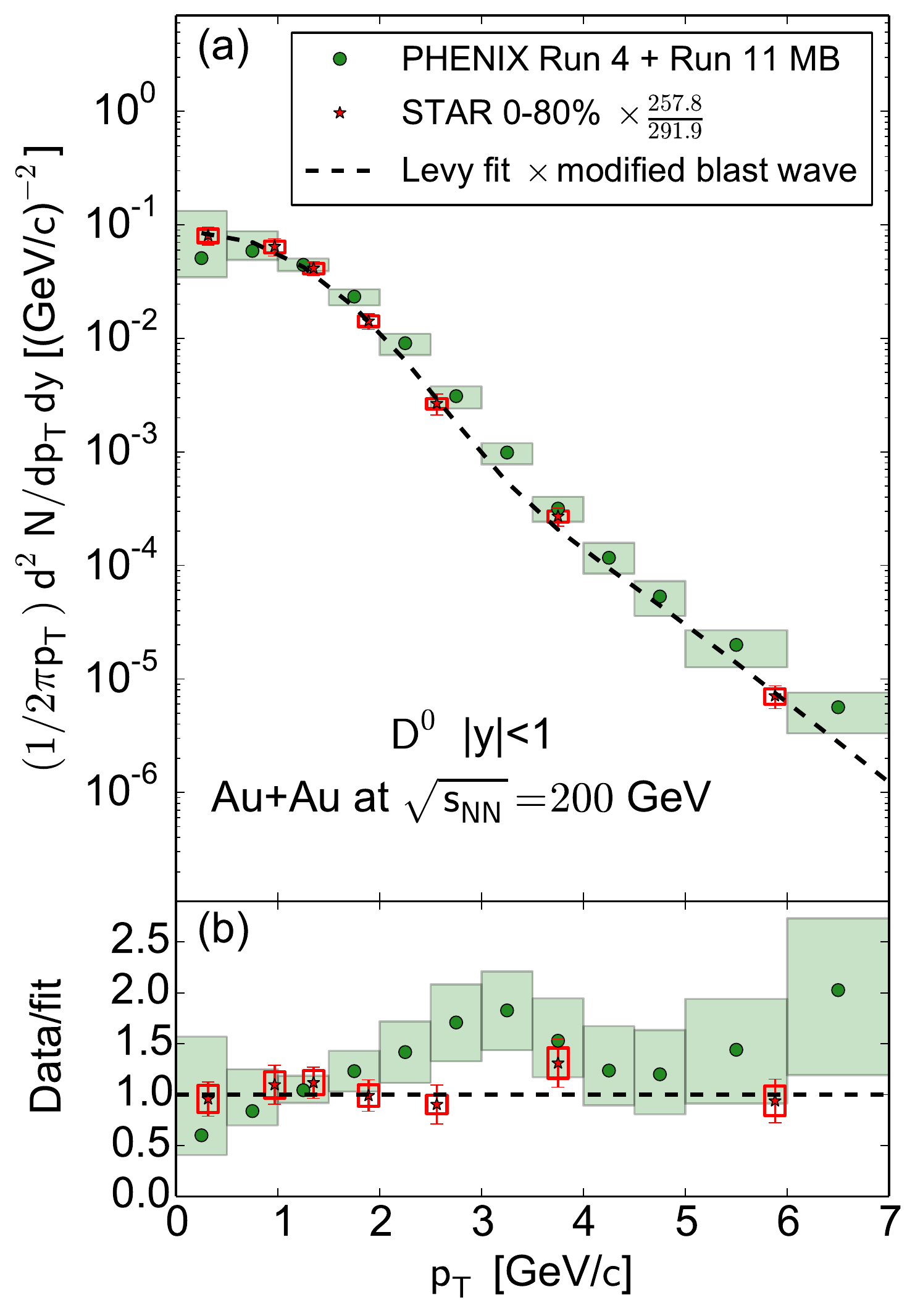}
\caption{(Color Online) 
The invariant yield of $D^0$ mesons as a function of \pt for $|y|<1$ 
inferred from the unfolded yield of charm hadrons integrated over all 
rapidity compared to measurements from STAR~\cite{Adamczyk:2014uip}. See 
the text for details on the calculation of the $D^0$ yield inferred from 
the unfolded result. To match the centrality intervals, the STAR result 
has been scaled by the ratio of \Ncoll values. The bottom panel shows the 
ratio of the data to a fit of the STAR $D^0$ yield.
}
\label{fig:STARD0}
\end{figure}

The yield of $D^0$ mesons over $|y|<1$ as a function of \pt has been 
previously published in \auau collisions at \sqsn=200~GeV by 
STAR~\cite{Adamczyk:2014uip}. In order to compare our unfolded charm 
hadron results over all rapidity to the STAR measurement, we use \pythia 
to calculate the fraction of $D^0$ mesons within $|y|<1$ compared to charm 
hadrons over all rapidity. Since the measurement by STAR is over a 
narrower centrality region (0\%--80\% vs 0\%-96\%), we scale the STAR 
result by the ratio of the \Ncoll values. This comparison is shown in 
Fig.~\ref{fig:STARD0}. For added clarity, we have fit the STAR measurement 
with a Levy function modified by a blast wave calculation given by
%\begin{widetext}
\begin{eqnarray}
f(p_T) &=& p_0 \left(1 - \frac{(1-p_1)p_T}{p_2} \right)^{1/(1-p_1)} \\
& \times & \left(1.3\sqrt{2\pi p_4^2} G(p_T, p_3, p_4) 
+ \frac{p_5}{1+e^{-p_T+3}}\right), \nonumber
\end{eqnarray}
%\end{widetext}
where $G(p_T, p_3, p_4)$ is a standard Gaussian function, and $p_i$ are 
the parameters of the fit. The ratio of the data to the fit is shown in 
the bottom panel of Fig.~\ref{fig:STARD0}. We find that, within 
uncertainties, the unfolded $D^0$ yield agrees with that measured by STAR 
over the complementary \pt range. The unfolded yield hints at a different 
trend than the STAR data for $\pt>5$ \gev. However, we note that the 
$\langle\pt\rangle$ of charm(bottom) hadrons which contribute electrons in 
the range $4.0<\pt\ [{\rm~GeV}/c]<5.0$ is 7.2(6.4) \gev. This means that 
the yields of charm and bottom hadrons have minimal constraint from the 
\DCAR measurements in the high-\pt regions, which is represented by an 
increase in the uncertainties.

%%%%%%%%%%%%%%%%%%%%%%%%%%%%%%%%%%%%%%%%%%%%%% Fig_16

	\subsection{The bottom electron fraction}
	\label{sec:bfraction}

%%%%%%%%%%%%%%%%%%%%%%%%%%%%%%%%%%%%%%%%%%%%%% Fig_17
\begin{figure}[!hbt]
\includegraphics[width=0.98\linewidth]{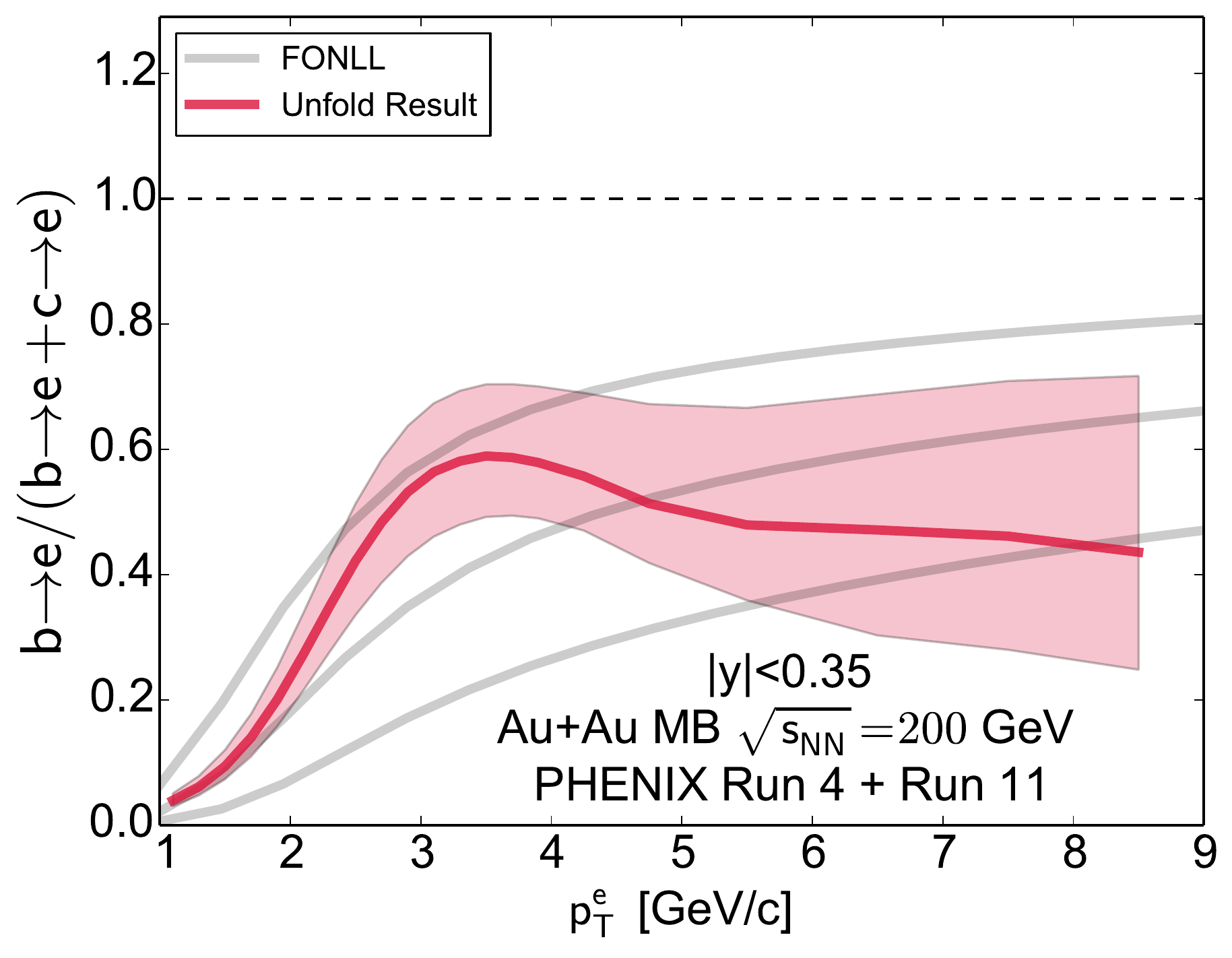}
\caption{(Color Online) 
The fraction of heavy flavor electrons from bottom hadron decays as a 
function of \pt from this work and from \fonll \pp 
calculations~\cite{Cacciari:2005rk}.
}
  \label{fig:bfrac}
\end{figure}

The fraction of heavy flavor electrons from bottom hadrons ($\frac{b 
\rightarrow e}{b \rightarrow e + c \rightarrow e}$) is computed by 
re-folding the charm and bottom hadron yields shown in 
Fig.~\ref{fig:hadron-pt} to get the invariant yield of electrons from 
charm and bottom decays at midrapidity ($|y|<0.35$). Here the electrons 
from bottom hadron decays include the cascade decay $b\rightarrow 
c\rightarrow e$. The resulting bottom electron fraction is shown as a 
function of \pt in Fig.~\ref{fig:bfrac}. The central values integrated 
over the \pt range of each \DCAR distribution are also quoted in 
Fig.~\ref{fig:dca-refold}. As in the hadron yields, the band represents 
the $1\sigma$ limits of the point-to-point correlated (Type B) 
uncertainties. 

Also shown in Fig.~\ref{fig:bfrac} is the bottom electron fraction 
predictions from \fonll~\cite{Cacciari:2005rk} for \pp collisions at 
\sqsn=200~GeV. We find a bottom electron fraction which is encompassed by 
the \fonll calculation uncertainties. The shape of the resulting bottom 
electron fraction shows a steeper rise in the region $2.0<\pt\ [{\rm 
GeV}/c]<4.0$ with a possible peak in the distribution compared to the 
central \fonll calculation.

%%%%%%%%%%%%%%%%%%%%%%%%%%%%%%%%%%%%%%%%%%%%%% Fig_18
\begin{figure}[!hbt]
	\includegraphics[width=0.98\linewidth]{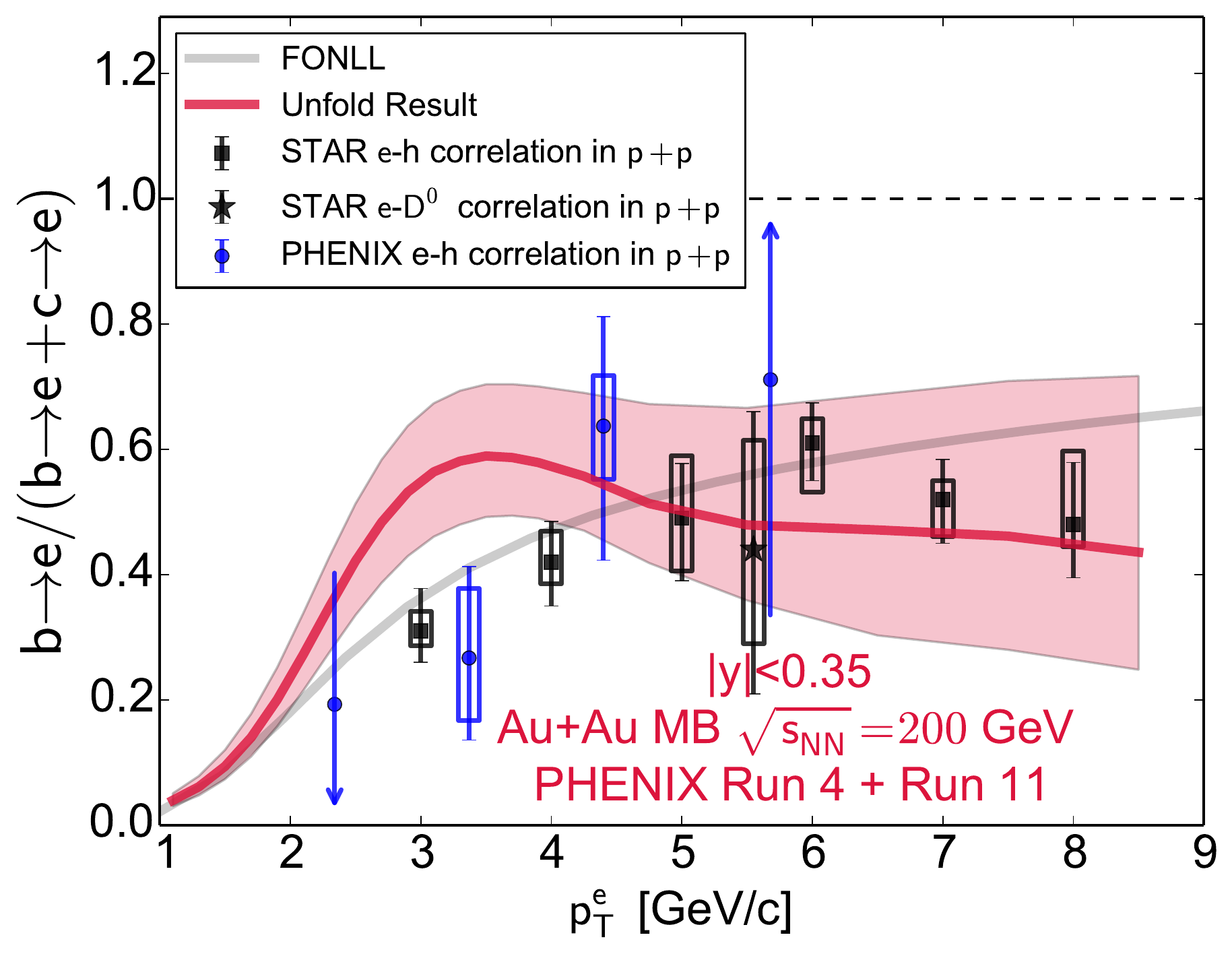}
	\caption{(Color Online) bottom electron fraction as a function of \pt compared to measurements in \pp collisions at \sqs=200~GeV from PHENIX~\cite{Adare:2009ic} and STAR~\cite{Aggarwal:2010xp}. Also shown are the central values for \fonll~\cite{Cacciari:2005rk} for \pp collisions at \sqsn=200~GeV.}
	\label{fig:bfrac-pp}
\end{figure}

The fraction of electrons from bottom decays has been previously measured 
in \pp collisions at \sqsn=200~GeV by both PHENIX~\cite{Adare:2009ic} and 
STAR~\cite{Aggarwal:2010xp}. These measurements are made through 
electron-hadron or electron-$D$ meson correlations. These are very 
different analyses than the one presented here, and have their own model 
dependencies. In Fig.~\ref{fig:bfrac-pp} we compare the bottom electron 
fraction between our unfolded \auau result and the electron-hadron 
correlation measurements in \pp. For $\pt>4$ \gev we find agreement 
between \auau and \pp within the large uncertainties on both measurements. 
This implies that electrons from bottom hadron decays are similarly 
suppressed to those from charm. For reference, included in 
Fig.~\ref{fig:bfrac-pp} is the central \fonll calculation which, within 
the large uncertainties, is consistent with the \pp measurements.

\begin{figure}[!hbt]
\includegraphics[width=0.98\linewidth]{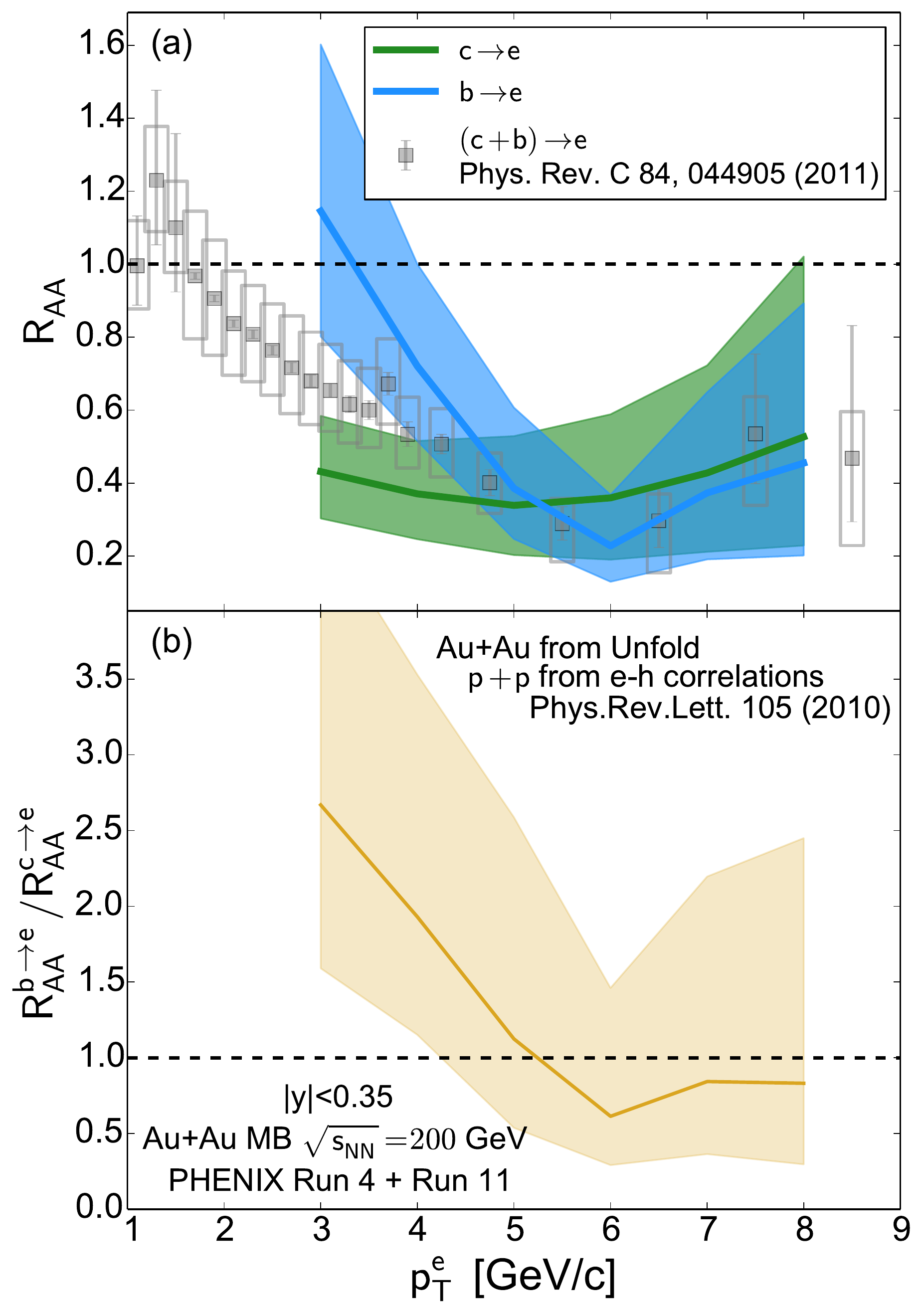}
\caption{(Color Online) 
(a) The \raa for $c\rightarrow e$, $b\rightarrow e$ and combined heavy 
flavor~\cite{Adare:2010de} as a function of \pte. The $c\rightarrow e$ and 
$b\rightarrow e$ \raa is calculated using Eq.~\ref{eq:craa}-\ref{eq:braa} 
where $F_{\rm AuAu}$ uses the unfolded result determined in this work and 
$F_{pp}$ determined from STAR $e-h$ correlations~\cite{Aggarwal:2010xp}. 
(b) The ratio $R_{AA}^{b\rightarrow e}/R_{AA}^{c\rightarrow e}$ as a 
function of \pte.
}
\label{fig:hfraa}
\end{figure}

With the additional constraints on the bottom electron fraction in \pp 
from the correlation measurements and the measured nuclear modification of 
heavy flavor electrons, we can calculate the nuclear modification of 
electrons from charm and bottom hadron decays separately. The nuclear 
modifications, $R_{AA}^{c\rightarrow e}$ and $R_{AA}^{b\rightarrow e}$, 
for charm and bottom hadron decays respectively are calculated using

\begin{eqnarray}
	R_{AA}^{c \rightarrow e} &= \frac{(1 - F_{\rm AuAu})}{(1 - F_{pp})}R_{AA}^{\rm HF}\label{eq:craa}\\
	R_{AA}^{b \rightarrow e} &= \frac{F_{\rm AuAu}}{F_{pp}}R_{AA}^{\rm HF},\label{eq:braa}
\end{eqnarray}
where $F_{\rm AuAu}$ and $F_{pp}$ are the fractions of heavy flavor electrons 
from bottom hadron decays in \auau and \pp respectively and $R_{AA}^{\rm HF}$ 
is the nuclear modification of heavy flavor electrons (combined charm and 
bottom). Rather than combining all measurements for the bottom electron 
fraction in \pp, which introduces a further extraction uncertainty, we 
have chosen to calculate $R_{AA}^{c\rightarrow e}$ and 
$R_{AA}^{b\rightarrow e}$ using only the six STAR electron-hadron $F_{pp}$ 
values. When performing the calculation we determine the full probability 
distributions assuming Gaussian uncertainties on $F_{\rm AuAu}$, $F_{pp}$ and 
$R_{AA}^{\rm HF}$. As when determining the charm and bottom hadron yields, we 
take the median of the distribution as the central value, and the 16\% and 
84\% of the distribution as the lower and upper $1\sigma$ uncertainties. 
The resulting values are shown in Fig.~\ref{fig:hfraa}(a). We find that 
the electrons from bottom hadron decays are less suppressed than electrons 
from charm hadron decays for $3<\pt\ \gev<4$. To further clarify this 
statement, we calculate the ratio of $R_{AA}^{b\rightarrow 
e}/R_{AA}^{c\rightarrow e}$, shown in Fig.~\ref{fig:hfraa}(b). In this 
ratio, the uncertainty on $R_{AA}^{\rm HF}$ cancels. Here again we calculate 
the full probability distributions and use the same procedure as above to 
determine the central values and uncertainties. We find that the 
probability distributions for $R_{AA}^{b\rightarrow 
e}/R_{AA}^{c\rightarrow e}$ are highly nonGaussian, which leads to the 
large asymmetric uncertainty band shown in Fig.~\ref{fig:hfraa}(b). It is 
clear from the ratio that $b\rightarrow e$ is less suppressed than 
$c\rightarrow e$ at the $1\sigma$ level up to $\pt\sim4$ \gev.

%%%%%%%%%%%%%%%%%%%%%%%%%%%%%%%%%%%%%%%%%%%%%% Fig_19

%%========================================================================
\section{Discussion}
\label{sec:discussion}

There are a number of theoretical calculations in the literature for the 
interaction of charm and bottom quarks with the QGP.  Many of these models 
have predictions for the nuclear modification factor $R_{AA}$ for 
electrons from charm decays and, separately, $R_{AA}$ for electrons from 
bottom decays.  For consistency, we have assumed the 
\fonll~\cite{Cacciari:2005rk} yields for electrons from charm (bottom) 
decays calculated for \pp at \sqsn=200~GeV and then scaled them by the 
heavy-ion model results for the $R_{AA}$ of electrons from charm (bottom).

%%%%%%%%%%%%%%%%%%%%%%%%%%%%%%%%%%%%%%%%%%%%%% Fig_20
\begin{figure*}[!hbt]
	\includegraphics[width=0.49\linewidth]{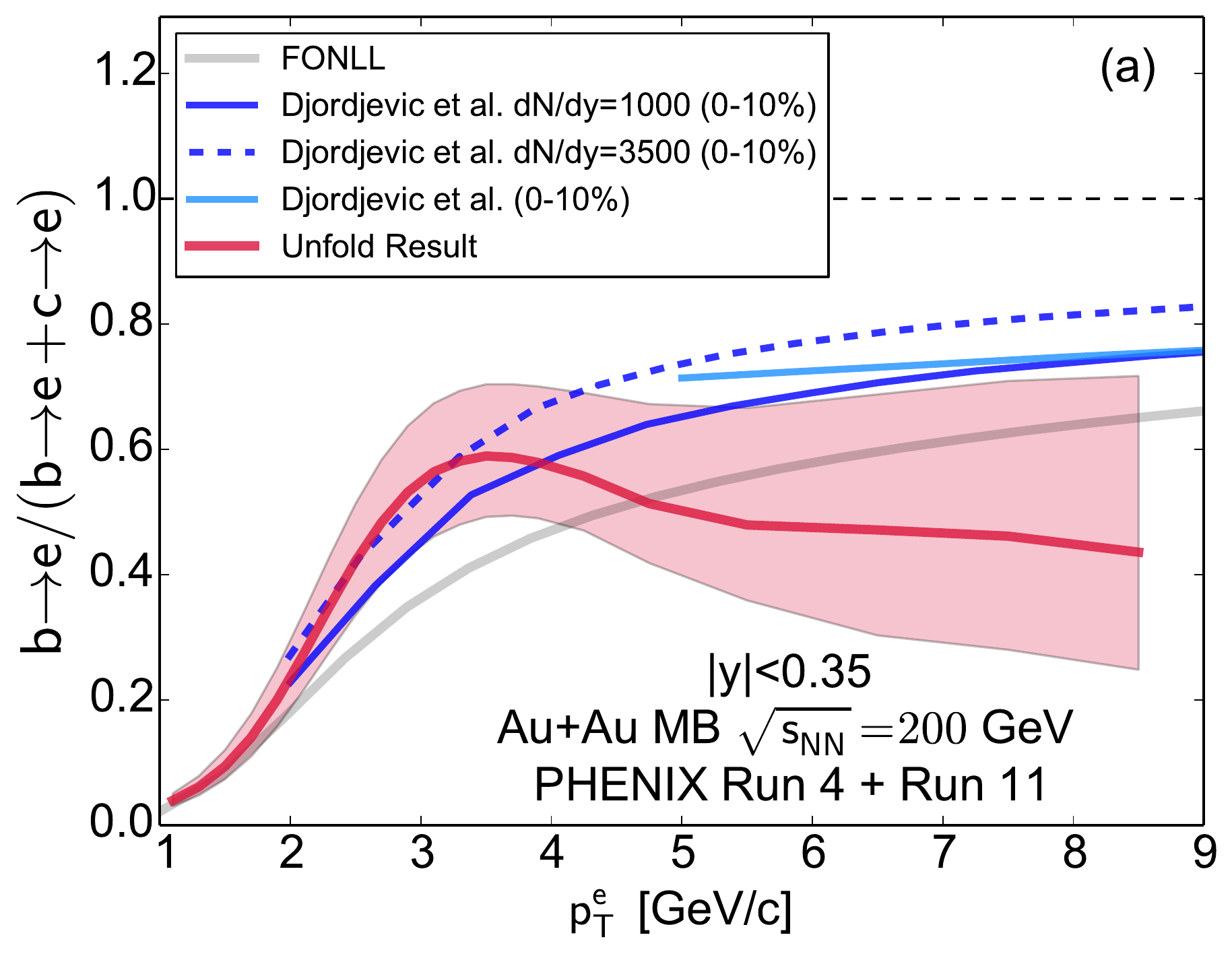}
	\includegraphics[width=0.49\linewidth]{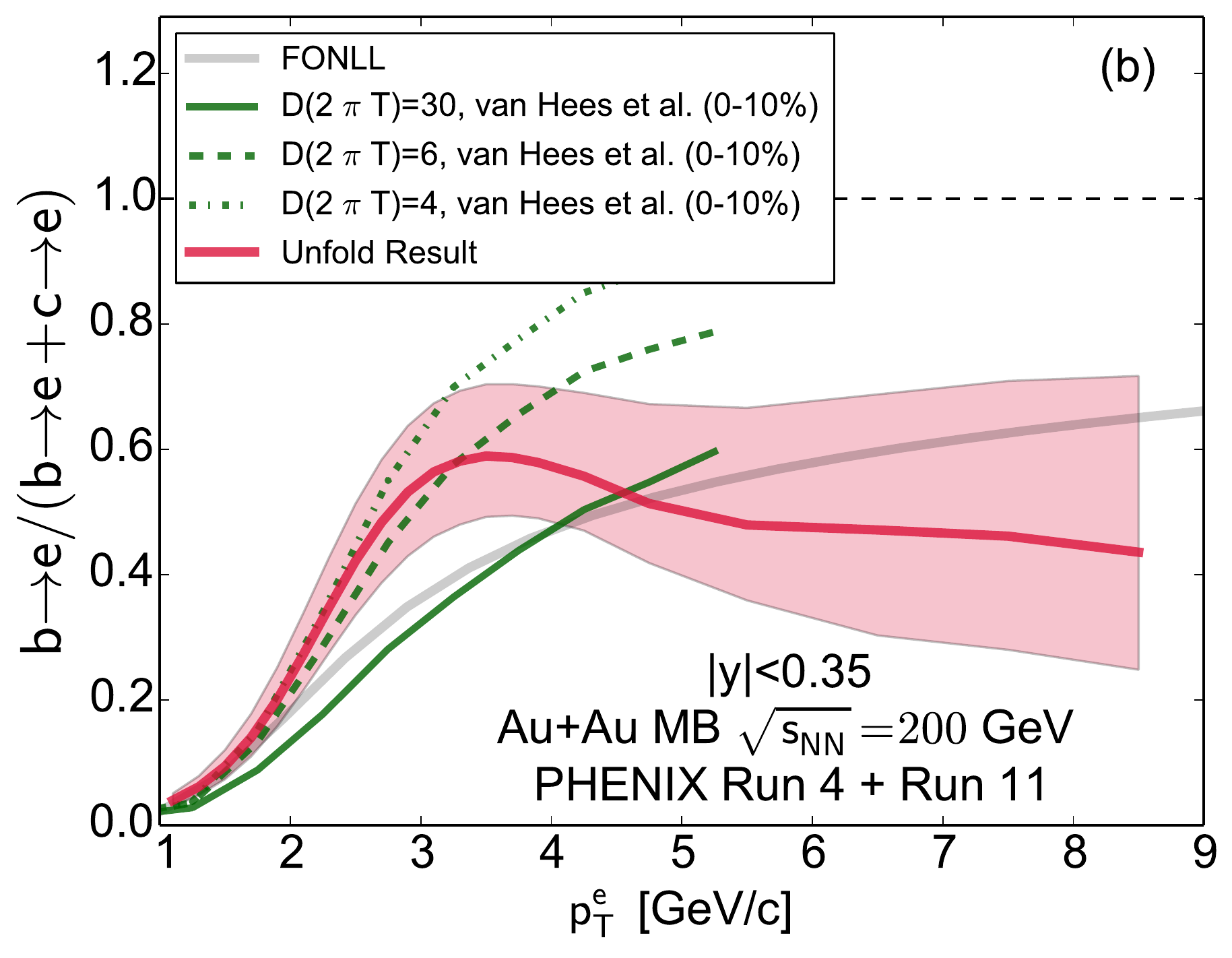}\\
	\includegraphics[width=0.49\linewidth]{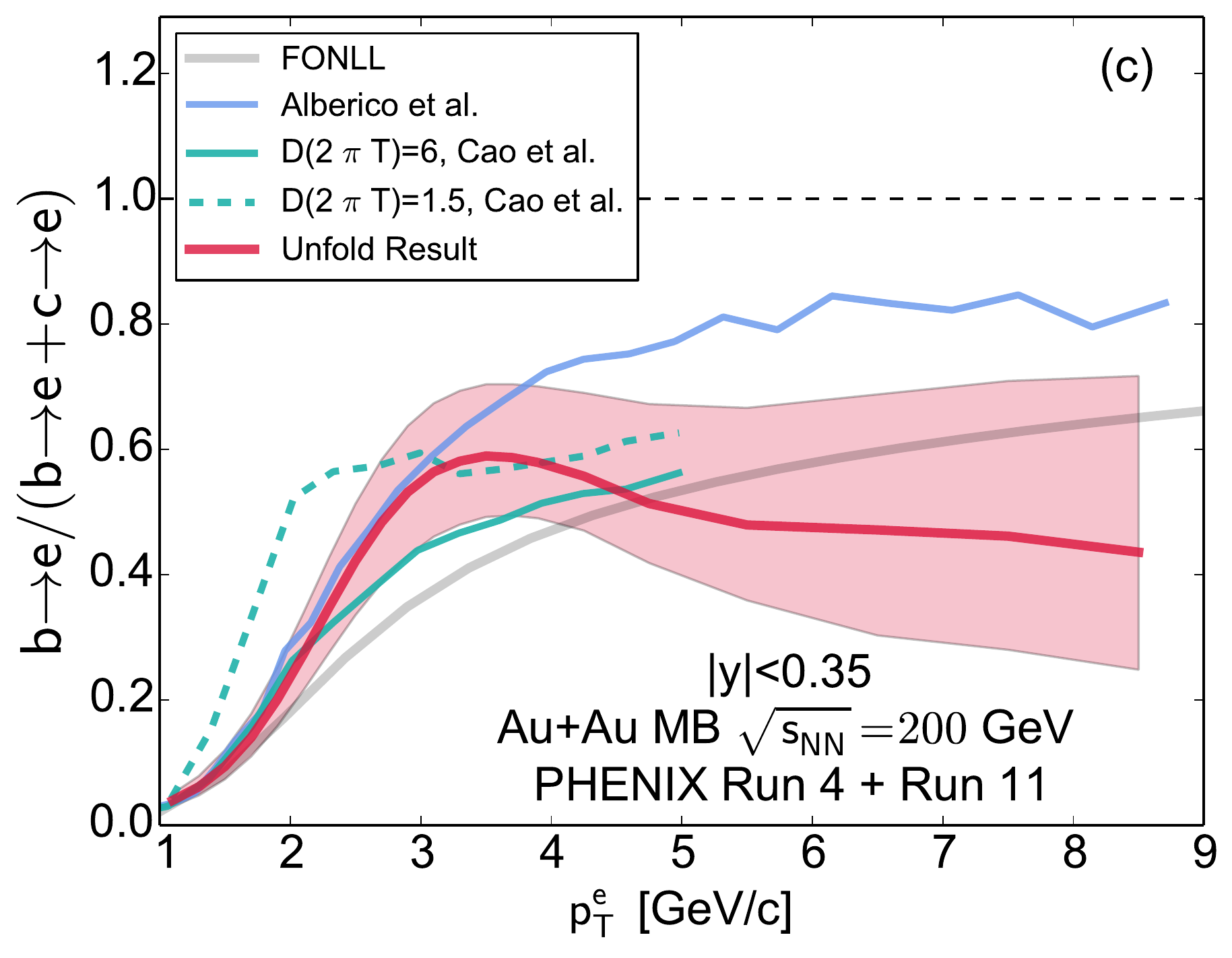}
	\includegraphics[width=0.49\linewidth]{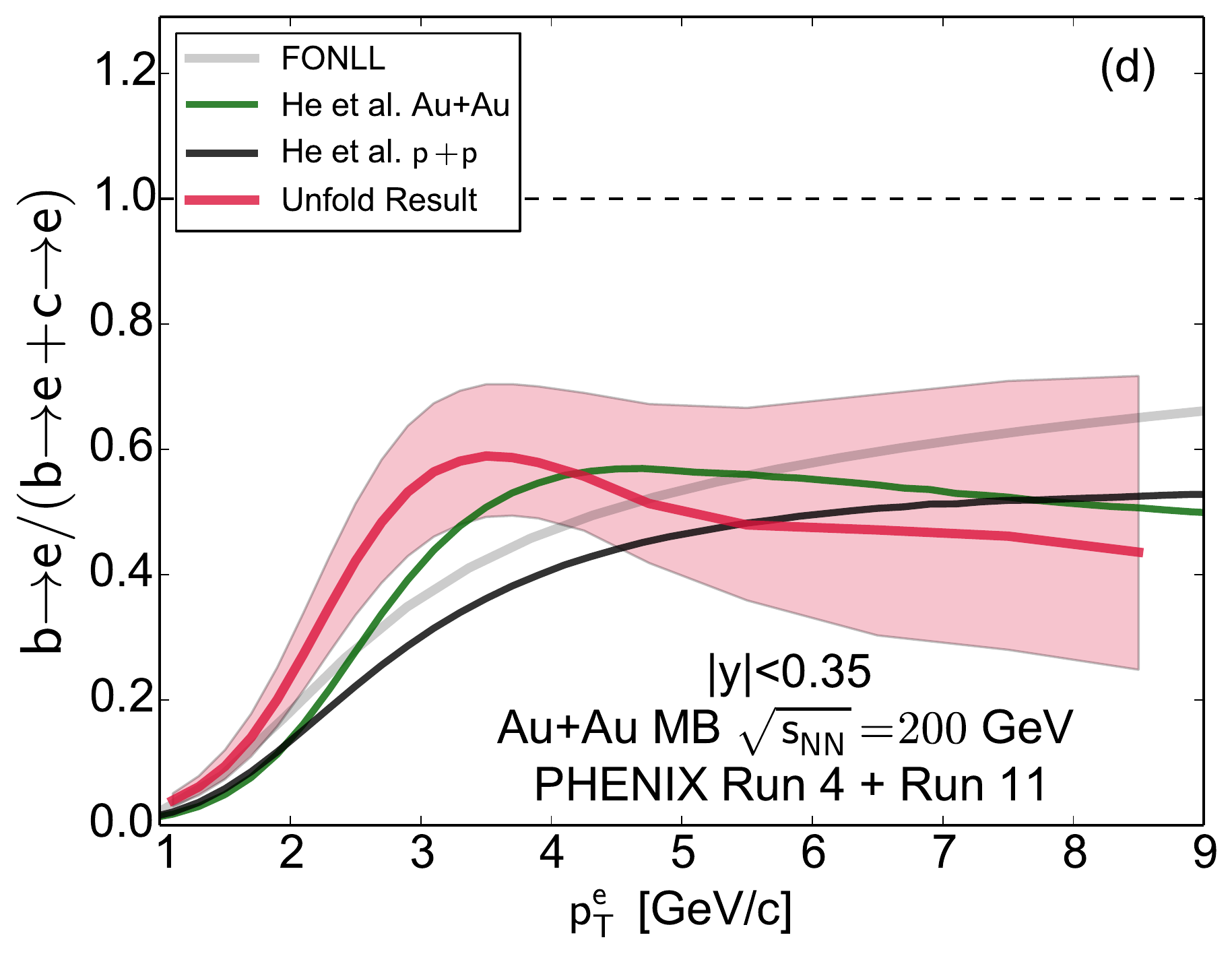}\\
\caption{(Color Online) 
Bottom electron fraction as a function of \pt compared to a series of 
model predictions detailed in the text.}
	\label{fig:bfrac-models}
\end{figure*}

Figure~\ref{fig:bfrac-models}(a) compares the bottom electron fraction 
from one class of calculations modeling only energy loss of these heavy 
quarks in medium.  In an early pQCD calculation by Djordjevic, Gyulassy,
Vogt, and Wicks~\cite{Djordjevic:2005db}, the authors apply the 
DGLV theory of radiative energy loss. They find that even for extreme opacities with 
gluon rapidity densities up to 3500, the bottom quark decay electrons 
dominate at high-\pt and that limits the single electron \raa to the range 
$0.5$--$0.6$ for $\pt>5$~\gev. Although this result is known to be higher 
than the PHENIX measured heavy flavor electron 
$R_{AA}$~\cite{Adare:2010de}, we show the $b \rightarrow e / (b 
\rightarrow e + c \rightarrow e)$ predictions for gluon rapidity densities 
of 1000 and 3500 in Fig.~\ref{fig:bfrac-models}(a). However, we do note 
that the calculations are for 0\%--10\% central collisions compared to the 
MB data, although the calculations span a factor of 3.5 range in the gluon 
density. We find that the calculations for both gluon rapidity densities 
are in good agreement with our results for $\pt<4$~\gev, but are slightly 
above and outside the uncertainty band on the unfolded result at higher 
\pt. More recent calculations in the same framework, but with the 
inclusion of collisional energy loss~\cite{Djordjevic:2014yka}, result in 
a heavy flavor electron high-\pt $R_{AA}$ closer to 0.3 and in reasonable 
agreement with previous PHENIX published results~\cite{Adare:2010de}. This 
updated prediction for the bottom electron fraction, also shown in 
Fig.~\ref{fig:bfrac-models}, gives a similar value to their previous 
result, but is only published for $\pt>5$ \gev.

Figure~\ref{fig:bfrac-models}(b) compares the bottom electron fraction 
from a calculation using a T-matrix approach by van Hees, Mannarelli, 
Greco, and Rapp~\cite{vanHees:2008gj}. The authors provided us with 
different results for 0\%--10\% central \auau collisions depending on the 
coupling of the heavy-quark to the medium. The coupling is encapsulated in 
the diffusion parameter $D$, where smaller values yield a stronger 
coupling.  Shown in Fig.~\ref{fig:bfrac-models}(b) are three results 
corresponding to three values of the parameter $D (2 \pi T) = 4, 6, 30$.  
The largest $D$ value, corresponding to the weakest coupling, yields 
almost no deviation from the \pp reference \fonll result, and the 
successively stronger coupling pushes the bottom fraction contribution 
higher and higher. We find that the calculations with $D (2 \pi T) = 4, 6$ 
are in good agreement with our result for $\pt<4$ \gev, but begin to 
diverge where the calculation stops at 5 \gev.

Figure~\ref{fig:bfrac-models}(c) compares the bottom electron fraction 
from another class of calculations which employ a combination of Langevin, 
or transport type modeling of heavy-quarks, in the bulk QGP with energy 
loss mechanisms that dominate at higher \pt.  In 
Ref.~\cite{Alberico:2011zy}, Alberico \textit{et al.} employ a Langevin 
calculation where a good match to the PHENIX heavy flavor electrons is 
found. It is notable that this calculation has a very strong suppression 
of charm decay electrons such that bottom contributions dominate even at 
modest \pt $\ge$ 2~\gev. The calculations are consistent with the data for 
$\pt<4$ \gev and over-predict the bottom contribution for higher \pt 
values.

Figure~\ref{fig:bfrac-models}(c) also compares the bottom electron 
fraction from another variant of the Langevin calculation by Cao 
\textit{et al.}, as detailed in Ref.~\cite{Cao:2012jt}. For this 
calculation, we show two results corresponding to two different input 
values $D (2 \pi T) = 1.5$ and 6. For the lower parameter, again stronger 
heavy-quark to medium coupling, there is a sharp rise in the bottom 
contribution which then flattens out. 
This feature is due to the increased collisional energy loss, which has a larger effect on the charm quarks, coupled with the strong radial flow effects enabling the heavier bottom quarks to dominate even at $\pt\sim2$ \gev.
%This feature is due to strong flow effects pushing the charm quarks to lower \pt, below 2~\gev, and enabling the heavier bottom quarks to dominate. 
These calculations use an impact 
parameter of $b=6.5$ fm, which should roughly correspond to MB collisions. 
We find that the calculation using the larger value of $D (2 \pi T) = 6.0$ 
is in reasonable agreement with the data across the calculated \pt range.

Lastly, Fig.~\ref{fig:bfrac-models}(d) shows a more recent calculation by 
He~\textit{et~al.} employing a T-matrix approach similar to that shown in 
Fig.~\ref{fig:bfrac-models}(b), but with a number of updates as described 
in Ref.~\cite{He:2014cla}. In this case the authors provided a calculation 
of the bottom electron fraction in both \pp and \auau at \sqsn=200~GeV, 
and we therefore do not calculate the bottom fraction using \fonll as a 
baseline. The calculation is performed for the 20\%--40\% centrality bin, 
which the authors find well represents MB. We find that the calculation 
under-predicts the bottom fraction for $\pt<3$ \gev, although it is worth 
noting that the calculation in \pp is also below the \fonll curve across 
the full \pt range. Above $\pt\sim3$ \gev the calculation is in agreement with 
the measurement. It is also worth noting that, of the models presented 
here, this is the only one that shows in \auau a slight decrease in the 
bottom fraction at high \pt.

There are numerous other calculations in the 
literature~\cite{Vitev:2007jj, Horowitz:2015dta, Horowitz:2009fw} that 
require mapping charm and bottom hadrons to electrons at midrapidity to 
make direct data comparisons. We look forward to soon being able to test 
these calculations with analysis of new PHENIX data sets.

%%========================================================================
	\section{Summary and Conclusions}
	\label{sec:conclusions}

This article has detailed the measurements of electrons as a function of 
\DCAR and \pt from \auau data taken at \sqsn=200~GeV in 2011 with the 
enhanced vertexing capabilities provided by the VTX detector. In 
conjunction with previous PHENIX results for the heavy flavor electron 
invariant yield as a function of \pt~\cite{Adare:2010de}, we perform an 
unfolding procedure to infer the parent charm and bottom hadron yields as 
a function of \pt. We find that this procedure yields consistent agreement 
between the heavy flavor electron invariant yield and the newly measured 
electron \DCAR distributions.

We find that the extracted $D^0$ yield vs \pt is in good agreement with 
that measured by STAR~\cite{Adamczyk:2014uip} over the complimentary \pt 
region. Without a proper \pp baseline extracted from a similar analysis it 
is difficult to make any quantitative statements about the charm or bottom 
hadron modification.

We compare the extracted bottom electron fraction to measurements in \pp 
collisions and find agreement between \auau and \pp for $\pt>4$ \gev 
within the large uncertainties on both measurements. The agreement between 
\auau and \pp coupled with the measured heavy flavor electron $R_{AA}$ 
strongly implies that electrons from charm and bottom hadron decays are 
suppressed. Using these components we calculate the nuclear modification 
for electrons from charm and bottom hadron decays and find that electrons 
from bottom hadron decays are less suppressed than those from charm hadron 
decays in the range \mbox{$3<\pt\ \gev<4$}. We further compare the bottom 
electron fraction to a variety of model calculations employing variously 
energy loss, Langevin transport, and T-matrix approaches. We find that 
there are a number of models which are in reasonable agreement with the 
extracted bottom electron fraction within the relatively large 
uncertainties.

We note that a significantly larger data set of \auau collisions at 
\sqsn=200~GeV was collected in 2014 with an improved performance of the 
VTX detector. The 2014 \auau data coupled with the \pp data taken in 2015 
should yield both an important baseline measurement of the bottom electron 
fraction and a more precise measurement in \auau.

%====================================================================
	\section*{ACKNOWLEDGMENTS}   % Run-14 long form for all journals

We thank the staff of the Collider-Accelerator and Physics
Departments at Brookhaven National Laboratory and the staff of
the other PHENIX participating institutions for their vital
contributions.  We acknowledge support from the
Office of Nuclear Physics in the
Office of Science of the Department of Energy,
the National Science Foundation,
Abilene Christian University Research Council,
Research Foundation of SUNY, and
Dean of the College of Arts and Sciences, Vanderbilt University
(U.S.A),
Ministry of Education, Culture, Sports, Science, and Technology
and the Japan Society for the Promotion of Science (Japan),
Conselho Nacional de Desenvolvimento Cient\'{\i}fico e
Tecnol{\'o}gico and Funda\c c{\~a}o de Amparo {\`a} Pesquisa do
Estado de S{\~a}o Paulo (Brazil),
Natural Science Foundation of China (P.~R.~China),
Croatian Science Foundation and
Ministry of Science, Education, and Sports (Croatia),
Ministry of Education, Youth and Sports (Czech Republic),
Centre National de la Recherche Scientifique, Commissariat
{\`a} l'{\'E}nergie Atomique, and Institut National de Physique
Nucl{\'e}aire et de Physique des Particules (France),
Bundesministerium f\"ur Bildung und Forschung, Deutscher
Akademischer Austausch Dienst, and Alexander von Humboldt Stiftung (Germany),
National Science Fund, OTKA, K\'aroly R\'obert University College,
and the Ch. Simonyi Fund (Hungary),
Department of Atomic Energy and Department of Science and Technology (India),
Israel Science Foundation (Israel),
Basic Science Research Program through NRF of the Ministry of Education (Korea),
Physics Department, Lahore University of Management Sciences (Pakistan),
Ministry of Education and Science, Russian Academy of Sciences,
Federal Agency of Atomic Energy (Russia),
VR and Wallenberg Foundation (Sweden),
the U.S. Civilian Research and Development Foundation for the
Independent States of the Former Soviet Union,
the Hungarian American Enterprise Scholarship Fund,
and the US-Israel Binational Science Foundation.

%================================================================
%	\appendix
	\section*{APPENDIX:  Detailed Normalization of electron 
background components}
	\label{app:norm}

This appendix details the calculation of the normalizations for the 
background components:
\begin{itemize}
	\item Photonic electrons
	\item Kaon decay electrons
	\item Heavy quarkonia decay electrons
\end{itemize}
using the bootstrap method described in Sec.~\ref{sec:norm}.  We first 
determine the fraction of nonphotonic electrons, $F_{\rm NP}$. We then 
calculate the normalization of Dalitz and conversion components followed 
by the normalization of $K_{e3}$ and quarkonia components.

\subsection{Fraction of nonphotonic electrons $F_{\rm NP}$}
\label{sec:FNP}

We first determine $F_{\rm NP}$, the fraction of nonphotonic electrons to 
inclusive electrons after the application of all analysis cuts, including 
the conversion veto cut. Note that nonphotonic electrons include 
contributions from heavy flavor semi-leptonic decays, quarkonia decays, 
and kaon decays. Photonic electrons are from $\pi^{0}$ and $\eta$ Dalitz 
decays and photon conversions.

$F_{\rm NP}$ in the 2011 data can be determined using the published 2004 
result~\cite{Adare:2010de} as follows. Let $Y_{\rm NP}$ be the yield of 
nonphotonic electrons and $Y_{Dalitz}$ the yield of electrons from Dalitz 
decays. Note that both $Y_{\rm NP}$ and $Y_{Dalitz}$ are independent of the 
year of data taking. In the PHENIX 2004 \auau data run, the ratio of the 
nonphotonic electron yield to the photonic electron yield 
($R_{\rm NP}^{2004}$) was measured.  The relation of $Y_{\rm NP}$ and $Y_{Dalitz}$ 
is as follows:
\begin{equation}
Y_{\rm NP}= R_{\rm NP}^{2004} (1 + R^{2004}_{CD}) \times Y_{Dalitz}, 
\end{equation}

where $R^{2004}_{CD}$ represents the ratio of conversion electron yield to 
Dalitz electron yield in the 2004 PHENIX detector. It is calculated as
\begin{equation}
R^{2004}_{CD}=
\sum_{i=\pi^0,\eta,\gamma} R_{CD}^{2004}(i)\cdot r_{Dalitz}(i)).
\end{equation}
Here $R_{CD}^{2004}(i)$ is the ratio of conversion electrons to electrons 
from Dalitz decays in the 2004 PHENIX detector calculated by a full \geant 
simulation. The factors
\begin{itemize}
        \item $r_{Dalitz}(\pi^0)$
        \item $r_{Dalitz}(\eta)$
        \item $r_{Dalitz}(\gamma)$
\end{itemize}
are the fractional contributions of $\pi^0$, $\eta$, and direct photon 
contribution to the total Dalitz decays, respectively\footnote{ Here we 
include internal conversion of direct photon in Dalitz decays. Note that 
the Dalitz decay of $\pi^0$ ($\eta$) is caused by internal conversion of 
one of two decay photons in $\pi^0 (\eta) \rightarrow \gamma\gamma$.}. We 
only consider the contributions of $\pi^0$, $\eta$, and $\gamma_{dir}$ 
(direct photon) since the sum of other contributions is small (5\% or 
less). Thus they are normalized such that

\begin{equation}
\sum_i r_{Dalitz}(i)=1. 
\end{equation}

\begin{figure}[!hbt]
\includegraphics[width=1.0\linewidth]{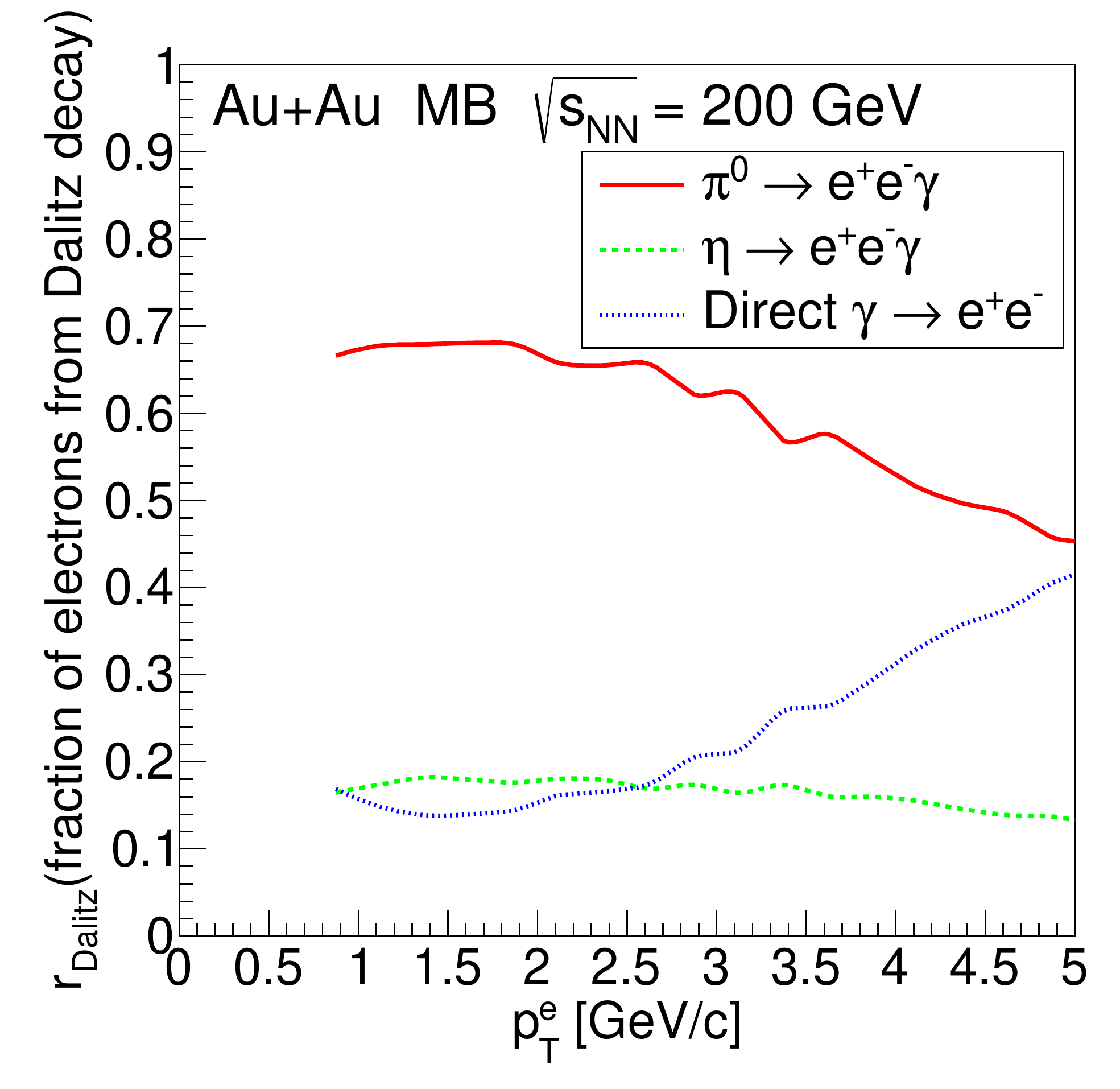}
\caption{\label{fig:dalitz_ratio} (Color Online) 
The fraction of $\pi^0$, $\eta$, and direct photon Dalitz decay electrons 
in all Dalitz electrons as a function of electron \pt (\pte). }
\end{figure}

Figure~\ref{fig:dalitz_ratio} shows $r_{Dalitz}$ for $\pi^0$, $\eta$, and 
direct photon as a function of transverse momentum of the electrons for 
MB \auau collisions at 200~GeV. The ratios are calculated from 
the invariant yield of $\pi^0$\cite{Adare:2008qa}, 
$\eta$\cite{Adare:2010dc}, and direct 
photons\cite{Afanasiev:2012dg,Adare:2008ab}.

%%%%%%%%%%%%%%%%%%%%%%%%%%%%%%%%%%%%%%%%%%%%%% Fig_21
In the 2011 data set the observed electron yields from conversion and 
Dalitz decays are modified by the electron survival probability after the 
conversion veto cut is applied. The yield of photonic electrons which pass 
the conversion veto $(Y_{P}^{2011})$ is
\begin{eqnarray}
Y_{P}^{2011} 
&=& R^{2011}_{PD}\times Y_{Dalitz}, \\
R^{2011}_{PD}&=&\sum_{i=\pi^0,\eta,\gamma}
         \left(S_{D}(i)+S_{C}\cdot R_{CD}^{2011}(i)\right)r_{Dalitz}(i), 
\end{eqnarray}
where $S_{C}$ is the survival probability of conversion electrons, 
$S_{D}(\pi^0), S_{D}(\eta), S_{D}(\gamma)$ are survival probabilities of 
Dalitz decay electrons from $\pi^0$, $\eta$, and direct photons, 
respectively, as shown in Fig.~\ref{fig:survival_rate}. $R_{CD}^{2011}(i)$ 
($i=\pi^0,\eta,\gamma$)is the ratio of conversion electrons to Dalitz 
electrons for particle $i$ in the 2011 PHENIX detector after the addition 
of the VTX and the replacement of the beam pipe. It is determined to be 
$R_{CD}^{2011}(i)\approx 1.10$ from full \geant simulations.

%%%%%%%%%%%%%%%%%%%%%%%%%%%%%%%%%%%%%%%%%%%%%% Fig_22
\begin{figure}[!hbt]
\includegraphics[width=1.0\linewidth]{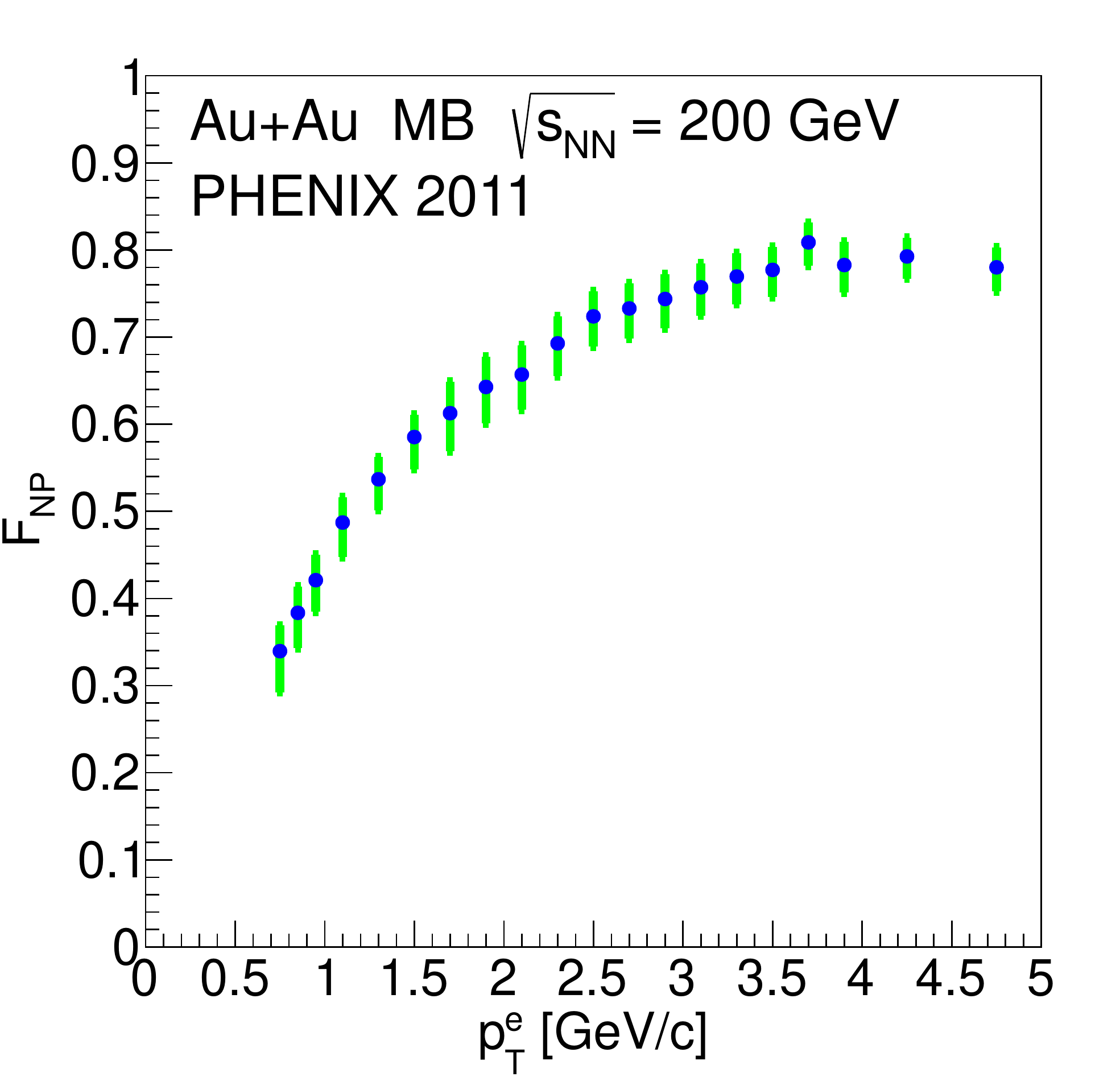}
\caption{\label{fig:FNP_method1} (Color Online) 
The fraction of nonphotonic electrons to inclusive electrons 
as a function of electron \pt (\pte).
}
\end{figure}

%%%%%%%%%%%%%%%%%%%%%%%%%%%%%%%%%%%%%%%%%%%%%% Fig_23
\begin{figure}[!hbt]
\includegraphics[width=1.0\linewidth]{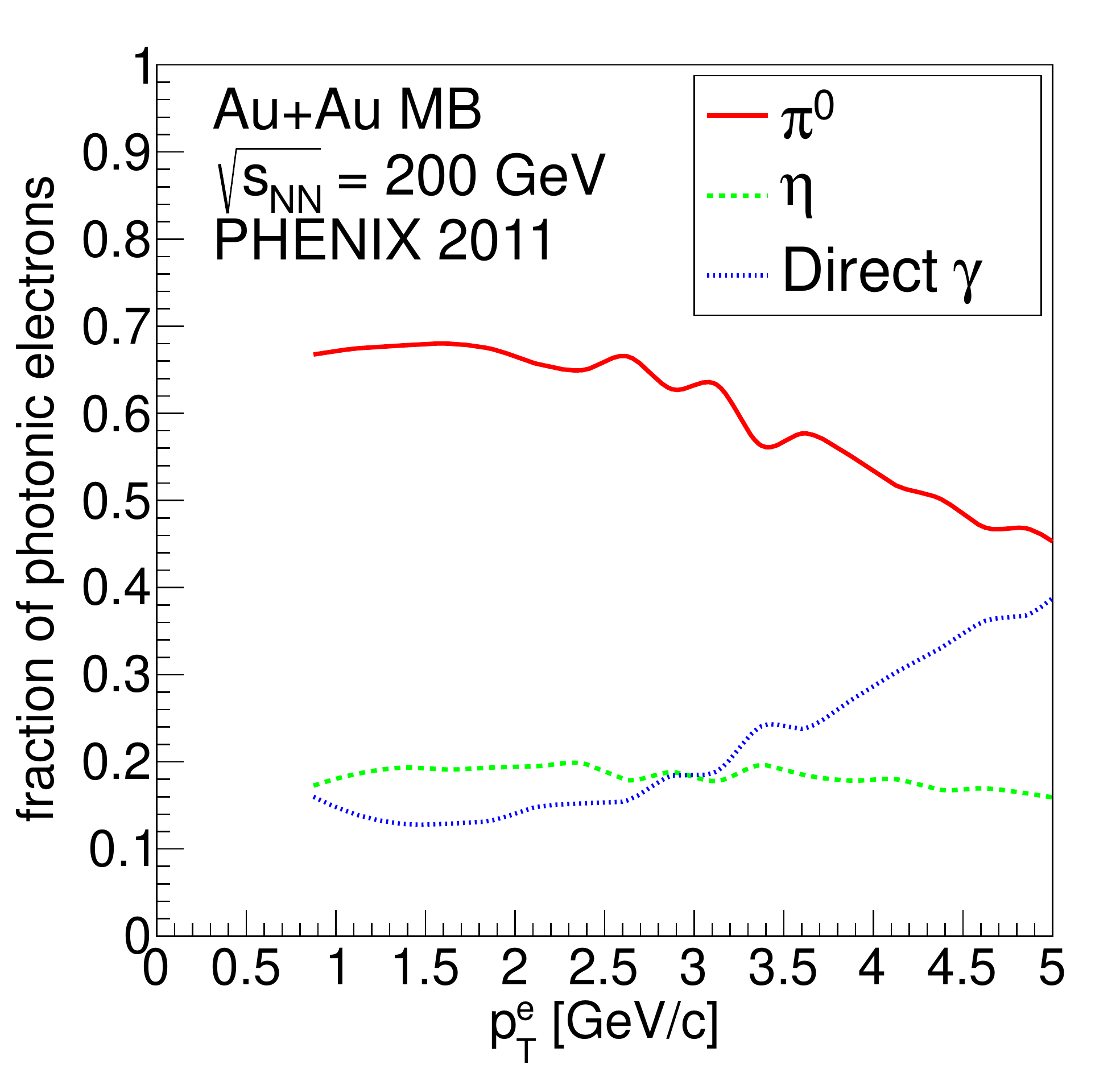}
\caption{\label{fig:photonic_ratio} (Color Online) 
The fraction of $\pi^0$, $\eta$, and direct photon electrons in all 
photonic electrons as a function of electron \pt (\pte).
}
\end{figure}

The fraction of nonphotonic electrons to inclusive electrons can then be 
calculated as

\begin{eqnarray}
F_{\rm NP} &=& \frac{Y_{\rm NP}}{Y_{\rm NP}+Y_{P}^{2011}}\\
&=& \frac{R^{2004}_{\rm NP}(1+R^{2004}_{CD})}{R^{2004}_{\rm NP}(1+R^{2004}_{CD})+R^{2011}_{PD}}
\end{eqnarray}

The resulting $F_{\rm NP}$ as a function of \pte and the calculated systematic 
uncertainties due to the uncertainties on the input yields is shown in 
Fig.~\ref{fig:FNP_method1}.  With $F_{\rm NP}$ in hand, we obtain the number 
of photonic electrons, $N^e_{P}$, and the number of nonphotonic 
electrons, $N^e_{\rm NP}$ as

\begin{eqnarray}
N_P^e &=& N_e (1-F_{\rm NP}) \\
N_{\rm NP}^e &=& N_e F_{\rm NP},
\end{eqnarray}
where $N^e$ is the number of electrons with conversion veto after the 
subtraction of the hadronic contamination and random background.

	\subsection{Normalization of Dalitz and conversion components}

In the previous section we obtained $N^e_{P}$, the number of photonic 
electrons in the data after the conversion veto cut. There are two 
components in the photonic electrons ($N_P^e$).

\begin{enumerate}
    \item Electrons from Dalitz decays ($\pi^0+\eta+\gamma$)
    \item Electrons from conversions in the beam pipe and B0
\end{enumerate}

In the next step, we determine the normalization of Dalitz and conversions 
separately. This is needed since the shape of \DCAR distribution of Dalitz 
and conversions are different.

After application of the conversion veto cut, we have

\begin{eqnarray}
N^e_C(i) &=& S_C R_{CD}^{2011}(i) (1-\delta_{random})\epsilon_A Y_{Dalitz},\\
N^e_D(i) &=& S_D(i) (1-\delta_{random}) \epsilon_A Y_{Dalitz},\\
(i &=& \pi^0,\eta,\gamma)
\end{eqnarray}
where $N_C^e(i)$ and $N_D^e(i)$ are the number of electrons from 
conversions and Dalitz from particle $i$ after the conversion veto cut, 
respectively; $\delta_{random}$ is the common reduction factor of tracks 
due to random hits in the windows of the conversion veto cut; and 
$\epsilon_A$ is the efficiency and acceptance without the conversion veto 
cut. Since the number of photonic electron is $N^e_P(i) = 
N^e_D(i)+N^e_C(i)$, the fraction of conversions and Dalitz decays in the 
photonic electrons are

\begin{eqnarray}
\frac{N^e_C(i)}{N^e_P(i)} &=& \frac{S_C R_{CD}(i)}{S_D(i) +S_C R_{CD}^{2011}(i)},\\
\frac{N^e_D(i)}{N^e_P(i)} &=& \frac{S_D(i)}{S_D(i) +S_C R_{CD}^{2011}(i)},
\end{eqnarray}

The fraction of electrons from conversions ($N^e_C/N^e_P$) and Dalitz 
($N^e_D/N^e_P$) is the average of these fractions, thus:

\begin{eqnarray}
N^e_{C}&=& N_P^{e}\sum_{i=\pi^0,\eta,\gamma}r^{ph}(i)\frac{S_C R_{CD}^{2011}(i)}{S_D(i)+S_C R_{CD}^{2011}(i)}\\
N^e_{D}&=& N_P^{e}\sum_{i=\pi^0,\eta,\gamma}r^{ph}(i)\frac{S_D(i)}{S_D(i)+S_C R_{CD}^{2011}(i)},
\end{eqnarray}
where $r^{ph}(i),(i = \pi^0, \eta, \gamma) $ is the relative contribution 
of electrons from (conversion + Dalitz decay) for particle $i$ after 
application of conversion veto cut. Figure~\ref{fig:photonic_ratio} shows 
$r^{ph}(i)$ $(i=\pi^0,\eta,\gamma)$ as a function of \pte.
The conversion contributions are nearly the same for $\pi^0, \eta$ and $\gamma$, 
and effectively cancel when calculating the ratio. 
Therefore, $r^{ph}$ (Fig.~\ref{fig:photonic_ratio}) is almost identical with  
$r_{Dalitz}$ (Fig.~\ref{fig:dalitz_ratio}).

	\subsection{Normalization of $K_{e3}$ and quarkonia components}

The ratio of electrons from kaons to all nonphotonic electrons before the 
application of the conversion veto cut, $\delta_K$, is calculated from the 
ratio of the nonphotonic electron yield to the electron yield from 
kaons~\cite{Adare:2010de}. Compared to Ref.~\cite{Adare:2010de}, we find 
that $\sim50\%$ of electrons from kaon decays are removed by \DCAR and 
\DCAZ cuts as well as the method used to subtract random background, which 
contains some real electrons from kaon decays.

The ratio of electrons from $J/\psi$ decays to all nonphotonic electrons 
before the application of the conversion veto cut, $\delta_{J/\psi}$, is 
taken from Ref.~\cite{Adare:2010de}. The survival rate for electrons from 
$J/\psi$ decays , $S_{J/\psi}$, is taken to be unity, while the survival 
rate for $K_{e3}$ decays, $S_{K}$, is taken to be the same value as that 
for electrons from charm and bottom decays (namely, $S_{\rm HF}$). See 
Sec.~\ref{sec:photonic_electrons} for details.

After application of conversion veto cut, the normalizations of these two 
nonphotonic electron components are described by
\begin{small}
\begin{eqnarray}
N_{J/\psi}^e &= N_{\rm NP}^e\frac{\delta_{J/\psi}S_{J/\psi}} 
{\delta_{J/\psi} S_{J/\psi}+\delta_{K}S_{K}+(1-\delta_{J/\psi}-\delta_{K})S_{\rm HF}} \\
N_K^e &=  N_{\rm NP}^e\frac{\delta_{K}S_{K}} 
{\delta_{J/\psi} S_{J/\psi}+\delta_{K}S_{K}+(1-\delta_{J/\psi}-\delta_{K})S_{\rm HF}}
\end{eqnarray}
\end{small}

%%%%%%%%%%%%%%%%%%%%%%%%%%%  References 

%\bibliography{ppg182x1}   

 
\end{document}